\documentclass[12pt,letterpaper]{article}
\usepackage[T1]{fontenc}
\usepackage[utf8]{inputenc}
\usepackage{lmodern,microtype}
\usepackage[letterpaper,margin=1in]{geometry}
\usepackage{amsmath,amssymb,amsthm,booktabs,tabularx,array}
\usepackage{enumitem}
\usepackage{graphicx}
\usepackage{setspace}
\usepackage{longtable,etoolbox,caption}
\usepackage{needspace}
\usepackage[authoryear,round]{natbib}
\usepackage[hidelinks]{hyperref}
\newcommand{\JBESbodyspacing}{\singlespacing\normalsize}
\newcommand{\BBPappendixspacing}{\singlespacing\normalsize}
\SetSinglespace{1}
\setdisplayskipstretch{1}
\newcommand{\BBPdisplayspacing}{%
  \setlength{\abovedisplayskip}{6pt plus 2pt minus 1pt}%
  \setlength{\belowdisplayskip}{6pt plus 2pt minus 1pt}%
  \setlength{\abovedisplayshortskip}{3pt plus 1pt}%
  \setlength{\belowdisplayshortskip}{4pt plus 1pt minus 1pt}}
\appto\normalsize{\BBPdisplayspacing}
\AtBeginDocument{%
  \JBESbodyspacing
  \setlength{\footnotesep}{7pt}%
  \setlength{\skip\footins}{10pt plus 2pt minus 2pt}%
  \typeout{JBES body-font=\csname f@size\endcsname; baseline=\the\baselineskip}%
  \typeout{JBES textwidth=\the\textwidth; textheight=\the\textheight}%
}
\AtBeginEnvironment{tabular}{\setstretch{1}\normalsize}
\AtBeginEnvironment{tabular*}{\setstretch{1}\normalsize}
\AtBeginEnvironment{tabularx}{\setstretch{1}\normalsize
  \ifdefined\BBProadmaptable\small\fi}
\AtBeginEnvironment{longtable}{\setstretch{1}\normalsize}
\DeclareCaptionFont{jbes}{\setstretch{1}\normalsize}
\newcommand{\BBPtablenotes}[1]{%
  \par\smallskip\begingroup\setstretch{1}\footnotesize
  \raggedright\noindent\textit{Notes:} #1\par\endgroup}
\setlist[enumerate]{itemsep=2pt,topsep=4pt}
\newtheoremstyle{plain}{6pt}{6pt}{\itshape}{}{\bfseries}{.}{.5em}{}
\newtheoremstyle{definition}{6pt}{6pt}{}{}{\bfseries}{.}{.5em}{}
\newtheoremstyle{remark}{6pt}{6pt}{}{}{\itshape}{.}{.5em}{}
\theoremstyle{plain}
\newtheorem{theorem}{Theorem}
\newtheorem{proposition}{Proposition}

\numberwithin{equation}{section}
\numberwithin{table}{section}
\numberwithin{figure}{section}
\theoremstyle{definition}
\newtheorem{assumption}{Assumption}
\newtheorem{algorithm}{Algorithm}
\theoremstyle{remark}
\newtheorem*{remark}{Remark}
\newcommand{\Pn}{P_n}
\newcommand{\E}{E}
\newcommand{\Var}{\operatorname{Var}}
\newcommand{\diag}{\operatorname{diag}}
\newcommand{\ind}{\mathbf 1}
\newcommand{\R}{\mathbb R}
\newcommand{\norm}[1]{\lVert#1\rVert}

\newcolumntype{L}{>{\raggedright\arraybackslash}X}
\title{Dealing with Logs and Zeros in Regression Models}
\author{David Benatia\textsuperscript{1} \and
  Christophe Bell\'ego\textsuperscript{2} \and
  Louis-Daniel Pape\textsuperscript{3}}
\newcommand{\BBPaffiliations}{%
  \textsuperscript{1}HEC Montr\'eal, D\'epartement d'\'Economie Appliqu\'ee\par
  \textsuperscript{2}CREST (UMR 9194), ENSAE, Institut Polytechnique de Paris\par
  \textsuperscript{3}CREST (UMR 9194), Telecom Paris, Institut Polytechnique de Paris\par}
\date{16 November 2026}
\ifdefined\BBPanonymous
  \author{}
  \renewcommand{\BBPaffiliations}{}
  \hypersetup{pdfauthor={},pdftitle={Dealing with Logs and Zeros in Regression Models}}
\else
  \hypersetup{pdfauthor={David Benatia, Christophe Bellego, Louis-Daniel Pape},
    pdftitle={Dealing with Logs and Zeros in Regression Models}}
\fi
\renewenvironment{abstract}{\par\noindent\textbf{Abstract}\par}{\par}
\makeatletter
\renewcommand{\maketitle}{%
  \begin{center}
    {\Large\bfseries\@title\par}
    \medskip
    \normalfont\def\and{\qquad}\@author\par
    \ifdefined\BBPanonymous\else
      \smallskip
      {\singlespacing\footnotesize\BBPaffiliations}%
    \fi
    \smallskip
    {\small\@date\par}
  \end{center}}
\makeatother
\newcommand{\BBPexternallabel}[3]{%
  \expandafter\gdef\csname BBPexternal@#1\endcsname{}%
  \newlabel{#1}{{#2}{#3}{}{}{}}}
\AtBeginDocument{%
  \NewCommandCopy\BBPoriginalref\ref
  \NewCommandCopy\BBPoriginalpageref\pageref
  \NewCommandCopy\BBPoriginalhyperref\hyperref
  \RenewDocumentCommand{\ref}{s m}{%
    \IfBooleanTF{#1}{\BBPoriginalref*{#2}}{%
      \ifcsname BBPexternal@#2\endcsname
        \BBPoriginalref*{#2}%
      \else\BBPoriginalref{#2}\fi}}%
  \RenewDocumentCommand{\pageref}{s m}{%
    \IfBooleanTF{#1}{\BBPoriginalpageref*{#2}}{%
      \ifcsname BBPexternal@#2\endcsname
        \BBPoriginalpageref*{#2}%
      \else\BBPoriginalpageref{#2}\fi}}%
  \RenewDocumentCommand{\hyperref}{o m}{%
    \IfNoValueTF{#1}{\BBPoriginalhyperref{#2}}{%
      \ifcsname BBPexternal@#1\endcsname
        #2%
      \else\BBPoriginalhyperref[#1]{#2}\fi}}}

\makeatletter
\newcommand{\BBPsupplementanchors}{%
  \global\Hy@linkcounter=10000\relax
  \setcounter{Hfootnote}{0}%
  \renewcommand{\theHHfootnote}{S.\arabic{Hfootnote}}}
\newcommand{\BBPresultpart}[1]{%
  \@for\BBPcounter:=theorem,proposition,corollary,lemma,assumption,algorithm\do{%
    \setcounter{\BBPcounter}{0}%
    \expandafter\edef\csname the\BBPcounter\endcsname{%
      #1\noexpand\arabic{\BBPcounter}}%
    \expandafter\edef\csname theH\BBPcounter\endcsname{%
      #1\noexpand\arabic{\BBPcounter}}}}
\newcommand{\BBPmainappendix}{%
  \clearpage
  \BBPappendixspacing
  \appendix
  \global\Hy@linkcounter=5000\relax
  \setcounter{footnote}{0}%
  \setcounter{Hfootnote}{0}%
  \renewcommand{\theHHfootnote}{A.\arabic{Hfootnote}}%
  \BBPresultpart{A}%
  \renewcommand{\thesection}{\Alph{section}}%
  \renewcommand{\theHsection}{\Alph{section}}}

\makeatother

\BBPexternallabel{app:application-all-coefficients}{S6.8}{S112}
\BBPexternallabel{app:application-half-power}{S6.6}{S109}
\BBPexternallabel{app:application-positive-only}{S6.7}{S109}
\BBPexternallabel{app:applications}{S6}{S105}
\BBPexternallabel{app:applications-inference}{S6.4}{S107}
\BBPexternallabel{app:applications-reestimate}{S6.2}{S105}
\BBPexternallabel{app:comparisons}{S1}{S2}
\BBPexternallabel{app:comparisons-estimating-equations}{S1.4}{S9}
\BBPexternallabel{app:comparisons-heterogeneity}{S1.3}{S8}
\BBPexternallabel{app:comparisons-supporting}{S1}{S2}
\BBPexternallabel{app:empirical-practice}{S7}{S122}
\BBPexternallabel{app:fe-cancel-identification}{S3.3.1}{S46}
\BBPexternallabel{app:fe-cancel-inference}{S3.1.3}{S41}
\BBPexternallabel{app:fe-computation}{S3.2.1}{S43}
\BBPexternallabel{app:fe-decision-tables}{S3.4.8}{S74}
\BBPexternallabel{app:fe-dyadic-scope}{S3.3.2}{S48}
\BBPexternallabel{app:fe-fixed-design}{S3.2.2}{S43}
\BBPexternallabel{app:fe-gamma-inference}{S3.1.4}{S41}
\BBPexternallabel{app:fe-gamma-twoway-bias}{S3.4.7}{S73}
\BBPexternallabel{app:fe-growing-covariance-refinement}{S3.4.3}{S57}
\BBPexternallabel{app:fe-growing-delete-covariance}{S3.4.2}{S55}
\BBPexternallabel{app:fe-growing-derivatives}{S3.4.1}{S48}
\BBPexternallabel{app:fe-growing-design}{S3.4.6}{S70}
\BBPexternallabel{app:fe-growing-mixed-scope}{S3.3.2}{S48}
\BBPexternallabel{app:fe-growing-selection}{S3.2.3}{S44}
\BBPexternallabel{app:fe-interactions}{S3.3.2}{S48}
\BBPexternallabel{app:fe-mixed}{S3.3.2}{S48}
\BBPexternallabel{app:fe-postestimation}{S3.6.2}{S82}
\BBPexternallabel{app:fe-power-cancellation}{S3.6.1}{S80}
\BBPexternallabel{app:fe-regime-decisions}{S3.4.5}{S70}
\BBPexternallabel{app:fe-short-panel}{S3.3.2}{S48}
\BBPexternallabel{app:fe-small-regions}{S3.5}{S78}
\BBPexternallabel{app:fe-supp-computation}{S3.2}{S43}
\BBPexternallabel{app:fe-supp-growing}{S3.4}{S48}
\BBPexternallabel{app:fe-supp-panels}{S3.3}{S46}
\BBPexternallabel{app:fe-supp-power}{S3.6}{S80}
\BBPexternallabel{app:fe-supplement}{S3}{S39}
\BBPexternallabel{app:fe-supporting}{S3.1}{S39}
\BBPexternallabel{app:fe-weighted-regularity}{S3.1.1}{S40}
\BBPexternallabel{app:gpml-completion}{S2.4.1}{S30}
\BBPexternallabel{app:gpml-roots}{S2.1}{S15}
\BBPexternallabel{app:gpml-specialization}{S2.7}{S38}
\BBPexternallabel{app:gpml-statistics}{S2.3}{S27}
\BBPexternallabel{app:gpml-units}{S2.7}{S38}
\BBPexternallabel{app:iols-completion-supplement}{S2.4}{S29}
\BBPexternallabel{app:iols-covariance}{S2.6}{S36}
\BBPexternallabel{app:iols-exact-curvature}{S2.1.3}{S18}
\BBPexternallabel{app:iols-fixed-design-example}{S2.4.2}{S31}
\BBPexternallabel{app:iols-generic-completion}{S2.4.1}{S30}
\BBPexternallabel{app:iols-ols-step-trials}{S2.4.2}{S31}
\BBPexternallabel{app:iols-phase1-supplement}{S2.2}{S19}
\BBPexternallabel{app:iols-supplement}{S2}{S15}
\BBPexternallabel{app:iols-supporting-results}{S2}{S15}
\BBPexternallabel{app:iols-tau-continuation}{S2.5}{S32}
\BBPexternallabel{app:iv}{S4}{S84}
\BBPexternallabel{app:iv-numerics}{S4.1}{S84}
\BBPexternallabel{app:iv-pilot}{S4.1.2}{S86}
\BBPexternallabel{app:iv-statistics}{S4.2}{S87}
\BBPexternallabel{app:iv-supporting}{S4.1}{S84}
\BBPexternallabel{app:log-like-supplement}{S1.5}{S11}
\BBPexternallabel{app:mean-log-derivation}{S1.3.1}{S8}
\BBPexternallabel{app:observation-influence}{S1.4.2}{S9}
\BBPexternallabel{app:popular-fix-derivation}{S1.5.1}{S11}
\BBPexternallabel{app:power-shift-local-response}{S1.5.4}{S14}
\BBPexternallabel{app:shift-log-roots}{S1.5.2}{S11}
\BBPexternallabel{app:shift-log-standardization}{S1.5.5}{S14}
\BBPexternallabel{app:sim-completed-four-way}{S5.1}{S89}
\BBPexternallabel{app:sim-design}{S5}{S89}
\BBPexternallabel{app:sim-fe-covariance-adjustments}{S5.2.1}{S101}
\BBPexternallabel{app:sim-proposed-fe}{S5.2}{S101}
\BBPexternallabel{app:sim-reproducibility}{S5.3}{S103}
\BBPexternallabel{app:standardization-derivation}{S1.1.4}{S5}
\BBPexternallabel{app:target}{S1}{S2}
\BBPexternallabel{app:target-additional}{S1.2}{S6}
\BBPexternallabel{app:target-supporting}{S1.1}{S2}
\BBPexternallabel{app:target-zero-mechanisms}{S1.2.1}{S6}
\BBPexternallabel{ass:fe-groups}{S1}{S41}
\BBPexternallabel{ass:fe-growing-covariance-refinement}{S2}{S58}
\BBPexternallabel{ass:iv-statistics}{S3}{S87}
\BBPexternallabel{cor:fe-growing-order-effects}{S4}{S81}
\BBPexternallabel{cor:fe-growing-regimes}{S3}{S70}
\BBPexternallabel{cor:fe-growing-tails}{S2}{S68}
\BBPexternallabel{cor:fe-mixed}{S3.1.4}{S42}
\BBPexternallabel{cor:iols-fixed-lambda-numerical}{S1}{S29}
\BBPexternallabel{eq:app-application-twoway-covariance}{S6.1}{S107}
\BBPexternallabel{eq:app-gpml-log-continuation-endpoint-expansion}{S1.27}{S12}
\BBPexternallabel{eq:app-gpml-log-continuation-jacobian}{S1.26}{S12}
\BBPexternallabel{eq:app-gpml-log-continuation-log-equation}{S1.28}{S12}
\BBPexternallabel{eq:app-gpml-log-continuation-map}{S1.25}{S12}
\BBPexternallabel{eq:app-iols-asymptotic-linearity}{S2.33}{S29}
\BBPexternallabel{eq:app-iols-asymptotic-normality}{S2.34}{S29}
\BBPexternallabel{eq:app-iols-compact-score-separation}{S2.32}{S27}
\BBPexternallabel{eq:app-iols-exact-criterion}{S2.1}{S16}
\BBPexternallabel{eq:app-iols-exact-gradient}{S2.3}{S17}
\BBPexternallabel{eq:app-iols-exact-hessian}{S2.4}{S17}
\BBPexternallabel{eq:app-iols-nate-common-ratio}{S2.51}{S37}
\BBPexternallabel{eq:app-iols-nate-expansion}{S2.50}{S37}
\BBPexternallabel{eq:app-iols-no-escape}{S2.2}{S16}
\BBPexternallabel{eq:app-iols-phase2-geometric}{S2.39}{S32}
\BBPexternallabel{eq:app-iols-power-standardized-expansion}{S2.53}{S38}
\BBPexternallabel{eq:app-iols-power-standardized-functional}{S2.52}{S38}
\BBPexternallabel{eq:app-iols-sandwich-bread}{S2.42}{S36}
\BBPexternallabel{eq:app-iols-sandwich-covariance}{S2.43}{S36}
\BBPexternallabel{eq:app-large-shift}{S1.23}{S11}
\BBPexternallabel{eq:app-log-nate-gap}{S1.1}{S4}
\BBPexternallabel{eq:app-nate-conditional-wedge}{S1.3}{S4}
\BBPexternallabel{eq:app-pilot-smooth-extension}{S2.15}{S20}
\BBPexternallabel{eq:app-potential-curvature}{S2.20}{S21}
\BBPexternallabel{eq:app-potential-gradient}{S2.17}{S21}
\BBPexternallabel{eq:app-potential-hessian}{S2.18}{S21}
\BBPexternallabel{eq:app-raw-contraction}{S2.26}{S24}
\BBPexternallabel{eq:app-raw-d}{S2.19}{S21}
\BBPexternallabel{eq:app-raw-descent}{S2.25}{S24}
\BBPexternallabel{eq:app-raw-global-rate}{S2.24}{S23}
\BBPexternallabel{eq:app-raw-jacobian-approx}{S2.14}{S20}
\BBPexternallabel{eq:app-raw-map-identity}{S2.12}{S19}
\BBPexternallabel{eq:app-raw-no-escape}{S2.21}{S21}
\BBPexternallabel{eq:app-raw-pilot-rate}{S2.28}{S25}
\BBPexternallabel{eq:app-raw-potential}{S2.16}{S21}
\BBPexternallabel{eq:app-raw-score-approx}{S2.13}{S20}
\BBPexternallabel{eq:app-raw-slowdown}{S2.30}{S26}
\BBPexternallabel{eq:app-raw-slowdown-whitened}{S2.31}{S27}
\BBPexternallabel{eq:app-raw-spectrum}{S2.27}{S25}
\BBPexternallabel{eq:app-raw-warm-start}{S2.29}{S25}
\BBPexternallabel{eq:app-small-shift}{S1.22}{S11}
\BBPexternallabel{eq:app-tau2-linear-lower}{S2.23}{S23}
\BBPexternallabel{eq:app-tau2-quadratic-tail}{S2.22}{S22}
\BBPexternallabel{eq:comparison-general-pilot-residual}{S1.29}{S12}
\BBPexternallabel{eq:comparison-gpml-working-outcome}{S1.24}{S11}
\BBPexternallabel{eq:conditional-log-mean-gap}{S1.17}{S8}
\BBPexternallabel{eq:continuous-nate-integrated}{S1.9}{S5}
\BBPexternallabel{eq:continuous-nate-standardization}{S1.8}{S5}
\BBPexternallabel{eq:fe-average-expansion}{S3.21}{S83}
\BBPexternallabel{eq:fe-cancel-expansion}{S3.1}{S41}
\BBPexternallabel{eq:fe-cancel-identification}{S3.5}{S47}
\BBPexternallabel{eq:fe-covariance-refinement-adjustment}{S3.11}{S59}
\BBPexternallabel{eq:fe-covariance-weight-second}{S3.10}{S58}
\BBPexternallabel{eq:fe-design-score}{S3.3}{S43}
\BBPexternallabel{eq:fe-gamma-twoway-bias}{S3.13}{S73}
\BBPexternallabel{eq:fe-gamma-twoway-expansion}{S3.14}{S73}
\BBPexternallabel{eq:fe-growing-covariance-refinement}{S3.12}{S62}
\BBPexternallabel{eq:fe-growing-delete-covariance}{S3.9}{S56}
\BBPexternallabel{eq:fe-growing-joint-curvature}{S3.7}{S49}
\BBPexternallabel{eq:fe-growing-tensor-bound}{S3.8}{S55}
\BBPexternallabel{eq:fe-growing-weight-derivative}{S3.6}{S49}
\BBPexternallabel{eq:fe-mixed-expansion}{S3.2}{S42}
\BBPexternallabel{eq:fe-power-cancel-score}{S3.18}{S81}
\BBPexternallabel{eq:fe-profiled-power-score}{S3.19}{S81}
\BBPexternallabel{eq:fe-ratio-difference-identity}{S3.4}{S46}
\BBPexternallabel{eq:fe-small-region-criterion}{S3.16}{S78}
\BBPexternallabel{eq:fe-small-region-limit}{S3.17}{S79}
\BBPexternallabel{eq:fe-small-region-product}{S3.15}{S78}
\BBPexternallabel{eq:fe-standardized-power-moment}{S3.20}{S83}
\BBPexternallabel{eq:gpml-pilot-outcome}{S2.54}{S38}
\BBPexternallabel{eq:gpml-score}{S1.18}{S9}
\BBPexternallabel{eq:iols-fixed-lambda-numerical-equivalence}{S2.35}{S29}
\BBPexternallabel{eq:iols-fixed-lambda-population-outward}{S2.36}{S29}
\BBPexternallabel{eq:iols-fixed-lambda-sample-separation}{S2.37}{S30}
\BBPexternallabel{eq:iols-generic-residual}{S2.38}{S30}
\BBPexternallabel{eq:iols-nate-gradient-score}{S2.44}{S36}
\BBPexternallabel{eq:iols-phase1-identity}{S2.10}{S19}
\BBPexternallabel{eq:iols-phase1-map}{S2.9}{S19}
\BBPexternallabel{eq:iols-phase1-response}{S2.8}{S19}
\BBPexternallabel{eq:iols-phase1-score-approximation}{S2.11}{S19}
\BBPexternallabel{eq:iols-power-mean-exact-hessian}{S2.5}{S18}
\BBPexternallabel{eq:iols-power-mean-no-global-cap}{S2.6}{S18}
\BBPexternallabel{eq:iols-power-residual-curvature-bound}{S1.31}{S14}
\BBPexternallabel{eq:iols-power-standardization-component}{S2.47}{S37}
\BBPexternallabel{eq:iols-power-standardization-gradient}{S2.48}{S37}
\BBPexternallabel{eq:iols-scalar-pilot-comparison}{S2.7}{S19}
\BBPexternallabel{eq:iols-standardized-nate-if}{S2.45}{S36}
\BBPexternallabel{eq:iols-standardized-nate-variance}{S2.46}{S36}
\BBPexternallabel{eq:iols-standardized-power-if}{S2.49}{S37}
\BBPexternallabel{eq:iols-tau-curvature-containment}{S2.40}{S32}
\BBPexternallabel{eq:iols-tau-root-certificate}{S2.41}{S33}
\BBPexternallabel{eq:iv-exact-hessian}{S4.4}{S86}
\BBPexternallabel{eq:iv-expansion}{S4.5}{S88}
\BBPexternallabel{eq:iv-fixed-map}{S4.1}{S84}
\BBPexternallabel{eq:iv-gmm-fixed-outcome}{S4.2}{S85}
\BBPexternallabel{eq:iv-gmm-jacobian-map}{S4.3}{S86}
\BBPexternallabel{eq:nate-individual-wedge}{S1.2}{S4}
\BBPexternallabel{eq:nate-positive-margin-decomposition}{S1.4}{S4}
\BBPexternallabel{eq:positive-sample-mean}{S1.10}{S6}
\BBPexternallabel{eq:power-mean-continuous-response}{S1.14}{S7}
\BBPexternallabel{eq:power-mean-covariance-wedge}{S1.12}{S7}
\BBPexternallabel{eq:power-mean-individual-response}{S1.15}{S7}
\BBPexternallabel{eq:power-mean-response-envelope}{S1.13}{S7}
\BBPexternallabel{eq:power-mean-response-sorting}{S1.16}{S8}
\BBPexternallabel{eq:power-mean-two-paths}{S1.11}{S7}
\BBPexternallabel{eq:power-score-case-weight}{S1.21}{S10}
\BBPexternallabel{eq:power-shift-first-order}{S1.30}{S14}
\BBPexternallabel{eq:ppml-score}{S1.19}{S9}
\BBPexternallabel{eq:saturated-mean-ratio}{S1.20}{S9}
\BBPexternallabel{eq:simulation-dyadic-mean}{S5.1}{S101}
\BBPexternallabel{eq:target-mean}{S1.7}{S5}
\BBPexternallabel{eq:target-standardization}{S1.6}{S5}
\BBPexternallabel{eq:truemodel-multiplicative}{S1.5}{S5}
\BBPexternallabel{fig:solver-stress-accuracy}{S5.1}{S93}
\BBPexternallabel{lem:fe-covariance-dense-rows}{S14}{S65}
\BBPexternallabel{lem:fe-covariance-expansions}{S13}{S59}
\BBPexternallabel{lem:fe-covariance-gamma-selection}{S15}{S66}
\BBPexternallabel{lem:fe-group-laws}{S8}{S40}
\BBPexternallabel{lem:fe-growing-derivatives}{S9}{S49}
\BBPexternallabel{lem:fe-growing-feasible}{S12}{S53}
\BBPexternallabel{lem:fe-growing-normalized}{S10}{S50}
\BBPexternallabel{lem:fe-growing-projection}{S16}{S70}
\BBPexternallabel{lem:fe-growing-quadratic}{S11}{S52}
\BBPexternallabel{lem:general-pilot-endpoints}{S1}{S12}
\BBPexternallabel{lem:iols-raw-bounded-sets}{S5}{S21}
\BBPexternallabel{lem:iols-raw-identity-approximation}{S3}{S19}
\BBPexternallabel{lem:iols-raw-potential}{S4}{S20}
\BBPexternallabel{lem:iols-raw-spectrum}{S6}{S25}
\BBPexternallabel{lem:iols-sample-event}{S2}{S15}
\BBPexternallabel{lem:iols-score-separation}{S7}{S27}
\BBPexternallabel{prop:common-mean}{S1.4}{S9}
\BBPexternallabel{prop:fe-fixed-design}{S10}{S43}
\BBPexternallabel{prop:fe-gamma-twoway-bias}{S14}{S73}
\BBPexternallabel{prop:fe-growing-covariance-refinement}{S13}{S62}
\BBPexternallabel{prop:fe-growing-delete-covariance}{S12}{S55}
\BBPexternallabel{prop:fe-growing-selection}{S11}{S44}
\BBPexternallabel{prop:fe-small-regions}{S15}{S78}
\BBPexternallabel{prop:fixed-shift}{S1.5.1}{S11}
\BBPexternallabel{prop:gmm-local}{S4.1.2}{S86}
\BBPexternallabel{prop:iols-asymptotic-linearity}{S2.3}{S28}
\BBPexternallabel{prop:iols-direct-continuation}{S7}{S32}
\BBPexternallabel{prop:iols-exact-curvature}{S2.1.3}{S18}
\BBPexternallabel{prop:iols-exact-sample-root}{S2}{S16}
\BBPexternallabel{prop:iols-generic-completion}{S6}{S30}
\BBPexternallabel{prop:iols-local-exact-completion}{S2.4.2}{S32}
\BBPexternallabel{prop:iols-phase1-global}{S3}{S23}
\BBPexternallabel{prop:iols-raw-delta-path}{S4}{S25}
\BBPexternallabel{prop:iols-raw-slowdown}{S5}{S26}
\BBPexternallabel{prop:iols-sandwich}{S2.6}{S36}
\BBPexternallabel{prop:iols-tau-finite-trials}{S8}{S34}
\BBPexternallabel{prop:iv-local}{S4.1.1}{S85}
\BBPexternallabel{prop:nate-coupling-invariance}{S1}{S2}
\BBPexternallabel{prop:power-mean-marginal-individual-paths}{S1.2.2}{S7}
\BBPexternallabel{prop:shift-limits}{S1.5.1}{S11}
\BBPexternallabel{tab:fe-decisions-additive}{S3.2}{S76}
\BBPexternallabel{tab:fe-decisions-interacted}{S3.3}{S77}
\BBPexternallabel{tab:fe-higher-order-pilot}{S5.7}{S104}
\BBPexternallabel{tab:fe-practical-choice}{S3.1}{S75}
\BBPexternallabel{tab:fe-wz-covariance-comparison}{S5.6}{S102}
\BBPexternallabel{tab:historical-aer-practice}{S7.1}{S122}
\BBPexternallabel{tab:seminar-zero-poll}{S7.2}{S123}
\BBPexternallabel{tab:solver-all-six-common-cap}{S5.4}{S99}
\BBPexternallabel{tab:solver-preliminary-appendix}{S5.1}{S94}
\BBPexternallabel{tab:solver-preliminary-stress}{S5.2}{S95}
\BBPexternallabel{tab:solver-safeguarded-appendix}{S5.3}{S96}
\BBPexternallabel{tab:solver-statistical-stress}{S5.5}{S100}
\BBPexternallabel{table:application-coefficients-bright-top1pct}{S6.9}{S118}
\BBPexternallabel{table:application-coefficients-bright-top1row}{S6.8}{S117}
\BBPexternallabel{table:application-coefficients-bright-top2pct}{S6.10}{S119}
\BBPexternallabel{table:application-coefficients-bright-top5pct}{S6.11}{S120}
\BBPexternallabel{table:application-coefficients-full}{S6.7}{S116}
\BBPexternallabel{table:application-coefficients-ols}{S6.4}{S113}
\BBPexternallabel{table:application-coefficients-ols-2}{S6.5}{S114}
\BBPexternallabel{table:application-coefficients-ols-3}{S6.6}{S115}
\BBPexternallabel{table:application-coefficients-positive-full}{S6.12}{S121}
\BBPexternallabel{table:application-full-grid}{S6.1}{S110}
\BBPexternallabel{table:application-full-intervals}{S6.2}{S110}
\BBPexternallabel{table:application-positive-only}{S6.3}{S111}
\BBPexternallabel{thm:fe-cancel-inference}{S9}{S41}
\BBPexternallabel{thm:iv-inference}{S4.2}{S88}

\begin{document}
\maketitle
\begin{abstract}
Log regression is widely used to express effects in percentage
terms, but zeros, heterogeneity and retransformation pose distinct difficulties. We unify the interpretation,
identification and estimation of proportional effects through weighted individual percentage changes. We connect these
parameters to pseudo-maximum likelihood estimation, characterize finite-sample
differences between consistent estimators, and establish relationships with log regression. We develop a new shifted-log transformation that
yields unit-invariant proportional-effect parameters and use it
to construct iterated ordinary least squares (IOLS), an algorithm
that estimates power-mean models through repeated OLS regressions. We prove global numerical convergence, consistency
and asymptotic normality, and extend the approach to endogenous
regressors. Furthermore, we develop fixed-effect cancellation and bias-correction methods
that address incidental-parameter bias in common panel settings
under general working variances, extending beyond the familiar
Poisson-PML case.
Finally, simulations document numerical reliability, competitive speed
and finite-sample performance,
while an application to economic development
illustrates their empirical relevance.
\end{abstract}

\medskip
\noindent\small
Software: \href{https://github.com/dvbn/powermean}{\texttt{github.com/dvbn/powermean}}.\\
Replication files: \href{https://github.com/dvbn/bbp-replication}{\texttt{github.com/dvbn/bbp-replication}}.
\par\normalsize\medskip
\section{Introduction}
\label{sec:introduction}

Log regression is widely used to express effects in percentage
terms, but raises three distinct difficulties. First, logarithms
are not defined at zeros: dropping these observations changes the represented
population, while adding a constant changes the estimand and
introduces dependence on outcome units
\citep{chen2024logs}. Second, under heterogeneous
responses, retransformed log effects generally differ from average
individual percentage changes \citep{zeng2026percentage}. Third,
economic counterfactuals often concern changes in population
means or aggregate outcomes, which log regression does not directly
recover \citep{manning2001estimating}. Addressing these issues
requires a coherent choice of target parameter, transformation and
estimator, together with reliable computation and appropriate
fixed-effect treatment.

This paper unifies the interpretation, estimation, inference, and computation
of proportional effects. We define the targets around power means
and weighted individual percentage changes, clarify identification,
and establish local relationships between log regression and
PML estimation. A new shifted-log construction
leads to a new algorithm: IOLS, with global
convergence guarantees and an extension to
endogenous regressors. Finally, we develop fixed-effect cancellation
and bias correction under general working variances to address the incidental parameter problem (IPP) beyond PPML.

Our survey of 341 empirical articles in the \emph{American Economic
Review} from January 2016 through October 2020 illustrates the
variety of existing practice. Of 133 articles using logarithmic
specifications or related alternatives, 48 encountered zeros.
Among these, 48\% added a constant, 35\% used Poisson or
negative-binomial regression, 31\% dropped observations, and
15\% used the inverse hyperbolic sine. Some used several methods.%
\footnote{Appendix~\ref{app:empirical-practice} reports the survey
and complementary evidence.}

Our first contribution concerns interpretation and identification.
\citet{chen2024logs} establish the tension
between averaging individual percent effects, identification from marginal
potential-outcome distributions and invariance to outcome units.
\citet{zeng2026percentage} develops heterogeneity-aware retransformation
and inference for an identified lower bound on average individual
percentage effects. We characterize the fixed positive weights
under which an arithmetic average of individual percentage effects
is identified from the marginal potential-outcome distributions.
Among weights depending only on baseline outcomes, the normalized
average treatment effect (NATE) is the
unique option. We also derive exact conditions under which
NATE and the harmonic ATE are respectively closer to the individual average
than the retransformed log effect, distinguishing discrepancies
due to outcome weighting from the Jensen gap generated by
response dispersion.

Power means provide a broader aggregation framework with welfare
foundations \citep{atkinson1970inequality}.
\citet{munroe_newton_shah_2026} derive their heterogeneous-elasticity
weights and develop flexible estimators for positive outcomes.
We connect power means to  proportional effects and
conditional power-moment models, allowing zeros at positive orders.
We derive an expansion showing how
conditional dispersion separates different power-mean specifications.
This places the familiar distinction between log regression and
exponential conditional means under heteroskedasticity 
within a broader framework \citep{manning1998logged,manning2001estimating,santos_silva_tenreyro_2006}.

Many estimators can consistently estimate the same population
parameter yet differ substantially in finite samples. We formally show how
residual weighting generates these differences and illustrate their
consequences by comparing PPML, GPML and log regression. PPML and
GPML consistently estimate the same correctly specified exponential
conditional-mean parameter without requiring correct working
distributions \citep{gourieroux_etal_1984}. We show that, within the power-score family, only GPML
shares log OLS's first-order response to changes in outcomes. 
These results explain both the similarities between GPML
and log regression and the potentially very different finite sample estimates
of PPML and GPML.

For nonnegative outcomes, the fixed shift \(\log(Y+1)\) is a
popular alternative to PML. \citet{mullahynorton2024} show how
transformation parameters change regression estimates, while
\citet{chen2024logs} establish the unit dependence of log-like
effects. Our shifted-log construction instead accommodates zeros
and connects log OLS to a selected PML estimator. For \(0<\lambda<1\), a simple case is
\(\log((1-\lambda)Y+\lambda\mu)\), the logarithm of a convex
combination of the outcome and its fitted mean. This combination
rescales with the outcome, preserving unit-invariant
proportional-effect parameters. Its shift parameter links the
retransformed log effect to NATE, with intermediate values
providing local second-order approximations to the corresponding
power-mean effects.

Generalized linear models are conventionally estimated by
iteratively reweighted least squares (IRLS). Our algorithm instead
uses the preceding transformation to construct a path of approximate
problems with a global curvature bound, solved through iterated
ordinary least squares (IOLS). Our contribution is the explicit
approximation path and its guarantees, combining established
continuation and majorization principles with reusable linear
algebra. Under the stated conditions, the procedure converges
from any finite initialization and delivers a \(\sqrt n\)-consistent,
asymptotically normal estimator. In simulations, IOLS reaches the
required \(10^{-8}\) numerical accuracy in every attempted run
within the common time budget, including designs where Newton
and IRLS frequently fail. An iterated two-stage
least-squares version (I2SLS) extends the approach to IV models
\citep{mullahy1997instrumental}.

The final contribution addresses the incidental-parameter problem
in panel data, enabling broader use of PML estimation beyond PPML.
We develop a centered family of ratio-difference moments, indexed
by the working variance, that cancels a multiplicative fixed-effect
block rather than estimating it. The family includes profiled
PPML as a special case
\citep{hausmanhallgriliches1984,wooldridge1999distributionfree},
whereas its other members generally differ from joint or
profiled PML. We establish consistency in short panels, with a
growing number of units and a fixed number of periods. When the
remaining fixed effects also grow, we characterize the coefficient
bias and adapt the empirical bias reduction of
\citet{kosmidislunardon2024empirical} to obtain feasible
bias-corrected inference. These results extend the consistency
and bias-corrected inference established by
\citet{weidner2021bias} for three-way PPML to the larger
weighted family. Simulations illustrate the resulting improvements.

The remainder of the paper is organized as follows.
Sections~\ref{subsec:targetparm} and~\ref{sec:comparison} study
target parameters and estimation. Section~\ref{sec:iols} develops
IOLS and its IV extension, and Section~\ref{sec:fixed-effects}
addresses fixed effects. Section~\ref{sec:application} presents
an application 
illustrating how transformations, residual weighting and
fixed-effect treatment affect empirical conclusions.
The main appendices contain the principal proofs; the supplemental
appendices are available separately with the arXiv record.\footnote{The
supplement develops further results on proportional effects and shifted-log
approximations, numerical convergence and inference, fixed-effect
cancellation and covariance refinements, and instrumental-variable
estimation. It also provides additional simulations, application
diagnostics, complete coefficient estimates and evidence on empirical practice.}

\section{Which proportional effect?}
\label{subsec:targetparm}

Proportional effects can be defined in different ways, depending on the research objective. In this section, we discuss the choice of target parameter as a decision that precedes the choice of transformations, estimators, and numerical solvers. We focus on a binary treatment to simplify exposition.

\subsection{The target parameters}
\label{subsec:individual-log-nate}

Let \(Y_i(0),Y_i(1)\geq0\) be potential outcomes with finite means
\(M_d=\E(Y_i(d))\) and \(M_0>0\). The average treatment effect is
\(\mathrm{ATE}=M_1-M_0\). The normalized average treatment effect (NATE) is
\begin{equation}
 \theta_N=\frac{M_1-M_0}{M_0}.
 \label{eq:ATEnorm}
\end{equation}
It measures the proportional change in the arithmetic mean, is invariant
to positive outcome rescaling, and allows zeros. Given the untreated mean,
it determines the treated mean through \(M_1=M_0(1+\theta_N)\), and for a
fixed population size, it also gives the proportional change in the
expected total outcome, such as aggregate expenditure or employment.

When \(Y_i(0)>0\) almost surely and the expectation is finite, the
average individual percentage effect is
\begin{equation}
 \theta_{\%}=\E\left(\frac{Y_i(1)}{Y_i(0)}-1\right).
 \label{eq:individual-percentage-effect}
\end{equation}
This generally does not determine the treated mean from the untreated
mean. Moreover, \(\theta_{\%}\) is not generally point-identified from the
marginal potential-outcome distributions
\citep{chen2024logs}. Average log changes are hence often used to approximate \(\theta_{\%}\). For strictly positive outcomes, define
\(\mathrm{LogATE}=\E(\log(Y_i(1))-\log(Y_i(0)))\) and
\begin{equation}
 \theta_L=\exp(\mathrm{LogATE})-1.
 \label{eq:log-nate-three-ratios}
\end{equation}
This is the proportional change in the geometric mean, not generally
\(\theta_{\%}\) or \(\theta_N\).
 Retransformation puts the log comparison on a proportional
scale, but does not turn geometric averaging into arithmetic averaging.
The accuracy of that approximation depends on the heterogeneity of
individual responses, as discussed below.

Both \(\theta_L\) and \(\theta_N\) belong to the power-mean
family. For \(\omega\ne0\), write
\(A_{\omega,d}=(\E(Y_i(d)^\omega))^{1/\omega}\), where negative orders require positive outcomes. Set
\(A_{0,d}=\exp(\E(\log(Y_i(d))))\). When \(A_{\omega,0}>0\), define
\begin{equation}
 \theta_\omega=\frac{A_{\omega,1}}{A_{\omega,0}}-1.
 \label{eq:power-mean-effect}
\end{equation}
Thus \(\theta_0=\theta_L\), \(\theta_1=\theta_N\), and the harmonic
effect is \(\theta_H=\theta_{-1}=\E(1/Y_i(0))/\E(1/Y_i(1))-1\).
Positive power orders allow zeros.\footnote{With zero outcomes, researchers
may separate extensive and intensive margins or combine them through NATE,
as we do here. The choice depends on the economic question and the exogeneity
and selection assumptions. We do not examine these margins separately.}
Scaling untreated outcomes by
\(1+\theta_\omega\) reproduces the treated power mean. Thus an effect of
20\% is equivalent, under the chosen aggregation, to raising every
untreated outcome by 20\%. 
For economic variables such as income or consumption, power orders below one have an equally distributed equivalent
interpretation under the corresponding isoelastic utility
\citep{atkinson1970inequality}. They can therefore capture distributional
changes that leave $\theta_N$ unchanged. Suppose half the
population initially receives one and half receives nine, and a reform
gives everyone five. Mean income stays at five, so NATE is zero. At
\(\omega=1/2\), equivalent income rises from
\(((\sqrt{1}+\sqrt{9})/2)^2=4\) to 5, giving a 25\% increase.
Choosing \(\omega\)
therefore expresses a substantive evaluation objective, which 
\citet{munroe_newton_shah_2026} relate to
welfare weights for aggregate elasticities.

\subsection{Aggregation and identification}

Under random assignment, we can identify each outcome
distribution separately, but not how \(Y_i(0)\) and \(Y_i(1)\) are
associated, because we do not observe both for the same individual.
This fundamental problem of causal inference implies that
\(\theta_{\%}\) is not generally point-identified. However,
identifying \(\theta_\omega\) requires only
the marginal distributions. 

More generally, for strictly positive outcomes and \(\omega\ne0\), with
the defining moments finite,
\begin{equation}
 \theta_\omega=
 \left(
 \E\left(\frac{Y_i(0)^\omega}{\E(Y_i(0)^\omega)}
       \left(\frac{Y_i(1)}{Y_i(0)}\right)^\omega\right)
 \right)^{1/\omega}-1.
 \label{eq:power-mean-weighted-ratios}
\end{equation}
This is a weighted power mean of the gross individual ratios.
Both the weights and the averaging rule change with the power order.
At order zero, equal weights and geometric
averaging give \(\theta_L\); at order one, baseline weights and arithmetic
averaging give \(\theta_N\); at order minus one, inverse-baseline weights
and harmonic averaging give \(\theta_H\). For $\omega\neq 0$, the weighted
numerator simplifies to \(\E(Y_i(1)^\omega)\), so the entire
family, including \(\theta_H\), \(\theta_L\), and \(\theta_N\),
is determined by the marginals rather than their joint distribution.

However, this identification property does not extend to arbitrary
arithmetic weighted averages of individual percentage effects. 
For a positive weight function \(w(\cdot)\), fixed across populations, define
the following effect when its expectations exist with a finite positive
denominator:
\begin{equation}
 \theta_w=
 \E\left(\frac{w(Y_i(0),Y_i(1))}
              {\E(w(Y_i(0),Y_i(1)))}
       \left(\frac{Y_i(1)}{Y_i(0)}-1\right)\right).
 \label{eq:weighted-proportional-effect}
\end{equation}
Equal weights then give \(\theta_{\%}\), rather than \(\theta_L\),
 baseline weights give \(\theta_N\), and inverse-treated weights give \(\theta_H\). Within the class of positive weight functions fixed
across populations, the only arithmetic weighted effects determined by
the marginals for every population use
\(w(y_0,y_1)=a y_0+b/y_1\), where \(a,b\geq0\) are not both zero.
They give \(\theta_w=\lambda\theta_N+(1-\lambda)\theta_H\), with
\(\lambda=aM_0/(aM_0+b\E(1/Y_i(1)))\).
Restricting weights to the baseline leaves NATE as the unique choice.\footnote{Appendix~\ref{app:target-supporting}
proves this characterization in
Proposition~\ref{prop:nate-coupling-invariance}.} Identification of other arithmetic averages requires additional
restrictions on the pairing beyond random assignment. For example, independence between the two
potential outcomes identifies \(\theta_{\%}\) as
\(\E(Y_i(1))\E(1/Y_i(0))-1\), when these expectations are finite.
If everyone instead has the same positive untreated outcome,
\(\theta_{\%}=\theta_N\).

Zeros introduce a separate composition issue. Write
\(p_d=\Pr(Y_i(d)>0)\) and
\(m_d=\E(Y_i(d)\mid Y_i(d)>0)\) when \(p_d>0\).
If both participation probabilities are positive, \(M_d=p_dm_d\), so
\(1+\theta_N=(p_1/p_0)(m_1/m_0)\). NATE combines the extensive-margin
change in participation with the change in positive-outcomes means.
The latter need not concern the same individuals: treatment can change
who has a positive outcome. Dropping zeros therefore does not generally help identify
an intensive-margin causal effect.\footnote{Appendix~\ref{app:target-zero-mechanisms} distinguishes
this composition issue from censoring and sample selection.}

\subsection{Why proportional effects differ}

If every individual experiences the same proportional change, all the
effects defined above coincide. With heterogeneous responses, two
features matter: dispersion in individual percentage effects and their
association with the potential outcomes in the weights.
Jensen's inequality gives
\(\theta_L\leq\theta_{\%}\), with equality only for a constant
ratio \(Y_i(1)/Y_i(0)\). NATE and the harmonic effect instead differ from
\(\theta_{\%}\) through the association between individual responses
and their outcome weights. In particular,
\begin{equation}
    \theta_N-\theta_{\%}
=\operatorname{Cov}(Y_i(0),Y_i(1)/Y_i(0))/M_0.
\end{equation}
The individual average  thus exceeds NATE when smaller baselines are
associated with larger percentage gains, and conversely. For example, 
equal-sized groups moving from 100 to 110 and from 10 to 12 give
\(\theta_{\%}=15\%\), but \(\theta_N\simeq10.9\%\).
The larger-baseline group receives more weight despite its smaller
percentage gain. The harmonic comparison instead
emphasizes smaller treated outcomes.

For small individual percentage changes, $\theta_{\%}$ and $\theta_{L}$
agree to first order. Yet, a small average change is not enough: large gains
and losses can offset in the average while remaining dispersed. For
example, if \(Y_i(1)/Y_i(0)-1=\varepsilon V_i\), with bounded \(V_i\),
then \(\theta_{\%}-\theta_L
=\varepsilon^2\operatorname{Var}(V_i)/2+O(\varepsilon^3)\).
The following result compares this averaging gap with the discrepancy
introduced by outcome-dependent weights.

\Needspace{3\baselineskip}
\begin{proposition}[When weighted effects are closer to individual ATE\%]
\label{prop:log-effect-nate-closeness}
Suppose both potential outcomes are strictly positive and the
expectations defining the effects below are finite. Then
\(|\theta_N-\theta_{\%}|\leq|\theta_L-\theta_{\%}|\) if and only if
\begin{equation}
 \left|\operatorname{Cov}(Y_i(0),Y_i(1)/Y_i(0))\right|
 \leq M_0(\theta_{\%}-\theta_L).
 \label{eq:log-nate-closeness}
\end{equation}
Likewise,
\(|\theta_H-\theta_{\%}|\leq|\theta_L-\theta_{\%}|\) if and only if
\begin{equation}
 \left|\operatorname{Cov}(1/Y_i(1),Y_i(1)/Y_i(0))\right|
 \leq \E(1/Y_i(1))(\theta_{\%}-\theta_L).
 \label{eq:log-harmonic-closeness}
\end{equation}
Each inequality is strict when the corresponding covariance
inequality is strict.
\end{proposition}

Either weighted effect can therefore improve on the log comparison when
its weighting discrepancy is smaller than the arithmetic--geometric gap.
Zero covariance gives equality with \(\theta_{\%}\) despite heterogeneous
responses, and small departures from zero can still leave the weighted effect
closer than the log comparison. 

Lognormal marginal distributions provide a useful illustration. Write
\(v_0\) and \(v_1\) for the untreated and treated log variances. Then
\(\theta_N=\exp(\mathrm{LogATE}+(v_1-v_0)/2)-1\), whereas
\(\theta_H=\exp(\mathrm{LogATE}-(v_1-v_0)/2)-1\).
If treatment increases log variance, the harmonic effect falls below
\(\theta_L\) and is necessarily farther from \(\theta_{\%}\): only NATE
can improve on the log comparison. If log variance decreases, the roles
reverse. Equal log variances give \(\theta_N=\theta_H=\theta_L\), but
with heterogeneous individual responses all three remain strictly below
\(\theta_{\%}\). Agreement among identified aggregates is therefore not
evidence that they recover the individual average. Their exact ranking
still depends on the unobserved pairing.

For a continuous treatment, a log response has a different interpretation
from a finite log comparison. Fix an interior dose \(d\), assume positive
differentiable outcomes with finite means, and define the individual semielasticity
\(s_i(d)=\partial_dY_i(d)/Y_i(d)\). Then
\(\partial_d\E(\log(Y_i(d)))=\E(s_i(d))\), whereas
\(\partial_d\log(\E(Y_i(d)))
=\E(Y_i(d)s_i(d))/\E(Y_i(d))\).
The local geometric-mean response thus equals the average individual
semielasticity exactly, and the arithmetic-mean response is outcome-weighted.\footnote{More generally, when the defining moments exist,
\(\partial_d\log(A_{\omega,d})
=\E(Y_i(d)^\omega s_i(d))/\E(Y_i(d)^\omega)\).
Thus power-mean semielasticities average individual semielasticities
with weights proportional to \(Y_i(d)^\omega\): order zero gives
the unweighted individual average, while order one gives
\(M'(d)/M(d)\), where \(M(d)=\E(Y_i(d))\).}
These are elasticities when \(d\) is log treatment.

\section{Conditional mean models}
\label{sec:comparison}

\subsection{Choosing a conditional mean model}

Let \(W_i\) contain exogenous covariates. Consistency
\(Y_i=Y_i(D_i)\), overlap \(0<\Pr(D_i=1\mid W_i)<1\), and conditional
mean independence
\(\E(Y_i(d)^\omega\mid D_i,W_i)=\E(Y_i(d)^\omega\mid W_i)\)
identify the marginal power moment by averaging
\(\E(Y_i^\omega\mid D_i=d,W_i)\). At order zero, the corresponding
restriction concerns log outcomes.

Allowing transformations and interactions of covariates, write \(X_i=x(D_i,W_i)\),
\(q_i(\beta)=X_i'\beta\), \(\mu_i(\beta)=\exp(q_i(\beta))\), and
\(U_i(\beta)=Y_i/\mu_i(\beta)\). For \(\omega\ne0\), specify
\begin{equation}
 \E(Y_i^\omega\mid D_i,W_i)
 =\exp(\omega x(D_i,W_i)'\beta_{\omega0}).
 \label{eq:power-mean-conditional-model}
\end{equation}
Equivalently, \(Y_i=\exp(X_i'\beta_{\omega0})U_{\omega i}\), with
\(\E(U_{\omega i}^{\omega}\mid D_i,W_i)=1\).
The positivity and moment restrictions of the chosen order remain
necessary. Standardizing over the covariate distribution gives
\begin{equation}
 \theta_\omega=
 \left(\frac{\E(\exp(\omega x(1,W_i)'\beta_{\omega0}))}
                  {\E(\exp(\omega x(0,W_i)'\beta_{\omega0}))}
 \right)^{1/\omega}-1.
 \label{eq:power-mean-standardization}
\end{equation}
Without treatment interactions this is \(\exp(\beta_{\omega,D})-1\).
At order zero, the model is
\(\E(\log(Y_i)\mid D_i,W_i)=X_i'\beta_{L0}\), and LogATE is
\(\E((x(1,W_i)-x(0,W_i))'\beta_{L0})\).
 Conditional
independence of treatment and potential outcomes identifies all existing
marginal moments, but the conditional-mean restriction alone does not
justify estimating the whole power-mean family. Changing the order thus
changes both the economic aggregate and the restriction used to estimate it.

At \(\omega=1\), the restriction \(\E(Y_i\mid X_i)=\mu_i(\beta_0)\),
equivalently \(\E(U_i\mid X_i)=1\), allows zeros and heteroskedasticity.
Write \(\mu_d(w;\beta)=\exp(x(d,w)'\beta)\).
Without interactions, \(\theta_N=\exp(\beta_{D,0})-1\); with
interactions, use \eqref{eq:power-mean-standardization}.
A corresponding potential-outcome model is
\(Y_i(d)=\mu_d(W_i;\beta_0)U_i(d)\), with
\(\E(U_i(d)\mid D_i,W_i)=1\).
Without interactions, positive potential outcomes satisfy
\(Y_i(1)/Y_i(0)=\exp(\beta_{D,0})U_i(1)/U_i(0)\):
a common conditional mean ratio therefore permits heterogeneous
individual percentage effects because \(U_i(1)/U_i(0)\) may vary
across individuals.
By comparison, the linear conditional mean
\(\E(Y_i\mid D_i,W_i)=a+bD_i+W_i'c\) combines the level effect
with the untreated mean to give \(\theta_N=b/(a+\E(W_i)'c)\).

\paragraph{Heteroskedasticity.} Log and exponential mean specifications are generally not interchangeable \citep{manning1998logged}.
More broadly, why do conditional power means respond differently to
the same regressor? For positive outcomes, write
\(A_\omega(X_i)=(\E(Y_i^\omega\mid X_i))^{1/\omega}\) for
\(\omega\ne0\), and \(A_0(X_i)=\exp(\E(\log(Y_i)\mid X_i))\).
The following result isolates the leading role of conditional dispersion.

\begin{proposition}[Heteroskedasticity and power means]
\label{prop:conditional-power-mean-dispersion}
Let \(h\geq0\) be a dispersion parameter and consider positive outcomes
\(Y_i(h)=\exp(m_h(X_i)+hV_i)\), where \(m_h(X_i)\) is finite and
\(\E(V_i\mid X_i)=0\). Assume that the conditional distribution of
\(V_i\) given \(X_i\) does not depend on \(h\), and its
moment-generating function is finite in a neighborhood of zero.
Suppressing dependence on \(h\), write \(Y_i=Y_i(h)\) and
\(\varepsilon_i=hV_i\). For any two fixed real orders \(\omega\) and
\(\nu\), as \(h\downarrow0\),
\begin{equation}
 \log\left(\frac{A_\omega(X_i)}{A_\nu(X_i)}\right)
 =\frac{\omega-\nu}{2}\Var(\varepsilon_i\mid X_i)+O(h^3),
 \label{eq:conditional-power-mean-gap}
\end{equation}
pointwise in \(X_i\). The same expansion holds with
\(\Var(U_i\mid X_i)\) in place of \(\Var(\varepsilon_i\mid X_i)\),
where \(U_i=Y_i/\E(Y_i\mid X_i)\) satisfies
\(\E(U_i\mid X_i)=1\).
\end{proposition}

Heteroskedasticity of the log error is therefore the leading mechanism
through which two power mean specifications can yield different coefficients.  The log--exponential comparison is the case \(\omega=1\), \(\nu=0\). Write a log model as
\(\log(Y_i)=X_i'\beta_{L0}+\varepsilon_i\), with
\(\E(\varepsilon_i\mid X_i)=0\). The arithmetic conditional mean is
\(\exp(X_i'\beta_{L0})\E(\exp(\varepsilon_i)\mid X_i)\).
As a regressor changes, this mean can change through both the log index
and the retransformation factor. Heteroskedasticity can therefore separate
the two specifications by more than an intercept. Recovering the arithmetic mean from a log model hence requires its
conditional retransformation factor
\citep{manning1998logged,manning2001estimating}. More
generally, heteroskedasticity can explain discrepancies across
specifications, but does not by itself determine which one should
be used. That choice should follow the parameter of interest and
the credibility of the corresponding restriction.

\subsection{Estimation and finite-sample comparisons}
\label{sec:residual-weighting}

For exposition, take \(\omega=1\). Other nonzero orders replace the
outcome by \(Y_i^\omega\) and rescale its log-mean coefficients by
\(1/\omega\). The conditional mean implies
\(\E(a(X_i)(Y_i-\mu_i(\beta_0)))=0\) for every integrable moment of
this form. Using the power working variance \(V_\tau(\mu)=\mu^\tau\)
gives the sample equation
\begin{equation}
 g_{\tau,n}(\beta)=\frac1n\sum_{i=1}^n
 X_i\mu_i(\beta)^{1-\tau}(Y_i-\mu_i(\beta))=0.
 \label{eq:iols-power-score}
\end{equation}
The choices \(\tau=0,1,2,3\) give nonlinear least squares in levels,
PPML, GPML and inverse Gaussian PML, respectively. Intermediate cases for $\tau\geq1$ admit a Tweedie interpretation. 
Under identification and regularity, all estimate the same correctly
specified mean parameter, regardless of the working-variance
\citep{gourieroux_etal_1984}. Efficient weighting depends on the actual
conditional variance. The two familiar forms GPML and PPML respectively use relative residuals
\(U_i(\beta)-1\) and level residuals \(Y_i-\mu_i(\beta)\).
Both equations allow zeros without requiring Gamma or Poisson sampling.

\paragraph{Local coefficient responses.}
\label{sec:local-tangency-influence}

Estimators with different residual weighting, like PPML and GPML, can give different estimates in the same sample while
consistently estimating the same population parameter. To isolate how
weighting produces these differences, we perturb a common exponential
relationship and compare the estimators' responses, without changing the
sample size or the design.

Fix a full-column-rank design \(X\), a coefficient \(b_*\), and
disturbances \(v_i\). Define 
\(\mu_{i,*}=\exp(X_i'b_*)\) and consider
\begin{equation}
 Y_i(t)=\mu_{i,*}\exp(t v_i),\qquad i=1,\ldots,n,
 \label{eq:positive-log-perturbation}
\end{equation}
where only the disturbance size \(t\) varies. Let \(h_\tau\) be the
coefficient from regressing \(v_i\) on \(X_i\) with weights
\(\mu_{i,*}^{2-\tau}\). Write \(e_i=v_i-X_i'h_2\), and let \(c\)
be the OLS coefficient from regressing \(e_i^2\) on \(X_i\).

\begin{proposition}[Residual weighting and local coefficient responses]
\label{thm:power-score-log-tangency}
For the outcome path in
\eqref{eq:positive-log-perturbation}, and each fixed
\(\tau\in[0,3]\), the power-score solution near \(b_*\) satisfies
\begin{equation}
 \widehat b_\tau(t)=b_*+t h_\tau+O(t^2),\qquad t\to0.
 \label{eq:local-weighted-coefficient-response}
\end{equation}
Log OLS gives \(\widehat b_{\log}(t)=b_*+t h_2\) exactly, and
GPML satisfies
\begin{equation}
 \widehat b_2(t)-\widehat b_{\log}(t)
 =\frac{t^2}{2}c+O(t^3).
 \label{eq:gpml-log-curvature}
\end{equation}
Among fixed power-score choices, only GPML \((\tau=2)\) shares
the first-order coefficient response of log OLS for every
full-rank $X$, reference coefficient $b_*$ and deviation vector $v$.
\end{proposition}

Thus, the estimators respond to the same log disturbances but
weight them differently. PPML uses fitted-mean weights, whereas
GPML uses the equal weights of log OLS. PPML and GPML can therefore
differ at first order whereas GPML and log OLS agree through first order. The quadratic adjustment
in \eqref{eq:gpml-log-curvature} is a sample counterpart of the dispersion
comparison in Proposition~\ref{prop:conditional-power-mean-dispersion}.

A second implication of Proposition~\ref{thm:power-score-log-tangency}
is that fitted-mean weighting can make PPML much more responsive
to proportional disturbances among high-outcome observations. In two equally sized,
equally treated groups with separate intercepts, common reference
treatment effect and baseline means 1 and 100, adding \(t\) to treated
log outcomes only in the low-baseline group gives first-order treatment
responses \(t/2\) for GPML and log OLS, but \(t/101\) for PPML.
Moving the disturbance to the high-baseline group gives \(100t/101\)
for PPML. PPML is therefore 100 times more responsive to this disturbance
in the high-baseline group, whereas GPML and log OLS respond equally at first order.\footnote{Appendix~\ref{app:observation-influence} gives the detailed derivation for this example.}

Zero outcomes provide a complementary illustration of this
result. At a common coefficient, a zero contributes
\(-X_i\mu_i\) to the PPML equation and \(-X_i\) to the GPML
equation. A small fitted mean therefore attenuates its PPML score
contribution. An empirical illustration appears in
\citet{santos_silva_tenreyro_2006}, who report very
similar PPML estimates with and without zero-trade observations. Zeros also mark the limit of the local GPML--log-OLS comparison. As \(Y_i/\mu_i\) approaches zero, the GPML residual approaches
minus one, whereas the log residual diverges to minus infinity.

\subsection{Log-like transformations when outcomes can be zero}
\label{sec:log-like-transformations}

A common way to keep zeros is to regress \(\log(Y_i+a)\) on
\(X_i\) for a fixed \(a>0\). Under the exponential
mean, its population OLS coefficient \(\gamma(a,b)\), after subtracting
an intercept adjustment \(b\), satisfies
\begin{equation}
 \gamma(a,b)-\beta_0
 =Q^{-1}\E\!\left(
 X_i\left(\log\left(U_i+\frac{a}{\mu_i(\beta_0)}\right)-b\right)
 \right).
 \label{eq:shift-displacement}
\end{equation}
The relative shift \(a/\mu_i(\beta_0)\) varies across observations,
which an intercept adjustment cannot generally remove. With zeros,
very small shifts emphasize the positive-outcome indicator; large
shifts approximate rescaled levels regressions
\citep{mullahynorton2024}. More generally, when treatment changes the
probability of a positive outcome, changing units can make log-like
transformed effects arbitrarily large in magnitude, precluding a
unit-invariant percentage interpretation \citep{chen2024logs}.

Under GPML weighting, a shift proportional to the fitted mean rescales
with the outcome.
For \(0\leq\lambda<1\), define
\begin{equation}
 H_{\lambda,\tau,i}(\beta)
 =X_i'\beta+\log\!\left(
 \frac{\lambda+(1-\lambda)Y_i\mu_i(\beta)^{1-\tau}}
      {\lambda+(1-\lambda)\mu_i(\beta)^{2-\tau}}
 \right).
 \label{eq:comparison-fitted-mean-shift}
\end{equation}
Here \(\tau\) determines residual weighting and \(\lambda\) controls
the approximation. At \(\tau=2\), this transformed outcome \(\log((1-\lambda)Y_i+\lambda\mu_i(\beta))\) corresponds to the log
of a convex combination of the outcome and its fitted mean. For positive outcomes, \(\lambda=0\) gives \(\log(Y_i)\), and every
interior \(\lambda\) allows zeros. Define 
\(r_{\lambda,\tau,i}=(H_{\lambda,\tau,i}-X_i'\beta)/(1-\lambda)\)
and obtain \(\widetilde\beta_{\lambda,\tau}\) as the root of 
\begin{equation}
 \frac1n\sum_{i=1}^n X_i r_{\lambda,\tau,i}(\beta)=0.
 \label{eq:gpml-log-continuation-equation}
\end{equation}
The continuous limit at \(\lambda=1\) is the power-score residual
\(\mu_i^{1-\tau}(Y_i-\mu_i)\). Therefore, with positive outcomes and full column rank of \(X\), each fixed
\(\tau\in[1,2]\) gives a unique finite solution that varies
continuously with \(\lambda\). The path connects log OLS at
\(\lambda=0\) to the power-score estimate at \(\lambda=1\),
including PPML for \(\tau=1\) and GPML for \(\tau=2\).

\begin{proposition}[Second-order approximation to power-mean coefficients]
\label{prop:gpml-log-continuation}
Fix a full-column-rank design \(X\) and \(Y_i\geq0\), with
\(Y_i>0\) at \(\omega=0\). Assume unique finite solutions
at each considered order \(\omega\in[0,1]\). 
Let \(\widehat\beta^P_\omega\) solve
\(\frac1n\sum_{i=1}^n X_i
((Y_i/\mu_i(\beta))^\omega-1)/\omega=0\),
where the fraction at \(\omega=0\) is interpreted as
\(\log(Y_i/\mu_i(\beta))\).
Set \(\widetilde\beta_\omega=\widetilde\beta_{\omega,2}\) and
\(\rho_\omega=\max_i|Y_i/\mu_i(\widehat\beta^P_\omega)-1|\).
There exist constants \(C_X<\infty\) and \(0<\rho_0<1\),
depending only on \(X\), such that, for every such \(\omega\)
with \(\rho_\omega\leq\rho_0\),
\begin{equation}
 \|\widetilde\beta_\omega-\widehat\beta^P_\omega\|_2
 \leq C_X\omega(1-\omega)\rho_\omega^3.
 \label{eq:shift-power-coefficient-approximation}
\end{equation}
The coefficients coincide exactly at \(\omega=0\) and \(\omega=1\)
whenever these endpoints are defined.
\end{proposition}

Matching \(\lambda=\omega\) under GPML weighting therefore approximates
power-mean coefficients through second order, with cubic error vanishing at the log-OLS and GPML endpoints. The next result gives the corresponding
approximation for proportional effects.

\begin{proposition}[Second-order approximation to power-mean effects]
\label{prop:iols-power-residual-calibration}
For \(d=0,1\), let
\(Y_i(d;t)=\mu_{d,*}\exp(t v_{di})\), with fixed 
\(\mu_{d,*}>0\) and bounded log disturbances \(v_{di}\).
Let \(A_{\omega,d}(t)\) be the corresponding power mean.
For \(0\leq\omega<1\), define \(\widetilde A_{\omega,d}(t)\)
as satisfying 
\(
 \E[\log((1-\omega)Y_i(d;t)+\omega A)]=\log(A).
\)
At \(\omega=1\), we take as limit  
\(\widetilde A_{1,d}(t)=\E[Y_i(d;t)]\).
Define
\(\theta_\omega(t)=A_{\omega,1}(t)/A_{\omega,0}(t)-1\) and
\(\widetilde\theta_\omega(t)=\widetilde A_{\omega,1}(t)/
\widetilde A_{\omega,0}(t)-1\). Then
\begin{equation}
 \widetilde\theta_\omega(t)-\theta_\omega(t)=O(t^3),
 \label{eq:shift-power-effect-second-order}
\end{equation}
uniformly over \(\omega\in[0,1]\) as \(t\to0\).
The effects coincide exactly at \(\omega=0\) and \(\omega=1\).
\end{proposition}

Under GPML weighting, matching the shift to the economic order therefore
approximates \(\theta_\omega\), with the log effect and NATE recovered
exactly at the endpoints. This need not require a small treatment effect but small dispersion within treatment states. The leading effect
error is cubic and depends on the difference in their third central log
moments.\footnote{Appendix~\ref{app:shift-power-approximation} gives the leading
cubic error. Appendix~\ref{app:shift-log-standardization} states the
additional conditions for the covariate-standardized comparison.}

This shifted-log transform also has a computational use. At a fixed interior approximation,
its residual has bounded derivatives in the index. Section~\ref{sec:iols}
 develops a new algorithm using these bounds.

\section{Iterated Ordinary Least Squares}
\label{sec:iols}
\label{sec:iols-fixed-tau}

The shifted-log construction makes it possible to compute power-mean
estimators using only ordinary least squares. Our iterated ordinary least
squares (IOLS) procedure is a numerical alternative or complement to solvers such as
IRLS: it first solves the approximate problem to obtain a starting value,
then solves the exact problem. Under the conditions below, the complete procedure is globally
convergent: it converges to the exact solution from any
finite initialization.

We use definitions from
Section~\ref{sec:comparison}. For $\omega=1$,
write \(r_{\tau i}(\beta)=\mu_i(\beta)^{2-\tau}(U_i(\beta)-1)\),
\(s_{\tau i}(\beta)=X_ir_{\tau i}(\beta)\),
\(g_\tau(\beta)=\E(s_{\tau i}(\beta))\) and
\(g_{\tau,n}(\beta)=n^{-1}\sum_i s_{\tau i}(\beta)\). Setting this
sample moment to zero gives \eqref{eq:iols-power-score}. Write
\(Q_n=X'X/n\), \(\|v\|_{Q_n}=(v'Q_nv)^{1/2}\) and
\(J_{\tau,n}(\beta)=-\partial g_{\tau,n}(\beta)/\partial\beta'\).

\subsection{Solving the shifted-log equation}
\label{sec:iols-power-mean-surrogate}

Fix a power-mean order \(\omega>0\) and a power-variance index
\(\tau\in[0,3]\). For the power-mean model, the fitted mean of
\(Y_i^\omega\) is \(\mu_i(\beta)^\omega=\exp(\omega X_i'\beta)\).
The exact moment is
\begin{equation}
 g_{\omega,\tau,n}(\beta)
 =\frac1{n\omega}\sum_iX_i\mu_i(\beta)^{\omega(2-\tau)}
 \left(\frac{Y_i^\omega}{\mu_i(\beta)^\omega}-1\right).
 \label{eq:iols-power-mean-exact-score}
\end{equation}
This is the power-score equation for \(Y_i^\omega\), with coefficient
\(\omega\beta\). Write \(s_{\omega,\tau,i}\) for its observation-level
contribution and
\(J_{\omega,\tau,n}=-\partial g_{\omega,\tau,n}/\partial\beta'\).
For a fixed numerical approximation \(0<\lambda<1\), use
\begin{equation}
 H^{(1)}_{\omega,\tau,\lambda,i}(\beta)
 =X_i'\beta+\frac1\omega\log\!\left(
 \frac{\lambda+(1-\lambda)Y_i^\omega\mu_i(\beta)^{\omega(1-\tau)}}
      {\lambda+(1-\lambda)\mu_i(\beta)^{\omega(2-\tau)}}
 \right).
 \label{eq:iols-power-mean-pilot-response}
\end{equation}
At \(\omega=1\), this is \(H_{\lambda,\tau,i}\) from
\eqref{eq:comparison-fitted-mean-shift}. The power order \(\omega\) and
weighting index \(\tau\) stay fixed as \(\lambda\uparrow1\) improves the
numerical approximation. OLS of this working outcome on \(X\) gives
\begin{equation}
 T^{(1)}_{\omega,\tau,\lambda,n}(\beta)
 =Q_n^{-1}\frac1n\sum_iX_iH^{(1)}_{\omega,\tau,\lambda,i}(\beta).
 \label{eq:iols-pilot-update}
\end{equation}
Only the working response changes, so a stable factorization of \(X\) can
be reused without forming a matrix inverse explicitly. We refer to
\eqref{eq:iols-pilot-update} as the Phase 1 map and denote its fixed
point by \(\widehat\beta^{(1)}_{\omega,\tau,\lambda,n}\).

\begin{assumption}[Positive-outcome variation]
\label{ass:iols-positive-overlap}
For the finite-sample numerical results, \(Y_i\geq0\), and:
\textnormal{(i)} \(X\) has full column rank;
\textnormal{(ii)} for \(D_Y=\operatorname{diag}(\mathbf1(Y_i>0))\)
and some \(\underline\eta>0\),
\(X'D_YX\succeq\underline\eta X'X\);
\textnormal{(iii)} at \(\tau=2\), every nonzero \(v\) satisfying
\(X_i'v\geq0\) at all positive outcomes also satisfies
\(\sum_iX_i'v>0\).
For a sample sequence, impose the applicable conditions with probability
approaching one.
\end{assumption}

Condition (ii) requires positive outcomes to keep variation in every
regressor direction; with an intercept and a binary regressor, it
excludes a group containing only zeros. Condition (iii) rules out a
direction along which the GPML criterion improves without a finite
coefficient limit. These sufficient conditions allow zeros. They concern
finite-root existence for the estimating equation, whatever solver is
used, and also make the OLS updates well defined.

Define the scalar pilot objective
\(\Psi^{(1)}_{\omega,\tau,\lambda,n}\), up to an additive constant,
through its gradient
\(\nabla\Psi^{(1)}_{\omega,\tau,\lambda,n}(\beta)
=-n^{-1}\sum_iX_i(H^{(1)}_{\omega,\tau,\lambda,i}(\beta)-q_i(\beta))\).\footnote{Let \(h_i(u)\) denote
the pilot working response evaluated at index \(q_i=u\) and write 
\(\Psi^{(1)}_{\omega,\tau,\lambda,n}(\beta)
=n^{-1}\sum_i\int_0^{q_i(\beta)}(u-h_i(u))\,du\),
normalized to zero at \(\beta=0\).}
Its first-order condition is exactly the pilot fixed-point equation.
\Needspace{3\baselineskip}
\begin{proposition}[Bounded curvature and convergence of Phase 1]
\label{thm:iols-bounded-power-surrogate}
Fix \(\omega>0\), \(\tau\in[1,2]\) and \(0<\lambda<1\).
Under Assumption~\ref{ass:iols-positive-overlap}, the Phase 1 map satisfies
\(T^{(1)}_{\omega,\tau,\lambda,n}(\beta)
=\beta-Q_n^{-1}\nabla\Psi^{(1)}_{\omega,\tau,\lambda,n}(\beta)\),
where the convex criterion satisfies
\begin{equation}
 0\preceq\nabla^2\Psi^{(1)}_{\omega,\tau,\lambda,n}(\beta)
 \preceq Q_n\qquad\text{for every }\beta\in\mathbb R^K.
 \label{eq:iols-power-mean-global-cap}
\end{equation}
The map is firmly nonexpansive in the \(Q_n\)-norm and has a unique
finite fixed point. From every finite \(\beta^{(0)}\), its iterates satisfy
\(\|\beta^{(t+1)}-\widehat\beta^{(1)}_{\omega,\tau,\lambda,n}\|_{Q_n}
\leq\kappa_1\|\beta^{(t)}-\widehat\beta^{(1)}_{\omega,\tau,\lambda,n}\|_{Q_n}\)
for all \(t\geq0\), with some \(\kappa_1\in(0,1)\) depending on the
sample, fixed indices and \(\beta^{(0)}\).
The corresponding exact equation also has a unique finite root
\(\widehat\beta_{\omega,\tau,n}\), with
\(\|\widehat\beta^{(1)}_{\omega,\tau,\lambda,n}
-\widehat\beta_{\omega,\tau,n}\|_{Q_n}
=O((1-\lambda)/\lambda)\)
as \(\lambda\uparrow1\), holding the sample, \(\omega\) and \(\tau\) fixed.
\end{proposition}

The curvature bound makes the OLS update a quadratic
majorization--minimization step, decreasing the pilot criterion without
a line search. Firm nonexpansiveness prevents increases in the
\(Q_n\)-distance between coefficient vectors.
Thus each fixed pilot is globally reachable from a finite starting value,
and refining its approximation brings its root arbitrarily close to the
exact root. The guarantee concerns this complete approximation procedure,
not a single large value of \(\lambda\) with an insufficient iteration budget.

The fixed point of \eqref{eq:iols-pilot-update} estimates the population
root of the transformed moment, which need not equal the coefficient of
the exact mean model. Define its contribution
\(\psi_{\lambda i}(\beta)=X_i(H^{(1)}_{\omega,\tau,\lambda,i}(\beta)
-X_i'\beta)\), suppressing the fixed \(\omega,\tau\), and write
\(\beta_\lambda\) for the population solution of
\(\E(\psi_{\lambda i}(\beta_\lambda))=0\). The following assumption is
stated for a generic moment contribution \(\psi_i(\beta)\) with
population root \(\beta_\star\), so it covers both \(\beta_\lambda\) in
Phase 1 and the exact equation in Phase 2.

\begin{assumption}[Sampling and moment regularity]
\label{ass:iols-sampling-design}
\label{ass:gpml-regularity}
For the specified \(\psi_i\) and \(\beta_\star\):
\textnormal{(i)} \(\{(Y_i,X_i)\}_{i=1}^n\) are independently and
identically distributed, with \(Y_i\geq0\), \(X_i\in\mathbb R^K\)
and fixed \(K\); \(X_i\) contains an intercept,
\(\E(X_iX_i')\) is finite and positive definite, and \(\beta_\star\)
is interior to a compact parameter set \(\mathcal B\);
\textnormal{(ii)} \(\E(\psi_i(\beta))=0\) has the unique solution
\(\beta_\star\) on \(\mathcal B\), \(\psi_i\) is continuously
differentiable, and
\(\E(\sup_{\beta\in\mathcal B}\|\psi_i(\beta)\|)<\infty\);
\textnormal{(iii)} on a neighborhood \(\mathcal N\) of \(\beta_\star\),
\(\E(\sup_{\beta\in\mathcal N}\|\partial\psi_i(\beta)/\partial\beta'\|)<\infty\),
\(\E(\sup_{\beta\in\mathcal N}\|\psi_i(\beta)\|^2)<\infty\), and
\(-\E(\partial\psi_i(\beta_\star)/\partial\beta')\) is positive definite.
\end{assumption}

For fixed \(\omega>0\), \(\tau\in[1,2]\) and \(0<\lambda<1\),
Proposition~\ref{prop:iols-fixed-lambda-inference} in
Appendix~\ref{app:iols-fixed-lambda-inference} establishes consistency
and asymptotic normality of the Phase 1 fixed point for
\(\beta_\lambda\), with the sandwich covariance of the transformed
moment, provided Assumption~\ref{ass:iols-positive-overlap} holds with
probability approaching one and
Assumption~\ref{ass:iols-sampling-design} holds with
\(\psi_i=\psi_{\lambda i}\) and \(\beta_\star=\beta_\lambda\).\footnote{Corollary~\ref{cor:iols-fixed-lambda-numerical} in
Appendix~\ref{app:iols-completion-supplement} gives the same first-order
distribution for a stopped iteration whose transformed sample-moment
residual is \(o_p(n^{-1/2})\).}
Phase 1 thus estimates \(\beta_\lambda\), not generally \(\beta_0\).
Under GPML weighting, the preceding section's local second-order
interpretation applies at power order \(\lambda\omega\).
To remove the approximation error, we refine the approximation and
solve the exact problem as follows.

\subsection{The complete procedure}
\label{sec:iols-ols-completion}
\label{sec:iols-inference}

The exact estimating equation can also be solved by repeated OLS,
with local convergence for a suitable initialization and step size.
For \(a_\rho=(1+\rho)^{-1}\), with \(\rho>0\), define the working response
\begin{equation}
 H_{\omega,\tau,\rho,i}(\beta)
 =X_i'\beta+\frac{a_\rho}{\omega}
 \mu_i(\beta)^{\omega(2-\tau)}
 \left(\frac{Y_i^\omega}{\mu_i(\beta)^\omega}-1\right).
 \label{eq:iols-phase2-working-outcome}
\end{equation}
Regressing this response on \(X\) by OLS gives the Phase 2 map:
\begin{equation}
 T_{\omega,\tau,\rho,n}(\beta)
 =\beta+a_\rho Q_n^{-1}g_{\omega,\tau,n}(\beta).
 \label{eq:iols-phase2-estimating-equation-form}
\end{equation}
Its fixed points are exactly the roots of
\(g_{\omega,\tau,n}(\beta)=0\).
At \(\omega=1\), the PPML and GPML working responses are
\(X_i'\beta+a_\rho(Y_i-\mu_i(\beta))\) and
\(X_i'\beta+a_\rho(Y_i/\mu_i(\beta)-1)\), respectively.

Two numerical properties shape the complete procedure: increasing
\(\lambda\) improves the root approximation but can slow Phase 1, and
decreasing \(a_\rho\) eventually makes the exact map locally contractive
but can slow Phase 2. Because suitable values are unknown in advance, the
guarantee below uses repeated finite trials. From any finite coefficient
vector, run Phase 1 along a prespecified sequence \(\lambda_j\uparrow1\),
optionally warm-starting each level at the preceding approximation, and
from each computed approximation try Phase 2 with a prespecified sequence
of steps \(a_{\rho_k}\downarrow0\) and a finite update budget. A trial
that stalls, diverges or exhausts its budget is skipped. The fixed pilot
candidates computed from the original start, paired with each fixed step,
are revisited with increasing budgets, and the procedure stops at the
first vector whose exact sample moment satisfies
\(\|g_{\omega,\tau,n}(\beta)\|_2\leq\varepsilon_n\).%
\footnote{Appendix~\ref{app:iols-fixed-design-example}
specifies the schedules.
Propositions~\ref{prop:iols-raw-delta-path} and
\ref{prop:iols-raw-slowdown} establish the pilot's properties.}
Our software implementation uses a finite warm-started
pilot grid, a curvature-scaled damping grid and Anderson acceleration.

\begin{theorem}[Two-phase convergence and inference]
\label{thm:iols-fixed-tau}
\label{cor:iols-local-corrector}
Fix \(\omega>0\), \(\tau\in[1,2]\), and assume 
Assumption~\ref{ass:iols-positive-overlap}.
\textnormal{(i) Computation.}
From every finite start, the repeated-trial two-phase procedure reaches
\(\|g_{\omega,\tau,n}(\beta)\|_2\leq\varepsilon\) for any
\(\varepsilon>0\) in finitely many OLS updates. Near the exact root,
the Phase 2 map converges Q-linearly when
\(0<a_\rho<2/\nu_{\max,n}\), where \(\nu_{\max,n}\) is the
largest eigenvalue of \(Q_n^{-1/2}J_{\omega,\tau,n}
(\widehat\beta_{\omega,\tau,n})Q_n^{-1/2}\).

\textnormal{(ii) Asymptotic properties.}
Suppose also that \(\E(Y_i^\omega\mid X_i)=\mu_i(\beta_0)^\omega\),
Assumption~\ref{ass:iols-sampling-design} holds for
\(\psi_i=s_{\omega,\tau,i}\), \(\beta_\star=\beta_0\). For a numerical tolerance satisfying
\(\varepsilon_n=o(n^{-1/2})\), the completed estimator
\(\overline\beta_{\omega,\tau,n}\) converges in probability to
\(\beta_0\) and satisfies
\begin{equation}
 \sqrt n(\overline\beta_{\omega,\tau,n}-\beta_0)
 =J_{\omega,\tau}^{-1}\frac1{\sqrt n}\sum_i
 s_{\omega,\tau,i}(\beta_0)+o_p(1),
 \label{eq:iols-asymptotic-linearity}
\end{equation}
where \(J_{\omega,\tau}=\E(X_iX_i'\mu_i(\beta_0)^{\omega(2-\tau)})\).
Consequently,
\begin{equation}
 \sqrt n(\overline\beta_{\omega,\tau,n}-\beta_0)
 \Longrightarrow N(0,\Omega_{\omega,\tau}),
 \label{eq:iols-asymptotic-normality}
\end{equation}
with \(\Omega_{\omega,\tau}=J_{\omega,\tau}^{-1}
\E(s_{\omega,\tau,i}(\beta_0)s_{\omega,\tau,i}(\beta_0)')
J_{\omega,\tau}^{-1}\), as $n\to\infty$.
\end{theorem}

Finite-tolerance completion therefore preserves the exact estimator's
consistency and first-order distribution.%
\footnote{Appendix~\ref{app:iols-covariance} gives the covariance
estimator and standardization to proportional effects, including
variation from the sampled covariates.}
These conclusions also extend to a fixed target
\(\tau_\star\in[0,3]\setminus[1,2]\) under the additional computational
and statistical conditions in
Appendix~\ref{app:iols-tau-continuation}.
Proposition~\ref{prop:iols-direct-continuation} establishes that,
under uniform curvature and containment on a specified region,
the solution at the nearest endpoint of \([1,2]\) can be continued
to \(\tau_\star\), holding \(\omega\) fixed, through Phase 2 OLS
updates initialized at the preceding solution.
The procedure follows the branch connected to that endpoint.
It does not require global convexity or uniqueness.

\label{sec:iols-starting-value}
Under the conditions of Theorem~\ref{thm:iols-fixed-tau}, IOLS combines
convergence from any finite start with a fixed matrix \(Q_n\) whose
factorization can be reused. The trade-off is a fixed metric that does
not adapt to changing curvature.

 Without damping or step control, even a zero starting value can cause
extreme overshooting. With outcomes \(0\) and \(2{,}000\), an
intercept-only PPML or GPML model fits mean \(1{,}000\). Poisson Newton
and Gamma Fisher scoring update the log mean as
\(q^{(t+1)}=q^{(t)}+1{,}000\exp(-q^{(t)})-1\).
Starting at \(q^{(0)}=0\) gives \(q^{(1)}=999\), at which direct
exponentiation overflows in double precision. Even avoiding overflow,
subsequent updates decrease the log mean by less than one, while the
solution is only \(\log(1{,}000)\simeq6.91\).
Line searches and trust regions address this difficulty. Undamped scoring iterations can also be
locally unstable and oscillate persistently, implementations may
encounter numerically singular systems or nonfinite working quantities,
and small coefficient changes can trigger a convergence declaration
while the original estimating equation remains far from satisfied.

\label{sec:iols-pilot-other-solvers}
Phase~1 promotes numerical stability through its curvature-controlled
surrogate and can also initialize Newton, scoring or quasi-Newton
refinement under the corresponding local convergence conditions instead
of Phase 2 IOLS. Alternatively, for
\(\tau\in[1,2]\), a regularized surrogate Hessian can precondition
search directions for the exact criterion, with step lengths chosen by a
line search.%

\subsection{Numerical comparison}
\label{sec:iols-numerical-comparison}

Table~\ref{tab:solver-preliminary-main} compares three numerical solvers
for the same PPML or GPML equation: IOLS and \texttt{statsmodels}'
observed-Hessian Newton and Fisher-scoring IRLS.%
\footnote{For GPML, an adapter evaluates the quasi-log-likelihood
\(-\sum_i(Y_i/\mu_i+\log(\mu_i))\), with a zero ratio contribution
when \(Y_i=0\). The adapter leaves the package's score, Hessian and IRLS
updates unchanged. Stopping uses coefficient changes,
rather than the recorded quasi-loss. Appendix~\ref{app:sim-completed-four-way}
documents the remaining numerical failures, which also occur in
strictly positive designs.}

The experiment uses 1,000 independent samples per simulation design at
\((n,K)=(1{,}000,10)\) and \((10{,}000,100)\), for sixteen ordinary
and four stress designs preserving the exponential conditional mean:
40,000 datasets in total. Stress designs add high dispersion, heavy
tails, rare large outcomes or near collinearity. Success requires
passing the \(10^{-8}\) accuracy check within 120 CPU and
300 wall seconds. Median and mean CPU times include all attempts,
with each recorded CPU time capped at 120 seconds.%
\footnote{Earlier uncapped runs lasted up to 38.1 minutes without
passing the accuracy check, motivating the common time limits.}
Timing summaries should be interpreted alongside failure rates,
since an unsuccessful attempt may terminate quickly.%

\begin{table}[htbp]
\centering
\begin{singlespace}
\small
\caption{Solver comparison under a common computational budget}
\label{tab:solver-preliminary-main}
\setlength{\tabcolsep}{2pt}
\begin{tabular}{@{}llrcrcrc@{}}
\toprule
 & & \multicolumn{2}{c}{IOLS}
   & \multicolumn{2}{c}{Newton}
   & \multicolumn{2}{c}{IRLS}\\
\cmidrule(lr){3-4}\cmidrule(lr){5-6}\cmidrule(l){7-8}
Design, \(n\) & Equation
 & Fail \% & \shortstack{Time (ms)\\med./mean}
 & Fail \% & \shortstack{Time (ms)\\med./mean}
 & Fail \% & \shortstack{Time (ms)\\med./mean}\\
\midrule
Ordinary, 1,000 & PPML
 & 0.00 & 2.4/2.7
 & 0.00 & 1.0/1.1
 & 0.00 & 1.2/1.3\\
 & GPML
 & 0.00 & 2.7/8.1
 & 1.94 & 1.0/28
 & 0.07 & 2.8/12\\
\addlinespace[2pt]
Ordinary, 10,000 & PPML
 & 0.00 & 106/111
 & 0.00 & 67/70
 & 0.00 & 200/206\\
 & GPML
 & 0.00 & 115/118
 & 0.32 & 67/102
 & 0.04 & 555/722\\
\addlinespace[2pt]
Stress, 1,000 & PPML
 & 0.00 & 3.4/4.8
 & 0.15 & 1.2/9.3
 & 0.15 & 1.5/7.8\\
 & GPML
 & 0.00 & 3.3/6.5
 & 14.62 & 1.5/436
 & 40.98 & 3.4/2,223\\
\addlinespace[2pt]
Stress, 10,000 & PPML
 & 0.00 & 158/202
 & 4.67 & 113/1,643
 & 0.35 & 397/496\\
 & GPML
 & 0.00 & 144/192
 & 27.75 & 105/14,831
 & 75.02 & 339/419\\
\bottomrule
\end{tabular}

\smallskip
\begin{minipage}{\linewidth}
\footnotesize
\raggedright
\textit{Notes:}
Methods use identical draws, with \(K=10\) at \(n=1{,}000\)
and \(K=100\) at \(n=10{,}000\).
Failure means no returned output passing the \(10^{-8}\)
accuracy check within both 120 CPU and 300 wall seconds.
Time entries report the median/mean CPU cost in milliseconds
over all attempts, including failures, with each recorded CPU
time capped at 120 seconds. Setup is excluded.
Certified structural exclusions are omitted from both summaries.
Appendix~\ref{app:sim-completed-four-way} reports failure types,
uncapped timings and software details.
\end{minipage}
\end{singlespace}
\end{table}

IOLS passes every attempted \(10^{-8}\) accuracy check within budget.
In the larger-stress GPML designs, Newton and IRLS fail this check
on 27.75 and 75.02 percent of attempts, respectively.
Newton's unsuccessful attempts include time-budget overruns and
inaccurate outputs despite reported convergence, while IRLS predominantly
reports failure. Neither method recovers additional accurate fits
in this cell when the tolerance is relaxed to \(10^{-6}\).\footnote{Appendix~\ref{app:sim-completed-four-way} also compares
SciPy's L-BFGS-B, which misses the strict check on 66.83 percent
of larger-stress GPML attempts despite reporting convergence.
Most of these misses disappear at \(10^{-6}\).
Scikit-learn's Newton--Cholesky solver with backtracking line search
achieves lower median CPU times; its seven strict misses across
the full experiment all satisfy the secondary \(10^{-6}\) tolerance.}

On draws accurately solved by all three main-table methods, the
treatment coefficients differ by less than \(5\times10^{-11}\).
For ordinary designs at \(n=10{,}000\), normalized-ATE bias is below
0.35 percentage points in absolute value, with standard deviations
of 1.49--6.07 points for a true effect of 8 percent.
The appendix also reports accurate IOLS estimates when one or both
full-step comparators fail, including extreme small-sample estimates.
Altogether, these results support IOLS as a reliable and
computationally competitive alternative to established solvers.

\subsection{Endogenous regressors}
\label{sec:iv}
\label{sec:extension-iv}

Our approach also extends to endogenous regressors under appropriate IV restrictions.
With instruments \(Z_i\), multiplicative IV uses
\(\E(Z_i(U_i(\beta_0)-1))=0\), implied by
\(\E(U_i(\beta_0)\mid Z_i)=1\)
\citep{mullahy1997instrumental}, whereas additive-error IV uses
\(\E(Z_i(Y_i-\mu_i(\beta_0)))=0\), implied by
\(\E(Y_i-\mu_i(\beta_0)\mid Z_i)=0\)
\citep{windmeijersantossilva1997}.

We pursue the multiplicative formulation because it is natural when endogeneity arises
from omitted determinants that, like the included regressors, affect
the outcome proportionally. It therefore treats observed and unobserved
determinants symmetrically within the multiplicative model
\citep{mullahy1997instrumental}. The additive formulation treats omitted determinants differently from
included regressors: the latter enter multiplicatively, whereas the former
enter additively. This is less natural when the
omitted variables are expected to operate in the same way as the
included ones. Consequently, orthogonality of the
instruments to the multiplicative error does not imply orthogonality to the additive error, and vice versa. Alternative power orders and
fitted-mean weights likewise require separately justified moments.

Under exact identification, replacing OLS by two-stage least squares
in both phases gives iterated 2SLS (I2SLS). The pilot approaches the
multiplicative sample equation as \(\lambda\uparrow1\).
Under the identification and regularity conditions in
Appendix~\ref{app:iv-statements}, the resulting IV estimator is consistent and
asymptotically normal. The appendix also establishes local
geometric convergence of both phases under its stability
conditions, with the pilot initializing exact completion.

With overidentification, we use the shifted-log pilot to initialize
minimization of a nonlinear GMM criterion. We solve successive approximate problems through repeated
linear regressions, increasing \(\lambda\) toward one and using
each solution to initialize the next, before completing the exact
GMM problem. Proposition~\ref{prop:iv-pilot-initialization} provides the local justification:
along a regular branch, the pilot solutions approach a
local GMM minimum and eventually provide suitable starting values
for exact completion. This construction offers a way to approach
a solution of a possibly nonconvex problem without requiring
global convexity or a globally unique minimum.
\label{sec:iv-gmm}

The three fixed-effect implementations remain available when the
instruments satisfy the moment restrictions for the corresponding
transformed error. Cancellation requires conditioning on the complete
group history. Absorption alone provides neither instrument validity
nor protection against incidental-parameter bias.

\section{Fixed effects and incidental parameters}
\label{sec:fixed-effects}
\label{sec:fe-ipp}

Fixed effects raise distinct computational and statistical issues.
Absorption reduces computational cost but does not resolve the
incidental-parameter problem (IPP). In a short panel, for example, adding
individuals adds individual effects, each still estimated from only a few
observations. Their estimation errors can contaminate the coefficients of
interest, preventing consistency or shifting confidence intervals.
Whether this occurs depends on the model, estimator and information per
effect \citep{hausmanhallgriliches1984,wooldridge1999distributionfree,
fernandez2016individual,fernandezvalweidner2018review}.
We show how to remove one multiplicative effect block without estimating it, using
centered moments valid for a specified positive smooth working variance.
When retained effects also grow, we establish consistency and apply an
estimating-equation correction to the remaining coefficient bias
\citep{kosmidislunardon2024empirical}.

Our illustrative case is a three-way gravity panel
with a fixed number of dates and a growing number of countries following \citet{weidner2021bias}:  
cancelling pair effects removes a growing block of poorly estimated
coefficients, but the retained country-date effects can still leave bias
of the same order as the standard error. Cancellation and correction
therefore address different problems.

\subsection{Power means with fixed effects}
\label{sec:fe-estimation}

We present cancellation for \(\omega=1\). As in
Section~\ref{sec:residual-weighting}, any fixed positive power
order uses the same construction with \(Y_i\) replaced by
\(Y_i^\omega\) and the linear index multiplied by \(\omega\),
under the corresponding conditional-mean restriction.

Let \(F_C\) contain indicators for the groups whose effects are to be
removed, and let \(M_C=I-F_C(F_C'F_C)^{-1}F_C'\) subtract group
means. The within design matrix \(B=M_CX\) has rows
\(B_i=X_i-\bar X_g\), where \(\bar X_g\) is the regressor mean
in group \(g\). Group-constant regressors, including the intercept,
disappear and we remove any further linearly dependent columns. We set \(\mu_i(\beta)=\exp(B_i'\beta)\) and condition on all group
regressors and the effect \(\alpha_g\), denoted  \(\mathcal H_g\):
\begin{equation}
 \E(Y_i\mid\mathcal H_g)=A_g\mu_i(\beta_0),\qquad i\in g,
 \label{eq:fe-cancel-mean}
\end{equation}
Here \(A_g=\exp(\alpha_g+\bar X_g'\beta_0)\) absorbs the
group mean of the original index.
Conditioning includes every group member's regressors because their
average enters \(B_i\).

For a positive working variance \(V(\mu)\), define
\(w_i(\beta)=\mu_i(\beta)^2/V(\mu_i(\beta))\).
Within each cancellation group, define
\(\bar B_{g,w}(\beta)=\sum_{i\in g}w_i(\beta)B_i/
\sum_{i\in g}w_i(\beta)\).
Write
\(\psi_{g,w}(\beta)=\sum_{i\in g}
w_i(\beta)(B_i-\bar B_{g,w}(\beta))Y_i/\mu_i(\beta)\)
for group \(g\)'s contribution. The estimator solves
\begin{equation}
 \Psi_{w,n}(\beta)=\frac1n\sum_g\psi_{g,w}(\beta)=0.
 \label{eq:fe-general-cancel-score}
\end{equation}

Since \(\E(Y_i/\mu_i(\beta_0)\mid\mathcal H_g)=A_g\),
\(\E(\psi_{g,w}(\beta_0)\mid\mathcal H_g)
=A_g\sum_{i\in g}w_i(\beta_0)
(B_i-\bar B_{g,w}(\beta_0))=0\).
Weighted centering therefore eliminates \(A_g\), implying that no consistent
estimate of an individual short-group scale is needed.\footnote{Appendix~\ref{app:fe-postestimation}
treats scale recovery and standardized averages.}

The baseline asymptotic setting has many short groups: the number of
cancelled effects grows, while the number of observations per group and
the dimension of the retained coefficient vector remain bounded. This is relaxed in the next subsection.

\begin{assumption}[Weighted group moments]
\label{ass:fe-weighted-groups}
\textup{(i) Sampling and moments.}
Groups are independent and their number satisfies \(G_n\to\infty\).
Group sizes are uniformly bounded, \(\sup_{n,g}n_g<\infty\), and
\(n/G_n\to c\in(0,\infty)\), where
\(n=\sum_{g=1}^{G_n}n_g\).
The dimension of \(\beta\) is fixed, and
\(\beta_0\in\operatorname{int}(\Theta)\), with \(\Theta\) compact.
The within-group deviations \(B_i=X_i-\bar X_g\) satisfy
\(\sup_{n,i}\|B_i\|_2\leq C_B\) almost surely for a fixed
\(C_B<\infty\), and
\(\sup_{n,i}\E(Y_i^{2+\epsilon})<\infty\) for some \(\epsilon>0\).

\textup{(ii) Working variance and information.}
The specified working variance
\(V:(0,\infty)\to(0,\infty)\) is continuously differentiable.
Let \(J_{w,n}(\beta)=-\partial\Psi_{w,n}(\beta)/\partial\beta'\).
As \(n\to\infty\),
\(\E(J_{w,n}(\beta_0))\to J_w\succ0\) and
\(n^{-1}\sum_g\E(\psi_{g,w}(\beta_0)\psi_{g,w}(\beta_0)')
\to\Sigma_w\), where \(\Sigma_w\) is finite.
\end{assumption}

Dependence within a group is unrestricted, and \(J_w\succ0\) requires
identifying within-group variation. Weights evaluated at \(\mu_i\)
need not be optimal for the full mean \(A_g\mu_i\) or its covariance.

\begin{proposition}[Consistency and inference after cancellation]
\label{prop:fe-general-cancellation}
Suppose \eqref{eq:fe-cancel-mean} and
Assumption~\ref{ass:fe-weighted-groups} hold. With probability
approaching one, \eqref{eq:fe-general-cancel-score} has a solution
in \(\Theta\). Let \(\widehat\beta_n\) denote the corresponding estimator. Then \(\widehat\beta_n\)
is consistent and, as \(n\to\infty\), satisfies
\begin{equation}
 \sqrt n(\widehat\beta_n-\beta_0)
 \Longrightarrow N(0,\Omega_w),
 \label{eq:fe-general-cancel-normality}
\end{equation}
where \(\Omega_w=J_w^{-1}\Sigma_w(J_w^{-1})'\).
The group-sandwich estimator consistently estimates \(\Omega_w\).
\end{proposition}

The number of cancelled effects can therefore grow without inducing the
usual short-panel bias: these effects are eliminated rather than
estimated, and the limiting distribution remains centered at \(\beta_0\).

For the power working variance \(V(\mu)=\mu^\tau\),
\(w_i=\mu_i^{2-\tau}\). At \(\tau=1\), cancellation yields
the same slope estimates as profiled one-way PPML
\citep{hausmanhallgriliches1984}.
At \(\tau=2\), the GPML equation is changed to 
\begin{equation}
 \frac1n\sum_i B_i\left(Y_i\exp(-B_i'\beta)-1\right)=0.
 \label{eq:fe-cancel-score}
\end{equation}
Here \(w_i=1\) and \(\bar B_{g,w}=0\). For \(\tau\neq1\),
cancellation differs from profiling, which keeps the joint
estimator and does not itself remove IPP.\footnote{Proposition~\ref{prop:fe-fixed-design} and
Appendix~\ref{app:fe-gamma-inference} treat computation and localization
of Gamma-weighted numerical outputs.}

\subsection{Cancellation and correction}
\label{sec:fe-mixed-applications}
\label{sec:fe-choice}
When the model contains both effects to be cancelled, \(F_C\), and effects to
be kept, \(F_R\), the cancellation route is applied to the full retained
parameter rather than to \(\beta\) alone. Use
\(B=M_C[X,F_R]\), remove collinear columns, and solve for
\(\theta=(\beta',\alpha')'\). The cancelled scale absorbs group
means of the complete index \(X\beta+F_R\alpha\). Yet, 
cancellation does not remove bias from estimating \(\alpha\).

In a two-way panel with \(G_n\to\infty\) and fixed \(T\),
Proposition~\ref{prop:fe-general-cancellation} gives consistency and
valid inference after cancelling individual effects and keeping
the \(T-1\) normalized time indicators. Cancelling time effects instead
leaves the growing individual block and does not generally resolve the
short-panel problem.

When several dimensions grow, the retained effects can become increasingly
well estimated. The coefficients of interest may then be consistent even
though their remaining bias is not negligible relative to their standard
errors. We use a second-order estimating-equation correction for this case.

Write \(d_n=\dim(\theta)\) for the total number of estimated
coefficients after normalization and any cancellation. The coefficients
of interest \(\beta=L_n\theta\) have fixed dimension, with
\(L_n=(I,0)\). Throughout, \(n\) denotes the total number of observations.
For independent short groups, keep
\(n/G_n\to c\in(0,\infty)\), as in
Assumption~\ref{ass:fe-weighted-groups}.
Define \(S(\theta)=\sum_{i=1}^n\psi_i(\theta)\), with 
\(S(\theta)=\sum_g\psi_g(\theta)\) for independent short groups, and
\(H=-\E(\nabla_\theta S(\theta_0))\), where \(\psi_i(\theta)\) is
observation \(i\)'s (\(\psi_g(\theta)\) group $g$'s) contribution to the chosen estimating equation and
\(\nabla_\theta S=\partial S/\partial\theta'\).

Assumption~\ref{ass:fe-growing-regularity} allows the number of estimated coefficients,
including the remaining fixed effects, to grow at
\(d_n=O(\sqrt n)\), provided identifying information is not
concentrated in a few observations or groups. At this rate, the
coefficient bias can vanish while remaining comparable to the
standard error, making bias correction important for inference.
Sampling permits independent short groups with within-group
dependence, or cells with a bounded-degree dependency graph. The remaining conditions require the conditional mean restriction
given the complete design, bounded \(Y_i^\omega\), regressors and
linear predictors at \(\theta_0\), a positive working variance
with four bounded derivatives, and a positive-definite
root-\(n\) target-variance limit. Only the feasible cell correction after
cancellation additionally requires growing cancellation groups
to estimate these scales uniformly, the short-group correction
uses group scores without estimating their scales.

\begin{theorem}[Growing effects and coefficient-bias correction]
\label{thm:fe-growing-general}
Under Assumption~\ref{ass:fe-growing-regularity}, the specified joint
or cancelled equation admits, with probability approaching one, a
locally unique solution \(\widehat\theta_n\) satisfying
\(\max_{i\le n}|B_i'(\widehat\theta_n-\theta_0)|=o_p(1)\) as \(n\to\infty\). The estimator
\(\widehat\beta_n=L_n\widehat\theta_n\) is consistent and satisfies
\begin{equation}
 \widehat\beta_n-\beta_0
 =L_nH^{-1}S(\theta_0)+b_n+o_p(n^{-1/2}),
 \label{eq:fe-growing-general-expansion}
\end{equation}
where \(b_n=O(d_n/n)\) is the second-order bias.
The equation-specific correction \(\widehat b_n\) in
\eqref{eq:fe-growing-group-correction} for independent short groups,
or Appendix~\ref{app:fe-growing-corrections} for the cell formulation,
satisfies \(\widehat b_n-b_n=o_p(n^{-1/2})\). Consequently, as
\(n\to\infty\),
\begin{equation}
 \sqrt n(\widehat\beta_n-\widehat b_n-\beta_0)
 \Longrightarrow N(0,\Omega),
 \label{eq:fe-growing-corrected-limit}
\end{equation}
where \(\Omega\succ0\) is the asymptotic covariance of the scaled
first-order term and is consistently estimable.
The same centered limit holds without correction if
\(d_n=o(\sqrt n)\) or \(b_n=o(n^{-1/2})\).
\end{theorem}

The result applies to joint estimation and to cancellation followed
by estimation of the remaining effects. When \(d_n\) is of order \(\sqrt n\), bias can be of standard-error order. Its correction recenters inference without changing first-order covariance.%
\footnote{Proposition~\ref{prop:fe-growing-selection} selects the local
solution using a Gamma-weighted preliminary estimate and Newton
completion. Appendix~\ref{app:fe-growing-delete-covariance} gives an
optional positive-semidefinite deletion covariance.}

For independent short groups, write
\(S(\theta)=\sum_g\psi_g(\theta)\),
\(\widehat H=-\nabla_\theta S(\widehat\theta_n)\), and
\(\widehat v_g=\widehat H^{-1}\psi_g(\widehat\theta_n)\).
The estimated bias is
\begin{equation}
 \widehat b_n=L_n\widehat H^{-1}\sum_g
 \left(\nabla_\theta\psi_g(\widehat\theta_n)\widehat v_g
 +\frac12\nabla_\theta^2 S(\widehat\theta_n)
       [\widehat v_g,\widehat v_g]\right).
 \label{eq:fe-growing-group-correction}
\end{equation}
The \(j\)th component of \(\nabla_\theta^2S(\theta)[v,v]\) is
\(v'\nabla_\theta^2S_j(\theta)v\): it measures the curvature of that
estimating equation in direction \(v\), held fixed.
This corresponds to the reduced-bias M-estimator of
\citet{kosmidislunardon2024empirical}, applied to the chosen equation
and restricted to \(\beta\). It removes leading coefficient bias,
not finite-sample covariance bias.\footnote{Appendix~\ref{app:fe-growing-corrections}
gives the Taylor derivation, exact mapping, nonsymmetric-Jacobian sandwich
and distinct cell formula.}
The correction accounts for two sources of bias: observations that change
the estimating equation also change its slope, and its curvature makes
the average nonlinear response differ from the response to an average
displacement. The correction removes
their leading contribution.

\begin{remark}[Three-way panels]
With \(N\) countries and fixed \(T\geq2\), cancelling \(O(N^2)\)
pair effects leaves \(O(N)\) country-date effects. Under
Assumption~\ref{ass:fe-growing-regularity}, independent pair histories
and complete or diagonal-omitted dense support,
Theorem~\ref{thm:fe-growing-general} gives consistency. Remaining bias
and standard errors can both be \(O(1/N)\). The group correction and
pair sandwich give centered inference. When two effect blocks remain
short, product moments require additional restrictions
\citep{jochmans2017two}.\footnote{Appendix~\ref{app:fe-decision-tables} gives practical decision rules for many common fixed-effects topologies.}
\end{remark}

At \(\tau=1\), cancellation coincides with profiled PPML, for
which \citet{weidner2021bias} derive both coefficient-bias and
covariance adjustments. Our results establish consistency and
feasible coefficient-bias correction for cancelled estimators
with general working variances and growing retained effects.
Under stronger regularity conditions, we also derive a
higher-order covariance adjustment for the bias-corrected
estimator.\footnote{Proposition~\ref{prop:fe-growing-covariance-refinement} in Appendix~\ref{app:fe-growing-covariance-refinement}
removes the leading \(O(d_n/n^2)\) bias of the covariance
estimator, leaving an \(o(d_n/n^2)\) remainder. The result
covers independent short groups and accounts for uncertainty
from estimating the coefficient correction.}

\subsection{Simulation evidence for cancellation and correction}
\label{sec:fe-simulation-design}

\begin{figure}[!htbp]
\centering
\includegraphics[width=\linewidth]{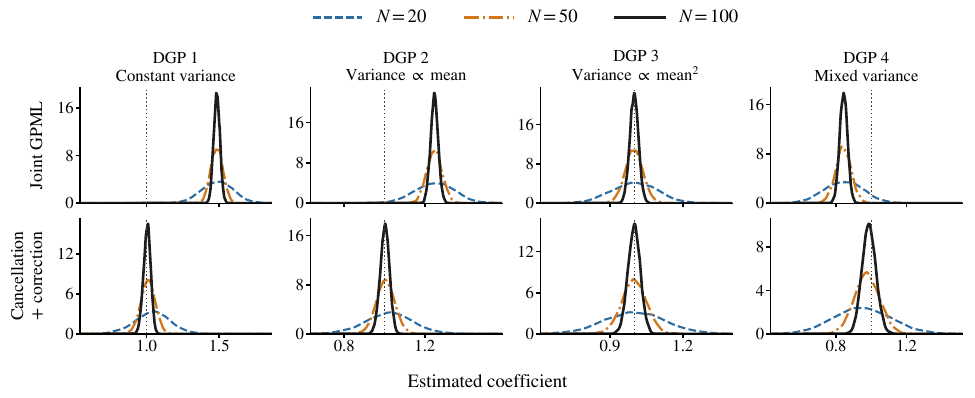}
\caption{GPML coefficient distributions, \(T=2\).}
\label{fig:mc-fe-gpml-T2}
\BBPtablenotes{Kernel densities use 5,000 estimates per curve at
\(N=20,50,100\). Dotted lines mark the true coefficient, \(\beta_0=1\).}
\end{figure}

We compare joint estimation, cancellation, and cancellation followed
by bias correction in the three-way panel design of
\citet{weidner2021bias}, using 5,000 replications at each
\(N=50,100,500\), with \(T=2\). The four DGPs share the exponential conditional mean. The first
three have \(\operatorname{Var}(Y_i\mid X_i)\propto\mu_i^\tau\),
with \(\tau=0\), \(\tau=1\) (PPML), and \(\tau=2\) (GPML);
the fourth allows additional covariate dependence.
Joint and cancelled PPML coincide; bias-corrected PPML uses
Weidner and Zylkin's coefficient correction. Both corrected PPML
and corrected GPML keep ordinary pair-cluster sandwich standard
errors.\footnote{Appendix~\ref{app:sim-proposed-fe} details the simulation
design and compares ordinary sandwich standard errors with
the PPML covariance adjustment of \citet{weidner2021bias}
and our higher-order covariance refinement for general working
variances.}
Table~\ref{tab:mc-fe-main-T2} reports bias, standard errors and
coverage. Figure~\ref{fig:mc-fe-gpml-T2} complements these summaries:
joint GPML concentrates away from the true coefficient in DGPs~1,
2 and~4, while cancellation followed by coefficient correction
substantially improves centering.

\begin{table}[!htbp]
\centering
\caption{Centering, standard errors and coverage, \(T=2\).}
\label{tab:mc-fe-main-T2}
\begingroup
\normalsize
\setstretch{1}
\setlength{\tabcolsep}{2pt}
\renewcommand{\arraystretch}{0.90}
\begin{tabular}{@{}rl*{4}{rrr}@{}}
\toprule
 & & \multicolumn{3}{c}{DGP 1} & \multicolumn{3}{c}{DGP 2} & \multicolumn{3}{c}{DGP 3} & \multicolumn{3}{c}{DGP 4} \\
\cmidrule(lr){3-5}\cmidrule(lr){6-8}\cmidrule(lr){9-11}\cmidrule(lr){12-14}
\(N\) & Method
 & Bias & SE & Cov.
 & Bias & SE & Cov.
 & Bias & SE & Cov.
 & Bias & SE & Cov. \\
\midrule
50 & GJ & 11.48 & 0.98 & 0.0 & 6.52 & 0.99 & 0.0 & -0.02 & 0.98 & 94.3 & -3.59 & 0.96 & 4.7 \\
 & GC & 0.63 & 0.88 & 84.3 & 0.36 & 0.91 & 89.8 & -0.02 & 0.90 & 91.8 & -0.57 & 0.83 & 83.3 \\
 & GC+BC & 0.20 & 0.85 & 89.1 & 0.09 & 0.88 & 91.4 & -0.02 & 0.86 & 90.9 & -0.34 & 0.77 & 84.5 \\
 & PJ & 0.58 & 0.89 & 85.5 & 0.35 & 0.92 & 90.3 & 0.01 & 0.90 & 92.0 & -0.55 & 0.83 & 84.3 \\
 & PJ+BC & 0.10 & 0.85 & 89.8 & 0.04 & 0.89 & 91.7 & -0.03 & 0.86 & 90.8 & -0.30 & 0.77 & 84.6 \\
\addlinespace[2pt]
100 & GJ & 22.64 & 0.96 & 0.0 & 13.16 & 0.99 & 0.0 & -0.02 & 0.99 & 94.5 & -7.19 & 0.97 & 0.0 \\
 & GC & 0.66 & 0.94 & 86.7 & 0.36 & 0.96 & 91.9 & -0.03 & 0.94 & 93.2 & -0.65 & 0.85 & 82.3 \\
 & GC+BC & 0.12 & 0.91 & 92.1 & 0.04 & 0.94 & 93.0 & -0.02 & 0.91 & 92.2 & -0.33 & 0.79 & 85.5 \\
 & PJ & 0.59 & 0.95 & 88.8 & 0.35 & 0.96 & 92.5 & 0.01 & 0.94 & 93.4 & -0.64 & 0.85 & 83.2 \\
 & PJ+BC & 0.04 & 0.92 & 92.6 & 0.00 & 0.95 & 93.6 & -0.03 & 0.91 & 92.4 & -0.29 & 0.79 & 86.0 \\
\addlinespace[2pt]
500 & GJ & 99.63 & 0.84 & 0.0 & 64.93 & 0.97 & 0.0 & -0.01 & 0.99 & 95.4 & -32.91 & 0.89 & 0.0 \\
 & GC & 0.70 & 0.97 & 87.7 & 0.39 & 0.98 & 92.5 & -0.02 & 0.97 & 94.4 & -0.84 & 0.92 & 81.7 \\
 & GC+BC & 0.05 & 0.95 & 93.7 & 0.01 & 0.97 & 94.4 & -0.02 & 0.97 & 94.3 & -0.26 & 0.87 & 89.9 \\
 & PJ & 0.59 & 0.98 & 89.5 & 0.36 & 0.98 & 92.9 & 0.01 & 0.98 & 94.6 & -0.85 & 0.92 & 81.9 \\
 & PJ+BC & 0.00 & 0.96 & 94.0 & -0.01 & 0.98 & 94.6 & -0.02 & 0.97 & 94.3 & -0.23 & 0.87 & 90.3 \\
\bottomrule
\end{tabular}
\endgroup

\BBPtablenotes{Bias and mean SE are reported in units of the
empirical SD. Cov. reports 95\% interval coverage (\%).
G/P: GPML/PPML; J: joint estimation; C: pair-effect cancellation;
BC: coefficient bias correction.
All 5,000 estimates and intervals per cell are valid.
\(N\) counts origins and destinations; \(n=N^2T\).
SEs cluster by pair; both corrected estimators use ordinary
sandwich SEs. PPML BC uses Weidner and Zylkin's (2021)
coefficient correction.}
\end{table}

Cancellation substantially reduces GPML bias, and coefficient
correction delivers performance comparable to WZ-corrected PPML.
At \(N=500\), corrected GPML coverage is 93.7\%, 94.4\%, 94.3\%
and 89.9\% across the four DGPs. Remaining undercoverage is most
pronounced in DGP~4, where mean SE is only 87.5\% of empirical SD.

\section{Political centralization and development}
\label{sec:application}
\label{subsec:political-centralization}

\citet{michalopoulos_etal_2013} relate pre-colonial political
centralization to contemporary African development. We revisit their
nighttime-light specification (Table~III, column~4). The unit is an
ethnic homeland within a country, and the main regressor is a
jurisdictional hierarchy index ranging from zero to four. Nighttime
light density \(Y\) serves as a proxy for economic development.
We use the original controls and cluster standard errors by
country and ethnic family.

Table~\ref{table:michalopoulos} compares PPML and GPML with OLS on
\(\log(Y+c)\) for three shifts, including \(c=0.01\), used by
\citet{michalopoulos_etal_2013}. For every specification, we report \(100(\exp(\gamma_h)-1)\).
For PPML and GPML, this measures the percentage association with
mean light density per hierarchy category. For shifted-log OLS,
it measures the percentage association with the 
geometric mean of \(Y+c\). Joint country-effect estimation is our main specification, and 
cancellation provides a sensitivity check.%
\footnote{Appendix~\ref{app:applications} gives specification and sensitivity details, and
Appendix~\ref{app:application-all-coefficients} reports all coefficients.}

The homeland with the largest outcome value, i.e. highest light density,
accounts for about 10\% of summed light
density, and the brightest 2\% for about half. We therefore compare
the full sample with upper-tail deletions and a positive-outcome-only
sample. These outcome-based restrictions diagnose sensitivity
rather than define a preferred sample.

\begin{table}[!htbp]
\centering
\begin{singlespace}
\normalsize
\caption{Political centralization and nighttime light density}
\label{table:michalopoulos}
\setlength{\tabcolsep}{3pt}
\newcommand{\BBPappmaincell}[2]{\(#1\;(#2)\)}
\begin{tabular*}{\textwidth}{@{\extracolsep{\fill}}lcccc@{}}
\toprule
Specification & Full & Drop 2\% & Drop 5\% & \(Y>0\) \\
Observations & 682 & 668 & 646 & 517 \\
\midrule
OLS, \(c=10^{-5}\) & \BBPappmaincell{19.90}{0.122} & \BBPappmaincell{25.29^{*}}{0.066} & \BBPappmaincell{19.93}{0.196} & \BBPappmaincell{16.12^{**}}{0.014} \\
OLS, \(c=0.01\) & \BBPappmaincell{19.32^{***}}{<0.001} & \BBPappmaincell{20.49^{***}}{<0.001} & \BBPappmaincell{16.20^{***}}{0.003} & \BBPappmaincell{17.18^{***}}{<0.001} \\
OLS, \(c=1\) & \BBPappmaincell{5.44^{***}}{<0.001} & \BBPappmaincell{4.86^{***}}{<0.001} & \BBPappmaincell{2.85^{***}}{0.001} & \BBPappmaincell{4.99^{**}}{0.011} \\
PPML & \BBPappmaincell{-10.46}{0.226} & \BBPappmaincell{21.43^{**}}{0.040} & \BBPappmaincell{8.24^{*}}{0.052} & \BBPappmaincell{-8.57}{0.360} \\
GPML (J) & \BBPappmaincell{16.04^{**}}{0.011} & \BBPappmaincell{19.80^{***}}{0.002} & \BBPappmaincell{15.95^{**}}{0.032} & \BBPappmaincell{17.27^{***}}{0.005} \\
GPML (C) & \BBPappmaincell{14.88^{**}}{0.019} & \BBPappmaincell{34.93^{***}}{<0.001} & \BBPappmaincell{15.94^{***}}{0.002} & \BBPappmaincell{11.56^{*}}{0.080} \\
\bottomrule
\end{tabular*}
\par\smallskip
\begin{minipage}{\textwidth}
\BBPtablenotes{Percentage associations; two-sided normal \(p\)-values in parentheses:
\(^{***}p<0.01\), \(^{**}p<0.05\), \(^{*}p<0.10\).
J: joint country effects; C: cancellation.
All models keep the original eleven controls and use signed country-and-family
two-way cluster-robust covariance. The full sample has 48 countries and 94 ethnic families.
Cuts remove the original-ranked 14 or 35 largest outcome values and keep
164 zeros; \(Y>0\) excludes zeros without a tail cut. Counts reflect singleton
exclusions. Appendix~\ref{app:applications} provides details.}
\end{minipage}
\end{singlespace}
\end{table}

Full-sample PPML and joint GPML give \(-10.5\%\) and \(16.0\%\),
with a statistically significant paired coefficient difference
(\(p=0.012\)). Removing the brightest homeland moves PPML to
\(0.8\%\), and further deletions produce positive estimates.
Joint GPML, by contrast, remains positive across all deletions,
with every 95\% interval excluding zero. Excluding zeros leaves
the full-sample signs unchanged. Shifted-log estimates also depend
on the chosen shift \(c\): their full-sample retransformed
coefficients range from \(5.4\%\) to \(19.9\%\).
These comparisons favor joint GPML as the more stable and
informative empirical summary. Although they do not establish
which mean specification is correct, PPML's negative full-sample
estimate depends too heavily on a single bright homeland to
provide a persuasive conclusion on its own. Joint and cancelled GPML instead
support a positive association that persists throughout these
sensitivity checks.%
\footnote{Joint Gamma-weighted half-power estimates are also
positive, ranging from \(13.2\%\) to \(18.2\%\) across the five
zero-inclusive samples, although all 95\% intervals include zero
(Appendix~\ref{app:application-half-power}).}

\section{Conclusion}
\label{sec:conclusion}

Dealing with logs and zeros requires aligning the proportional effect
of interest with its transformation, estimating equation and numerical
implementation. Our framework connects these choices through power
means, shifted logarithms and repeated least-squares computation,
and develops cancellation and bias correction for fixed effects.
The simulations and application illustrate why these choices matter
for numerical reliability, inference and  interpretation.
Further work will focus on  model selection.

\par
\label{bbp:main-end}
\phantomsection
\label{bbp:article-end}

\BBPmainappendix
\begingroup
\def\BBProadmapmain{}
\section*{\centering Appendices}
\ifdefined\BBProadmapmain
The main appendices, Appendices~\ref{app:proofs-target-comparisons}--\ref{app:iv-pilot-proof},
follow the article in this file and provide its main proofs.
The supplemental appendices,
Appendices~\ref{app:comparisons}--\ref{app:empirical-practice}, are in a
separate PDF in the ancillary files of
\href{https://arxiv.org/abs/2203.11820}{arXiv:2203.11820}.
They contain supporting assumptions and results, simulation
details, application results and evidence on empirical practice.
\else
This file contains the supplemental appendices,
Appendices~\ref{app:comparisons}--\ref{app:empirical-practice}.
The main proof appendices,
Appendices~\ref{app:proofs-target-comparisons}--\ref{app:iv-pilot-proof},
follow the article in the main PDF of
\href{https://arxiv.org/abs/2203.11820}{arXiv:2203.11820}.
The supplement contains supporting
assumptions and results, simulation details, application results and
evidence on empirical practice.
\fi
The contents below cover both files. Pages beginning with S use the
supplement's separate pagination.

\begin{center}
\begin{singlespace}
\small
\def\BBProadmaptable{}
\begin{tabularx}{\textwidth}{@{}p{0.25\textwidth}Lr@{}}
\toprule
Appendix & Contents & Page \\
\midrule
\multicolumn{3}{@{}l}{\emph{Proof appendices and supporting statements}} \\
\addlinespace
\ref{app:proofs-target-comparisons}. Proportional effects and comparisons &
Weighted effects and individual ATE\%, conditional power means, local
tangency of log OLS and GPML, and the shifted-log approximations.
& \pageref{app:proofs-target-comparisons} \\
\addlinespace
\ref{app:iols-fixed-tau-proofs}. Computation &
Phase 1 curvature and convergence, inference at a fixed approximation,
and two-phase completion and inference.
& \pageref{app:iols-fixed-tau-proofs} \\
\addlinespace
\ref{app:fixed-effects}. Fixed effects &
Cancellation with short groups; growing retained effects, their
conditions, feasible correction and proof.
& \pageref{app:fixed-effects} \\
\addlinespace
\ref{app:iv-pilot-proof}. Endogenous regressors &
Two-phase I2SLS and initialization of local GMM completion.
& \pageref{app:iv-pilot-proof} \\
\addlinespace
\multicolumn{3}{@{}l}{\emph{Supplemental appendices}} \\
\addlinespace
\ref{app:comparisons}. Proportional effects and log-like regressions &
Identification and weighting identities, zeros and power means,
retransformation, estimating equations and influence, the
fixed log shifts and the shifted-log continuation.
& \pageref{app:comparisons} \\
\addlinespace
\ref{app:iols-supplement}. Computation and inference &
Intermediate results  \& additional results
& \pageref{app:iols-supplement} \\
\addlinespace
\ref{app:fe-supplement}. Fixed effects &
Intermediate results \& additional results
& \pageref{app:fe-supplement} \\
\addlinespace
\ref{app:iv}. Endogenous regressors &
Intermediate results \& additional results
& \pageref{app:iv} \\
\addlinespace
\ref{app:sim-design}. Simulations &
Design and detailed results.
& \pageref{app:sim-design} \\
\addlinespace
\ref{app:applications}. Application &
Descriptives, equations,  inference, sensitivity.
& \pageref{app:applications} \\
\addlinespace
\ref{app:empirical-practice}. Additional content &
Article survey, seminar poll and discussion across disciplines.
& \pageref{app:empirical-practice} \\
\bottomrule
\end{tabularx}
\end{singlespace}
\end{center}

\endgroup
\clearpage
\section{Proofs for proportional effects and comparisons}
\label{app:proofs-target-comparisons}

The proofs appear following the order in the manuscript. The five subsections prove the propositions on
weighted effects (\ref{prop:log-effect-nate-closeness}), conditional
dispersion (\ref{prop:conditional-power-mean-dispersion}), local coefficient
responses (\ref{thm:power-score-log-tangency}), coefficient approximation
(\ref{prop:gpml-log-continuation}) and proportional-effect approximation
(\ref{prop:iols-power-residual-calibration}).
Identification identities, retransformation calculations, shifted-log root
properties and further comparisons are collected in
Appendix~\ref{app:comparisons}. All proofs are preceded by a proof description and its main intuitions to help readability.

\subsection{Weighted effects and individual ATE\%}
\label{app:target-main-proofs}
Proposition~\ref{prop:log-effect-nate-closeness} compares each weighted effect's distance from
individual ATE\% with the corresponding distance of the log effect.
In this subsection and in Appendix~\ref{app:comparisons}, abbreviate
\(R_i=Y_i(1)/Y_i(0)\),
\(L=1+\theta_L\), \(N=1+\theta_N\),
\(H=1+\theta_H\), and \(P=1+\theta_{\%}=\E(R_i)\).

\begin{samepage}
\noindent The following proof proceeds in three steps.
Subtracting one leaves distances unchanged, so it suffices to compare the
gross ratios with \(P\). Jensen's inequality identifies the log-effect
distance. Covariance decompositions then turn each comparison
into its stated necessary and sufficient condition.
\begin{enumerate}[label=\textup{(\Roman*)},leftmargin=*,itemsep=1pt,topsep=3pt,parsep=0pt]
\item \emph{Log-effect distance.} Jensen's inequality gives \(P\geq L\),
so the absolute distance from the individual average is \(P-L\).
\item \emph{NATE condition.} Writing \(N-P\) as a covariance divided
by the positive untreated mean makes comparison with Step~(I)
equivalent to the stated NATE condition.
\item \emph{Harmonic condition.} Weighting by inverse treated outcomes
gives the analogous covariance expression and the second condition.
Positive denominators preserve the strict comparisons as well.
\end{enumerate}
\end{samepage}

\begin{proof}[Proof of Proposition~\ref{prop:log-effect-nate-closeness}]
\emph{(I) Log-effect distance.}
Since \(L=\exp(\E(\log(R_i)))\) and \(P=\E(R_i)\),
convexity of the exponential gives
\(\exp(\E(\log(R_i)))\leq\E(\exp(\log(R_i)))=P\).
Thus \(|\theta_L-\theta_{\%}|=|L-P|=P-L\).
This identifies the nonnegative quantity against which both weighted
effects will be compared.

\emph{(II) NATE condition.}
\(N=\E(Y_i(0)R_i)/M_0\), because \(Y_i(1)=Y_i(0)R_i\).
Expanding the numerator into a covariance and a product of means gives
\begin{equation}
 N=P+\frac{\operatorname{Cov}(Y_i(0),R_i)}{M_0}.
 \label{eq:app-log-nate-weighted-ratio}
\end{equation}
Consequently \(|\theta_N-\theta_{\%}|\leq|\theta_L-\theta_{\%}|\)
is \(|N-P|\leq P-L\). By
\eqref{eq:app-log-nate-weighted-ratio}, the left side is
\(|\operatorname{Cov}(Y_i(0),R_i)|/M_0\).
Multiplication by the positive mean \(M_0\) gives precisely
\eqref{eq:log-nate-closeness}, in both directions.

\emph{(III) Harmonic condition.}
For the harmonic effect, \(R_i/Y_i(1)=1/Y_i(0)\), so
\(H=\E(R_i/Y_i(1))/\E(1/Y_i(1))\). Expanding this numerator
into its covariance and product of means gives
\(H-P=\operatorname{Cov}(1/Y_i(1),R_i)/\E(1/Y_i(1))\).
Consequently \(|H-P|\leq P-L\) is equivalent to
\(|\operatorname{Cov}(1/Y_i(1),R_i)|
\leq\E(1/Y_i(1))(P-L)\), which is
\eqref{eq:log-harmonic-closeness}. Both denominators are positive,
so strict inequalities are preserved. The covariance identities require
no additional second-moment assumption: their product expectations are
the assumed finite quantities \(\E(Y_i(1))\) and \(\E(1/Y_i(0))\),
respectively.
\end{proof}

\Needspace{16\baselineskip}
\subsection{Conditional power means and dispersion}
\label{app:comparisons-main-proofs}
\label{app:conditional-power-mean-dispersion}
The conditional power means are defined in Section~\ref{sec:comparison}. The following proof proceeds in three steps.
The common expected-log term cancels when log power means are compared,
so controlling their order-dependent terms through second order suffices.
Showing that the two conditional error variances differ only by a
third-order remainder then permits either variance in the approximation.
\begin{enumerate}[label=\textup{(\Roman*)},leftmargin=*,itemsep=1pt,topsep=3pt,parsep=0pt]
\item \emph{Exponential-moment expansion.} Centering removes the linear
term, and the exponential-moment bound controls the remainder by
\(O(h^3)\), leaving conditional variance as the leading term.
\item \emph{Power-mean gap.} Inserting Step~(I)'s expansion into
\eqref{eq:power-mean-log-error-identity} adds only a fourth-order error.
Subtracting the fixed orders cancels their common expected log and gives
\eqref{eq:conditional-power-mean-gap}.
\item \emph{Multiplicative-error variance.} Applying Step~(I) at orders
one and two shows that this variance differs from the log-error variance
by \(O(h^3)\), while substitution in Step~(II) preserves the claimed precision.
\end{enumerate}

\begin{proof}[Proof of Proposition~\ref{prop:conditional-power-mean-dispersion}]
We work conditionally on \(X_i\) and let \(h\to0\), holding the orders
and the conditional law of \(V_i\) fixed. All remainder bounds are
pointwise at covariate values where the stated assumptions hold; their
constants may depend on that value and on the fixed orders.

\emph{(I) Exponential-moment expansion.}
We express the power mean in terms of the log error.
By definition,
\(\log(Y_i)=\E(\log(Y_i)\mid X_i)+\varepsilon_i\), so
\(Y_i^\omega
=\exp(\omega\E(\log(Y_i)\mid X_i))\exp(\omega\varepsilon_i)\).
Taking conditional expectations and using the definition of
\(A_\omega(X_i)\), for \(\omega\ne0\), gives
\begin{equation}
 \log(A_\omega(X_i))
 =\E(\log(Y_i)\mid X_i)
  +\frac{1}{\omega}
    \log\left(\E(\exp(\omega\varepsilon_i)\mid X_i)\right).
 \label{eq:power-mean-log-error-identity}
\end{equation}
Thus we need to expand the exponential of the log error.

Expanding \(\exp(\omega\varepsilon_i)\) around
\(\varepsilon_i=0\) and taking conditional expectations gives
\begin{equation}
 \E(\exp(\omega\varepsilon_i)\mid X_i)
 =1+\omega\E(\varepsilon_i\mid X_i)
   +\frac{\omega^2}{2}\E(\varepsilon_i^2\mid X_i)
   +O(h^3).
 \label{eq:conditional-exponential-expansion}
\end{equation}
To justify the \(O(h^3)\) remainder, apply Taylor's bound
\(|\exp(z)-1-z-z^2/2|\le |z|^3\exp(|z|)/6\)
with \(z=\omega hV_i\).
The conditional expectation of this upper bound is
\(\frac{|\omega|^3|h|^3}{6}
\E(|V_i|^3\exp(|\omega h||V_i|)\mid X_i)\).
Choose \(0<s<t\) within the conditional moment-generating-function
neighborhood. For some finite \(C_{s,t}\),
\(|v|^3\exp(s|v|)\leq C_{s,t}\exp(t|v|)\), and
\(\E(\exp(t|V_i|)\mid X_i)
\leq\E(\exp(tV_i)+\exp(-tV_i)\mid X_i)<\infty\).
The expectation in the bound is therefore bounded when
\(|\omega h|\leq s\). Thus the expected absolute
remainder is \(O(h^3)\), pointwise in \(X_i\).
The linear term in
\eqref{eq:conditional-exponential-expansion} vanishes because
\(\E(\varepsilon_i\mid X_i)=0\); for the same reason,
\(\E(\varepsilon_i^2\mid X_i)=\Var(\varepsilon_i\mid X_i)\).
Hence
\(\E(\exp(\omega\varepsilon_i)\mid X_i)
=1+\omega^2\Var(\varepsilon_i\mid X_i)/2+O(h^3)\).

\emph{(II) Power-mean gap.}
Since
\(\Var(\varepsilon_i\mid X_i)=h^2\Var(V_i\mid X_i)\),
the expression inside the logarithm in
\eqref{eq:power-mean-log-error-identity} differs from one by
\(O(h^2)\). Using \(\log(1+z)=z+O(z^2)\) therefore adds only
an \(O(h^4)\) error, which is absorbed into \(O(h^3)\).
Substitution into \eqref{eq:power-mean-log-error-identity} gives
\(\log(A_\omega(X_i))
=\E(\log(Y_i)\mid X_i)
+\omega\Var(\varepsilon_i\mid X_i)/2+O(h^3)\).
At order zero, \(\log(A_0(X_i))=\E(\log(Y_i)\mid X_i)\)
by definition. Subtracting the expressions for orders \(\omega\)
and \(\nu\) cancels the common expected-log term and proves
\eqref{eq:conditional-power-mean-gap}.

\emph{(III) Multiplicative-error variance.}
The normalization by the conditional arithmetic mean gives
\(U_i=\exp(\varepsilon_i)/\E(\exp(\varepsilon_i)\mid X_i)\)
because the common factor
\(\exp(\E(\log(Y_i)\mid X_i))\) cancels. Since
\(\E(U_i\mid X_i)=1\), its conditional variance is
\(\E(\exp(2\varepsilon_i)\mid X_i)/
(\E(\exp(\varepsilon_i)\mid X_i))^2-1\).
Applying \eqref{eq:conditional-exponential-expansion} at orders
two and one gives a numerator of
\(1+2\Var(\varepsilon_i\mid X_i)+O(h^3)\)
and a denominator of
\(1+\Var(\varepsilon_i\mid X_i)+O(h^3)\).
Their difference is
\(\Var(\varepsilon_i\mid X_i)+O(h^3)\).
The denominator is bounded away from zero for small \(h\), and its
reciprocal is \(1+O(h^2)\). Thus division changes the leading
\(O(h^2)\) variance term by only \(O(h^4)\), yielding
\(\Var(U_i\mid X_i)=\Var(\varepsilon_i\mid X_i)+O(h^3)\).
Replacing the variance in
\eqref{eq:conditional-power-mean-gap} therefore changes only
the remainder.
\end{proof}

\subsection{Local tangency of log OLS and GPML}
\label{app:log-gpml-local-proof}
\begin{samepage}
\noindent The following proof proceeds in four steps.
We express local coefficient responses as weighted regressions,
determine when they coincide with log OLS, and show that GPML
satisfies this condition for every design $X$. A further expansion
quantifies its remaining discrepancy.

\begin{enumerate}[label=\textup{(\Roman*)},leftmargin=*,itemsep=1pt,topsep=3pt,parsep=0pt]
\item \emph{Local response.}
Starting from \(Y_i(0)=\mu_{i,*}\), differentiate the estimating
equation along the perturbation path
\eqref{eq:positive-log-perturbation}. We characterize the coefficient's first-order response as
the weighted regression of \(v_i\) on \(X_i\), with weights
\(\mu_{i,*}^{2-\tau}\). Log OLS's coefficient response corresponds to the unweighted
regression.

\item \emph{When weighting leaves the response unchanged.}
For a given $X$, the two regressions agree for every disturbance
vector precisely under the following condition. Writing
\(W_\tau=\diag(\mu_{i,*}^{2-\tau})\), we show that this is equivalent
to \(W_\tau X=XR\) for some square matrix \(R\).
Thus each weighted regressor column must remain a linear
combination of the original columns.

\item \emph{Agreement and remaining discrepancy for GPML.}
At \(\tau=2\), \(W_2=I\), so Step~(II)'s condition holds for
every design and the first-order responses coincide.
Expanding the GPML equation to second order then gives
\eqref{eq:gpml-log-curvature}.

\item \emph{Why the agreement is specific to GPML.}
For each \(\tau\ne2\), a three-observation design violates Step~(II)'s
condition, so some perturbation \(v\) gives different first-order responses.
\end{enumerate}
\end{samepage}
\begin{proof}[Proof of Proposition~\ref{thm:power-score-log-tangency}]
\emph{(I) Local response.}
Define \(W_\tau=\diag(\mu_{i,*}^{2-\tau})\),
\(C_\tau=(X'W_\tau X)^{-1}X'W_\tau\),
\(C_L=(X'X)^{-1}X'\), and \(P_X=XC_L\).
The coefficient \(h_\tau=C_\tau v\) is the weighted least-squares
coefficient from regressing \(v\) on \(X\), with weights
\(\mu_{i,*}^{2-\tau}\). We show that the nearby power-score root satisfies
\begin{equation}
 \widehat b_\tau(t)=b_*+t h_\tau+O(t^2),
 \label{eq:power-score-first-order}
\end{equation}
as \(t\to0\), holding \(n,X,b_*,v\) fixed.

For the path in \eqref{eq:positive-log-perturbation}, define
\(F_\tau(b,t)=\sum_iX_i\mu_i(b)^{1-\tau}(Y_i(t)-\mu_i(b))\).
At \((b,t)=(b_*,0)\), the fit is exact:
\(Y_i(0)=\mu_{i,*}\). In the derivative with respect to \(b\), the
term differentiating the weight \(\mu_i(b)^{1-\tau}\) therefore
vanishes because it multiplies \(Y_i(0)-\mu_{i,*}=0\).
Differentiating the remaining term and the outcome path gives
\(\partial F_\tau/\partial b'=-X'W_\tau X\) and
\(\partial F_\tau/\partial t=X'W_\tau v\). The first matrix is
nonsingular because \(X\) has full column rank and \(W_\tau\) is
positive definite. The finite sum \(F_\tau\) is smooth on an open
neighborhood of \((b_*,0)\), so the implicit-function theorem
\citep[Corollary~6.7.13]{corbae_stinchcombe_zeman2009} gives a nearby
continuously differentiable root. Its derivative formula and the
smoothness of \(F_\tau\) imply smoothness of this branch. Differentiating
\(F_\tau(\widehat b_\tau(t),t)=0\) at zero gives
\(-X'W_\tau X\widehat b_\tau'(0)+X'W_\tau v=0\), so that
\(\widehat b_\tau'(0)=C_\tau v\), proving
\eqref{eq:power-score-first-order}. This argument concerns the unique
branch in a neighborhood of \(b_*\), not all roots away from that
neighborhood. Since
\(\log(Y_i(t))=X_i'b_*+tv_i\), the stated log-OLS coefficient follows from
the OLS normal equations.

\emph{(II) When weighting leaves the response unchanged.}
Agreement for every perturbation is equivalent to \(C_\tau=C_L\). If
this equality holds, multiplication by \(X'W_\tau X\) gives
\(X'W_\tau=X'W_\tau P_X\). Transposing yields
\((I-P_X)W_\tau X=0\), so
\(W_\tau\operatorname{col}(X)\subseteq
\operatorname{col}(X)\). Conversely, suppose
\(W_\tau X=XR\) for a square matrix \(R\). The matrix \(R\)
is nonsingular because \(W_\tau X\) has full column rank. Symmetry
of \(W_\tau\) gives \(X'W_\tau=R'X'\), and hence
\(X'W_\tau X=R'X'X\). Substitution into \(C_\tau\) gives
\(C_\tau=(R'X'X)^{-1}R'X'=(X'X)^{-1}X'=C_L\),
proving the equivalence.
In particular, \(X'W_\tau(I-P_X)=0\) says that every ordinary
least-squares residual remains orthogonal to the weighted regressor
columns. Weighting then leaves the fitted coefficient unchanged for
every \(v\). By symmetry of \(W_\tau\), this is exactly the requirement
that weighting each regressor column keeps it in the original column space.

\emph{(III) Agreement and remaining discrepancy for GPML.}
At \(\tau=2\), the power-score equation becomes
\(\sum_iX_i(\exp(tv_i-X_i'd(t))-1)=0\), where
\(d(t)=\widehat b_2(t)-b_*\). With \(h=C_Lv\) and
\(e=v-Xh\) as in the main text, define \(q=(e_1^2,\ldots,e_n^2)'\),
so that \(c=C_Lq\). Smoothness of the implicit root permits
the expansion \(d(t)=th+t^2a+O(t^3)\). Expanding the exponential in the
GPML equation gives
\(X'[t e+t^2(q/2-Xa)]+O(t^3)=0\). The remainder remains \(O(t^3)\) after summation because the sample
and design are fixed. The first-order term vanishes because \(X'e=0\).
Dividing by \(t^2\) and taking the limit gives
\(X'Xa=X'q/2\), hence \(a=c/2\). Subtracting the exact
log-OLS expression \(b_*+th\) from \(b_*+d(t)\) therefore leaves
\(t^2c/2+O(t^3)\), proving \eqref{eq:gpml-log-curvature}.

\emph{(IV) Why the agreement is specific to GPML.}
For any fixed
\(\tau\in[0,3]\setminus\{2\}\), take
the three rows of \(X\) to be \((1,0)\), \((1,1)\), and \((1,2)\), and
set \(b_*=(0,1)'\). Then
\(W_\tau\mathbf1=(1,a,a^2)'\), where
\(a=\exp(2-\tau)\). If the column space of \(X\) were invariant, this
vector would be affine in its second coordinate. Its second difference is
\((a-1)^2\), which is positive because \(\tau\ne2\). Invariance
therefore fails. By the equivalence just proved, \(C_\tau\ne C_L\), so
there is a perturbation \(v\) for which their first-order responses differ.
\end{proof}

\subsection{The shifted-log coefficient approximation}
\label{app:shift-log-continuation-proof}

\begin{samepage}
\noindent The following proof proceeds in five steps.
We first bound the shifted-log moment evaluated at the power root.
A local curvature bound then establishes a nearby root and converts
that moment bound into a coefficient bound. Global convexity makes
the localized root the unique solution of the shifted-log equation.
\begin{enumerate}[label=\textup{(\Roman*)},leftmargin=*,itemsep=1pt,topsep=3pt,parsep=0pt]
\item \emph{Residual gap.} Matching the residuals through second order
leaves a cubic difference bounded uniformly in the power order.
\item \emph{Moment at the power root.} The power moment vanishes at
its root, so Step~(I)  is used to bound the shifted-log moment.
\item \emph{Local curvature and convexity.} The criterion is globally
convex and has a positive curvature lower bound near the power root,
providing control for localization and the coefficient-distance bound.
\item \emph{Root localization.} The bounds in Steps~(II)--(III) make
the criterion higher on a ball's boundary than at its center. Its minimum
over the ball is therefore an interior root. Local positive curvature
and global convexity establish uniqueness.
\item \emph{Coefficient bound and endpoints.} Integrating the curvature
between the localized roots converts Step~(II)'s moment bound into the
stated coefficient-distance bound. Exact agreement of the endpoint
residuals gives equality there without a small-residual condition.
\end{enumerate}
\end{samepage}

\begin{proof}[Proof of Proposition~\ref{prop:gpml-log-continuation}]
We fix the sample throughout the proof.\footnote{The root properties of the whole continuation path are not used here and
are verified separately in Appendix~\ref{app:shift-log-roots}.} For \(\omega\in[0,1]\) and \(u>-1\), and also at
\(u=-1\) when \(\omega>0\), define
\(p_\omega(u)=((1+u)^\omega-1)/\omega\), with \(p_0(u)=\log(1+u)\), and
\(s_\omega(u)=\log(1+(1-\omega)u)/(1-\omega)\), with \(s_1(u)=u\).
Write \(u_i(\beta)=Y_i/\mu_i(\beta)-1\). The power equation in the
proposition is \(\Pn X_ip_\omega(u_i(\beta))=0\). With \(\tau=2\) and
\(\lambda=\omega\), the normalized residual in
\eqref{eq:gpml-log-continuation-equation} is
\(r_{\omega,2,i}(\beta)=\log(1+(1-\omega)u_i(\beta))/(1-\omega)
=s_\omega(u_i(\beta))\), including its continuous limit at
\(\omega=1\). Define \(d_\omega=s_\omega-p_\omega\).

\emph{(I) Residual gap.} Let \(\rho_0\in(0,1)\); its value is chosen in
Step~(IV). Both functions
are smooth on \(u>-1\), vanish at zero, have unit first derivative there,
and have second derivative \(-(1-\omega)\) there. Taylor's theorem with
integral remainder therefore gives
\(d_\omega(u)=\int_0^u(u-v)^2d_\omega'''(v)\,dv/2\).
Direct differentiation gives
\(s_\omega'''(v)=2(1-\omega)^2(1+(1-\omega)v)^{-3}\) and
\(p_\omega'''(v)=(1-\omega)(2-\omega)(1+v)^{\omega-3}\), so
\(d_\omega'''(v)=(1-\omega)\phi(\omega,v)\), where
\(\phi(\omega,v)=2(1-\omega)(1+(1-\omega)v)^{-3}
-(2-\omega)(1+v)^{\omega-3}\).
Since \(\phi(0,v)=0\), the mean-value theorem in \(\omega\) gives
\(|\phi(\omega,v)|\leq\omega\sup_{0\leq\omega'\leq1}
|\partial_\omega\phi(\omega',v)|\), where
\(\partial_\omega\phi(\omega,v)=-2(1+(1-\omega)v)^{-3}
+6(1-\omega)v(1+(1-\omega)v)^{-4}
+(1+v)^{\omega-3}(1-(2-\omega)\log(1+v))\).
For \(|v|\leq\rho_0\), the bounds \(1+(1-\omega)v\geq1-\rho_0\),
\((1+v)^{\omega-3}\leq(1-\rho_0)^{-3}\) and
\(|\log(1+v)|\leq|\log(1-\rho_0)|\) give
\(|\partial_\omega\phi(\omega,v)|\leq K(\rho_0)\), with
\(K(\rho_0)=(3+2|\log(1-\rho_0)|)(1-\rho_0)^{-3}
+6\rho_0(1-\rho_0)^{-4}\). Consequently,
\(|d_\omega(u)|\leq K(\rho_0)\omega(1-\omega)|u|^3/6\) for every
\(\omega\in[0,1]\) and \(|u|\leq\rho_0\). The constant
\(K(\rho_0)\) does not depend on the sample or on \(\omega\).
The integral remainder gives the same absolute bound for negative \(u\)
by reversing the integration limits. The factor \(\omega(1-\omega)\)
preserves the exact agreement at both orders zero and one.

\emph{(II) Moment at the power root.} Write
\(\beta^P=\widehat\beta^P_\omega\) and
\(F_\omega(\beta)=\Pn X_is_\omega(u_i(\beta))\), so that
\(\widetilde\beta_\omega\) solves \(F_\omega(\beta)=0\). Suppose
\(\rho_\omega\leq\rho_0\). Then \(|u_i(\beta^P)|\leq\rho_0<1\) for every
observation, so every outcome is positive. Because \(\beta^P\) solves the
power equation, \(F_\omega(\beta^P)=\Pn X_id_\omega(u_i(\beta^P))\), and
Step~(I) gives \(\|F_\omega(\beta^P)\|_2\leq \Delta_\omega\), where
\(\Delta_\omega=\bar x K(\rho_0)\omega(1-\omega)\rho_\omega^3/6\) and
\(\bar x=n^{-1}\sum_i\|X_i\|_2\). A small moment alone does not yet
bound coefficient distance: Steps~(III)--(IV) provide curvature and
show that a root lies in the region where that curvature is controlled.

\emph{(III) Local curvature and convexity.} Write
\(U_i=U_i(\beta)=1+u_i(\beta)\).
The coefficient derivative of \(s_\omega(u_i(\beta))\) is
\(-w_iX_i\), with \(w_i=U_i/(\omega+(1-\omega)U_i)\geq0\).
Hence \(\partial F_\omega/\partial\beta'=-\Pn w_iX_iX_i'\), the
\(\tau=2\) case of \eqref{eq:app-gpml-log-continuation-jacobian}. This
matrix is symmetric and negative semidefinite at every \(\beta\).
Explicitly, \(F_\omega=-\nabla\mathcal Q_\omega\) for the separable criterion
\(\mathcal Q_\omega(\beta)=\Pn G_i(X_i'\beta)\), where
\(G_i(\eta)=-\int_0^\eta s_\omega(Y_ie^{-t}-1)\,dt\) has second
derivative \(w_i\geq0\); hence \(\mathcal Q_\omega\) is convex, and the
roots of \(F_\omega\) are exactly its minimizers. Let \(x_{\max}=\max_i\|X_i\|_2\),
\(r_X=1/x_{\max}\), and let \(\mathcal B\) be the closed Euclidean ball
of radius \(r_X\) about \(\beta^P\). On \(\mathcal B\),
\(|X_i'(\beta-\beta^P)|\leq1\), so
\(U_i(\beta)=U_i(\beta^P)\exp(-X_i'(\beta-\beta^P))\geq(1-\rho_0)e^{-1}\).
The weight \(w_i\) is increasing in \(U_i\) and at least
\(\min(U_i,1)\). Every weight on \(\mathcal B\) is therefore at least
\(c_0=(1-\rho_0)e^{-1}\). With \(\nu_X\) the smallest eigenvalue of
\(X'X/n\), positive by full column rank,
\(\nabla^2\mathcal Q_\omega\succeq m_XI\) on \(\mathcal B\), where
\(m_X=c_0\nu_X\). Indeed, for any direction \(v\),
\(v'\nabla^2\mathcal Q_\omega v
=\Pn w_i(X_i'v)^2\geq c_0v'(X'X/n)v
\geq m_X\|v\|_2^2\). This lower bound holds throughout the ball,
which permits integration along the coefficient segments below.

\emph{(IV) Root localization.} Choose \(\rho_0\in(0,1)\) so
small that \(\bar xK(\rho_0)\rho_0^3/24<m_Xr_X/2\). This is possible
because the left side tends to zero as \(\rho_0\downarrow0\), whereas the
right side tends to \(\nu_X/(2ex_{\max})>0\); the choice depends only on
\(X\). Since \(\omega(1-\omega)\leq1/4\), it gives
\(\Delta_\omega<m_Xr_X/2\). For \(\beta\) on the boundary of \(\mathcal B\),
the curvature bound along the segment and
\(\nabla\mathcal Q_\omega(\beta^P)=-F_\omega(\beta^P)\) give
\(\mathcal Q_\omega(\beta)\geq\mathcal Q_\omega(\beta^P)
-F_\omega(\beta^P)'(\beta-\beta^P)+m_Xr_X^2/2
\geq\mathcal Q_\omega(\beta^P)-\Delta_\omega r_X+m_Xr_X^2/2
>\mathcal Q_\omega(\beta^P)\). The minimum of
\(\mathcal Q_\omega\) over the compact ball is consequently attained in
its interior, where the gradient vanishes. This interior point is a root
of \(F_\omega\). It is the only minimizer of \(\mathcal Q_\omega\) on
\(\mathbb R^K\): by convexity, the segment joining two minimizers
consists of minimizers, but \(\mathcal Q_\omega\) is strictly convex on
the part of that segment inside \(\mathcal B\). The root in the ball is
therefore \(\widetilde\beta_\omega\).

\emph{(V) Coefficient bound and endpoints.}
Define \(v=\widetilde\beta_\omega-\beta^P\). Both endpoints and
their segment belong to \(\mathcal B\) by Step~(IV). The integrated
derivative identity gives
\(F_\omega(\beta^P)
=\int_0^1\nabla^2\mathcal Q_\omega(\beta^P+tv)v\,dt\),
because \(F_\omega(\widetilde\beta_\omega)=0\).
Premultiplying by \(v'\), applying Step~(III)'s curvature bound, and
then Cauchy--Schwarz give
\(m_X\|\widetilde\beta_\omega-\beta^P\|_2^2\leq
F_\omega(\beta^P)'(\widetilde\beta_\omega-\beta^P)
\leq \Delta_\omega\|\widetilde\beta_\omega-\beta^P\|_2\).
If the coefficient difference is zero, the desired bound is immediate;
otherwise, divide by its norm to obtain
\(\|\widetilde\beta_\omega-\widehat\beta^P_\omega\|_2\leq \Delta_\omega/m_X
=C_X\omega(1-\omega)\rho_\omega^3\), with
\(C_X=\bar xK(\rho_0)/(6c_0\nu_X)\). Both \(C_X\) and \(\rho_0\) depend
only on \(X\), including its sample size.

At \(\omega=0\) and \(\omega=1\), \(d_\omega\) vanishes identically:
\(p_0(u)=s_0(u)=\log(1+u)\) and \(p_1(u)=s_1(u)=u\). The power root then
solves the shifted-log equation, and the assumed uniqueness gives
\(\widetilde\beta_\omega=\widehat\beta^P_\omega\) whenever the endpoint
is defined, without a condition on \(\rho_\omega\). The endpoint
equations are log OLS at \(\omega=0\) and GPML at \(\omega=1\).
\end{proof}

\subsection{Approximation to power means and effects}
\label{app:shift-power-approximation}

Throughout this subsection, write \(\epsilon=t\), \(V_{di}=v_{di}\)
and \(a_d=\log(\mu_{d,*})\), so that
\(\log(Y_i(d;t))=a_d+\epsilon V_{di}\). Write
\(A_\omega(d;\epsilon)=A_{\omega,d}(t)\) and
\(\widetilde M_\omega(d;\epsilon)=\widetilde A_{\omega,d}(t)\).
For a log residual \(z\), the exact power residual is
\(B_\omega(z)=(e^{\omega z}-1)/\omega\), with \(B_0(z)=z\).
The proof establishes the following sharper expansions, to which the
main text refers.
For the means and effects in Proposition~\ref{prop:iols-power-residual-calibration},
write \(\kappa_d=\E((V_{di}-\E(V_{di}))^3)\). The precise mean comparison is
\begin{equation}
 \log(\widetilde M_\omega(d;\epsilon))-\log(A_\omega(d;\epsilon))
 =-\frac{\omega(1-\omega)}6\epsilon^3\kappa_d+O(\epsilon^4).
 \label{eq:shift-power-mean-approximation}
\end{equation}
For the proportional effects, it follows that
\begin{equation}
 \widetilde\theta_\omega-\theta_\omega
 =-\frac{e^{a_1-a_0}\omega(1-\omega)}6
     \epsilon^3(\kappa_1-\kappa_0)+O(\epsilon^4).
 \label{eq:shift-power-effect-approximation}
\end{equation}
Both expansions are uniform in \(\omega\in[0,1]\) as \(\epsilon\to0\).

\begin{samepage}
\noindent The following proof proceeds in four steps.
Agreement of the residual expansions does not yet compare the means,
because each mean solves a different equation. Uniform expansions of
those roots give the log-mean gap. Differencing across treatment states
and exponentiating then gives the effect gap, with exact endpoint
equations completing the result.
\begin{enumerate}[label=\textup{(\Roman*)},leftmargin=*,itemsep=1pt,topsep=3pt,parsep=0pt]
\item \emph{Uniform residual expansions.} Removing the endpoint
singularities provides uniform derivative and Taylor bounds, so the
endpoint orders can be treated together with the interior orders.
\item \emph{Log-mean difference.} Uniqueness and a nonzero derivative at
zero dispersion turn Step~(I)'s smoothness into a uniform root expansion.
Matching lower-order terms leaves the stated cubic log-mean gap.
\item \emph{Proportional-effect difference.} Differencing Step~(II)
across treatment states and exponentiating preserves a fourth-order
remainder, proving the effect approximation and second-order agreement.
\item \emph{Exact endpoints.} The normalized endpoint equations give
the geometric and arithmetic means directly, completing the exact
endpoint identities as well as the approximation.
\end{enumerate}
\end{samepage}

\begin{proof}[Proof of Proposition~\ref{prop:iols-power-residual-calibration}]
\emph{(I) Uniform residual expansions.}
For \(0\leq\omega<1\), write
\begin{equation}
 R_\omega(z)
 =\frac{\log(\omega+(1-\omega)e^z)}{1-\omega},
 \label{eq:iols-power-residuals}
\end{equation}
with its continuous extension at one. On a neighborhood of \(z=0\),
this function and \(B_\omega(z)\) extend smoothly across both parameter
endpoints. To see why the apparent divisions by zero are removable,
write \(B_\omega(z)=z\int_0^1 e^{r\omega z}\,dr\) and
\(R_\omega(z)=(e^z-1)\int_0^1
[1+r(1-\omega)(e^z-1)]^{-1}\,dr\).
For small \(|z|\), the latter denominator stays uniformly away from
zero for \(\omega\in[0,1]\), and both representations have bounded
derivatives of every fixed order on a smaller compact neighborhood.
Their Taylor expansions are
\begin{equation}
 R_\omega(z)
 =z+\frac{\omega}{2}z^2
       +\frac{\omega(2\omega-1)}6z^3+O(z^4)
 \label{eq:iols-power-pilot-expansion}
\end{equation}
and
\begin{equation}
 B_\omega(z)
 =z+\frac{\omega}{2}z^2+\frac{\omega^2}{6}z^3+O(z^4).
 \label{eq:iols-power-exact-expansion}
\end{equation}
Smoothness on a compact neighborhood makes both remainders uniform
in \(\omega\in[0,1]\). Subtraction gives
\begin{equation}
 R_\omega(z)-B_\omega(z)
 =-\frac{\omega(1-\omega)}6z^3+O(z^4).
 \label{eq:iols-power-residual-gap}
\end{equation}

\emph{(II) Log-mean difference.}
Fix a treatment state, suppress \(d\), and define \(m=\E(V)\),
\(\sigma^2=\Var(V)\), and \(\kappa=\E((V-m)^3)\). Boundedness of \(V\)
makes its cumulant-generating function smooth near zero. Expanding
\(\omega^{-1}\log(\E(e^{\omega\epsilon V}))\), with its limit at zero,
gives
\begin{equation}
 \log(A_\omega)
 =a+\epsilon m+\frac{\omega\epsilon^2}{2}\sigma^2
       +\frac{\omega^2\epsilon^3}{6}\kappa+O(\epsilon^4).
 \label{eq:power-mean-small-dispersion}
\end{equation}
In detail, if \(K(r)=\log(\E(e^{rV}))\), then
\(K'(0)=m\), \(K''(0)=\sigma^2\), and \(K'''(0)=\kappa\).
The remainder in \(K(\omega\epsilon)\) is bounded by a constant times
\(|\omega\epsilon|^4\). Dividing by \(\omega>0\) leaves a bound of
order \(\omega^3\epsilon^4\), hence \(O(\epsilon^4)\) uniformly over
\([0,1]\); at zero the log mean is exactly \(a+\epsilon m\).
For the shifted-log mean, write \(q=\log(\widetilde M_\omega)\).
Dividing the argument of the logarithm in the defining equation of
\(\widetilde A_{\omega,d}(t)\) by \(A\) turns it into
\(\E(\log(\omega+(1-\omega)Y/A))=0\). Since
\((1-\omega)R_\omega(z)=\log(\omega+(1-\omega)e^z)\) and
\(\log(Y)=a+\epsilon V\), the root \(q\) solves
\(\E(R_\omega(a+\epsilon V-q))=0\).
At \(\omega=1\) we use this normalized equation's continuous
extension, not the unnormalized logarithmic equation, which becomes
an identity at that endpoint.
The function \(\E(R_\omega(a+\epsilon V-q))\) is strictly decreasing
in \(q\), positive below the support of \(\log(Y)\), and negative
above it. Hence its root is unique and finite. At \(\epsilon=0\)
the root is \(q=a\), with derivative \(-1\) with respect to \(q\),
uniformly in \(\omega\). Bounded \(V\) permits differentiation under
expectation, so the function is smooth jointly in
\((q,\epsilon,\omega)\) on an open neighborhood of
\(\{(a,0)\}\times[0,1]\). The implicit-function theorem
\citep[Corollary~6.7.13]{corbae_stinchcombe_zeman2009} applies at every
point of this set. Its derivative formula inherits the map's higher
smoothness, and a finite cover of \([0,1]\) gives a common
\(\epsilon\)-neighborhood on which the implicit branch exists, coincides
with the unique root by strict monotonicity, and has derivatives bounded
uniformly in \(\omega\). Differentiating
at zero first gives \(\partial_\epsilon q(0)=m\).

Set \(q=a+\epsilon m+c_2\epsilon^2+c_3\epsilon^3+O(\epsilon^4)\).
Its residual argument is
\(z=\epsilon(V-m)-c_2\epsilon^2-c_3\epsilon^3+O(\epsilon^4)\).
Consequently,
\(\E(z)=-c_2\epsilon^2-c_3\epsilon^3+O(\epsilon^4)\),
\(\E(z^2)=\epsilon^2\sigma^2+O(\epsilon^4)\), and
\(\E(z^3)=\epsilon^3\kappa+O(\epsilon^4)\).
The potentially cubic cross-term in \(\E(z^2)\) vanishes because it
is \(-2c_2\epsilon^3\E(V-m)=0\). Boundedness of \(V\) also makes
\(|z|=O(|\epsilon|)\) uniformly over its support and over the orders,
so the expected fourth-order residual remainder is \(O(\epsilon^4)\).
Substitution in \eqref{eq:iols-power-pilot-expansion} gives
\(c_2=\omega\sigma^2/2\) and \(c_3=\omega(2\omega-1)\kappa/6\). Thus
\begin{equation}
 \log(\widetilde M_\omega)
 =a+\epsilon m+\frac{\omega\epsilon^2}{2}\sigma^2
       +\frac{\omega(2\omega-1)\epsilon^3}{6}\kappa+O(\epsilon^4).
 \label{eq:shift-mean-small-dispersion}
\end{equation}
Subtracting \eqref{eq:power-mean-small-dispersion} proves
\eqref{eq:shift-power-mean-approximation}.

\emph{(III) Proportional-effect difference.}
Define the log-ratio gap
\(D_\omega=[\log(\widetilde M_\omega(1;\epsilon))
-\log(A_\omega(1;\epsilon))]
-[\log(\widetilde M_\omega(0;\epsilon))
-\log(A_\omega(0;\epsilon))]\).
Subtracting \eqref{eq:shift-power-mean-approximation} between treatment
states replaces \(\kappa_d\) by \(\kappa_1-\kappa_0\), with a
uniform \(O(\epsilon^4)\) remainder. By the definitions of the effects,
\(\widetilde\theta_\omega-\theta_\omega
=(1+\theta_\omega)(e^{D_\omega}-1)\).
Here \(D_\omega=O(\epsilon^3)\), so
\(e^{D_\omega}-1=D_\omega+O(\epsilon^6)\), while the exact ratio is
\(1+\theta_\omega=e^{a_1-a_0}(1+O(\epsilon))\).
Thus exponentiation multiplies the cubic term by \(e^{a_1-a_0}\)
and changes the remainder by at most \(O(\epsilon^4)\). This proves
\eqref{eq:shift-power-effect-approximation}, uniformly in \(\omega\).
It implies \eqref{eq:shift-power-effect-second-order}; the calculation uses
only the two marginal laws.

\emph{(IV) Exact endpoints.}
At \(\omega=0\), \(R_0(z)=z\), so the root equation gives
\(q=\E(\log(Y))\) and \(e^q=A_0\). At \(\omega=1\),
\(R_1(z)=e^z-1\), so the continuously extended equation gives
\(e^{-q}\E(Y)-1=0\) and \(e^q=A_1\).
The two means, and therefore their ratios across treatment states,
coincide exactly at these endpoints for every \(\epsilon\).
\end{proof}

\section{Proofs for computation and inference}
\label{app:iols-fixed-tau-proofs}
\label{app:gpml}
\label{app:iols-main-proofs}

The main proofs establish Proposition~\ref{thm:iols-bounded-power-surrogate}
on bounded curvature and global Phase 1 convergence,
Proposition~\ref{prop:iols-fixed-lambda-inference} on inference at a
fixed approximation, and Theorem~\ref{thm:iols-fixed-tau} on two-phase
completion and inference for the chosen exact equation. The main text
summarizes the fixed-approximation result informally. Its formal statement
is given in Appendix~\ref{app:iols-fixed-lambda-inference} beside its
proof. The proofs use the
following supporting results, stated and proved in
Appendix~\ref{app:iols-supplement}: existence, uniqueness and positive
curvature of the exact sample root
(Proposition~\ref{prop:iols-exact-sample-root}); existence, uniqueness,
global convergence and \(O(\delta^{-1})\) accuracy of the fixed pilot
(Propositions~\ref{prop:iols-phase1-global}
and~\ref{prop:iols-raw-delta-path}); finite completion by repeated
trials (Proposition~\ref{prop:iols-generic-completion} and
Appendix~\ref{app:iols-ols-step-trials}); and consistency and asymptotic
linearity of approximate roots (Lemma~\ref{lem:iols-score-separation} and
\eqref{eq:app-iols-asymptotic-linearity}). Variance-index continuation,
covariance estimation and standardized effects are treated there as
further results.

For compactness in the proofs, write \(P_nf_i=n^{-1}\sum_i f_i\).
We keep \(Q_n=X'X/n\) and
\(\|v\|_{Q_n^{-1}}=(v'Q_n^{-1}v)^{1/2}\).
The direct numerical results concern \(\tau\in[1,2]\).
The curvature calculation and its sample conditions concern the particular
transformation used here. The fixed-metric descent and contraction
arguments below, and the continuation argument in
Appendix~\ref{app:iols-tau-continuation}, use established optimization principles
\citep{boydvandenberghe2004,hunter_lange_2004,allgower_georg_2003}.
Local convergence of a generic corrector, e.g. damped Newton, is stated separately. Outside that
interval, the continuation guarantee requires the additional local 
conditions stated in Appendix~\ref{app:iols-tau-continuation}.

\subsection{Bounded curvature and convergence of Phase 1}
\label{app:iols-phase1-proof}

We first set out the fixed-order pilot and its approximation bound, then
prove Proposition~\ref{thm:iols-bounded-power-surrogate}.

\label{app:iols-pilot-construction}
Fix \(\omega>0\) and \(\tau\), keeping
\(\mu_i(\beta)^\omega=\exp(\omega X_i'\beta)\) from
Section~\ref{sec:iols-power-mean-surrogate}.
For the numerical construction, define
\(a_{\omega,\tau,i}(\beta)=Y_i^\omega\mu_i(\beta)^{\omega(1-\tau)}\)
and \(b_{\omega,\tau,i}(\beta)=\mu_i(\beta)^{\omega(2-\tau)}\).
The positive scalar \(\delta=\lambda/(1-\lambda)\) controls approximation
accuracy, not the economic order or score weighting. In this parameterization,
the working outcome \eqref{eq:iols-power-mean-pilot-response} is
\(H^{(1)}_{\omega,\tau,\delta,i}(\beta)
=X_i'\beta+\omega^{-1}\log((\delta+a_{\omega,\tau,i}(\beta))/
(\delta+b_{\omega,\tau,i}(\beta)))\).
Its OLS map is
\(T^{(1)}_{\omega,\tau,\delta,n}(\beta)
=Q_n^{-1}P_nX_iH^{(1)}_{\omega,\tau,\delta,i}(\beta)\).
For each fixed \(\delta\), these are the main-text objects indexed by
\(\lambda=\delta/(1+\delta)\); the alternative subscript here is only
a reparameterization for the approximation proofs.

At \(\omega=1\), this response equals \(H_{\lambda,\tau,i}\) in
\eqref{eq:comparison-fitted-mean-shift} for every \(\tau\).
At \(\omega=1,\tau=2\), it simplifies to
\(\log((Y_i+\delta\mu_i)/(1+\delta))\), hence equals
the GPML specialization \(H_{\lambda,2,i}\) in
\eqref{eq:comparison-gpml-working-outcome} for
\(\lambda=\delta/(1+\delta)\).
At any fixed \(\omega>0\) and \(\tau=2\), it instead applies the
same shifted log to \(Y_i^\omega\), with its index divided by \(\omega\).
Increasing \(\delta\) then approaches the exact order-\(\omega\)
equation. This refinement holds \(\omega\) fixed; it is distinct from
moving the matched economic order along the original fitted-mean-shift
curve. For general \(\tau\), the two terms inside the logarithm
also incorporate the specified power-score weights.

Write \(\ell_i=H^{(1)}_{\omega,\tau,\delta,i}-X_i'\beta\).
The supporting proofs use the scaled pilot moment
\(g^{(1)}_{\omega,\tau,\delta,n}=\delta P_nX_i\ell_i\) and
potential \(\Phi_{\omega,\tau,\delta,n}\), with
\(\nabla\Phi=-g^{(1)}\). The main text uses the normalized
criterion \(\Psi^{(1)}_{\omega,\tau,\lambda,n}=\Phi/\delta\), which gives
the same OLS update and root. Under the conditions of
Proposition~\ref{thm:iols-bounded-power-surrogate}, the quantitative
approximation bound is
\begin{equation}
 \|\widehat\beta^{(1)}_{\omega,\tau,\delta,n}
       -\widehat\beta_{\omega,\tau,n}\|_{Q_n}
 \leq C_n/\delta
 \quad\text{for every }\delta\geq\bar\delta_n,
 \label{eq:iols-power-mean-root-rate}
\end{equation}
for finite sample-dependent constants \(C_n\) and
\(\bar\delta_n>0\). Here the first coefficient vector is the
fixed-pilot root and the second solves the original moment.
Equivalently, its error is at most \(C_n(1-\lambda)/\lambda\) for
\(\lambda\) sufficiently close to one.
The following proof establishes this bound together with the proposition's
global convergence statement.

The following proof proceeds in four steps.
The curvature bound controls each OLS update. The change of coordinates
then lets us use the supporting existence and convergence results to
identify its limit and bound its distance from the exact root.

\begin{enumerate}[label=\textup{(\Roman*)},leftmargin=*,itemsep=1pt,topsep=3pt,parsep=0pt]
\item \emph{Curvature.} Differentiate the residuals to express the
criterion's Hessian relative to the fixed design matrix. The resulting
bound holds at every coefficient value and controls the OLS update.
\item \emph{Descent and stability.} Integrating (I) gives a quadratic
upper bound minimized by the OLS step, proving descent. The same Hessian
bound also prevents expansion of distances between starts.
\item \emph{Equivalent power-order problem.} The change to \(Y_i^\omega\)
and \(\gamma=\omega\beta\) preserves the sample conditions and maps entire
iterations and roots to the order-one problem, not just their equations.
\item \emph{Existence, convergence and approximation.} This equivalence
transfers the supporting convergence theorem and root-distance bound.
Thus the pilot converges to its unique root from any finite start, and
refinement makes that root arbitrarily close to the exact one.
\end{enumerate}

\begin{proof}[Proof of Proposition~\ref{thm:iols-bounded-power-surrogate}]
\emph{(I) Curvature.}

Abbreviate \(a_i=a_{\omega,\tau,i}(\beta)\),
\(b_i=b_{\omega,\tau,i}(\beta)\), and
\(\ell_i(\beta)=\omega^{-1}\log((\delta+a_i)/(\delta+b_i))\).
Differentiating with respect to \(q_i=X_i'\beta\) gives
\(\partial\ell_i/\partial q_i=-d_{\omega,\tau,\delta,i}(\beta)\), where
\begin{equation}
 d_{\omega,\tau,\delta,i}(\beta)
 =(\tau-1)\frac{a_i}{\delta+a_i}
 +(2-\tau)\frac{b_i}{\delta+b_i}.
 \label{eq:iols-power-mean-pilot-weight}
\end{equation}
The fractions lie in \([0,1)\), and the nonnegative coefficients sum to
one. Thus \(0\leq d_i<1\).
Each residual depends on \(\beta\) only through \(q_i\), so integrating
its negative with respect to that scalar index gives a criterion.
The resulting scaled potential satisfies
\(\nabla\Phi_{\omega,\tau,\delta,n}=-g^{(1)}_{\omega,\tau,\delta,n}\)
and has Hessian
\begin{equation}
 \nabla^2\Phi_{\omega,\tau,\delta,n}(\beta)
 =\delta P_nX_iX_i'd_{\omega,\tau,\delta,i}(\beta).
 \label{eq:iols-power-mean-pilot-hessian}
\end{equation}
For every vector \(v\), the Hessian quadratic form is
\(\delta P_nd_i(X_i'v)^2\), and it is bounded between zero and
\(\delta P_n(X_i'v)^2=\delta v'Q_nv\). Dividing by \(\delta\) proves
\eqref{eq:iols-power-mean-global-cap} for
\(\Psi^{(1)}_{\omega,\tau,\lambda,n}=\Phi/\delta\). The OLS normal equations give
\(T^{(1)}(\beta)=\beta-\delta^{-1}Q_n^{-1}\nabla\Phi(\beta)
=\beta-Q_n^{-1}\nabla\Psi^{(1)}_{\omega,\tau,\lambda,n}(\beta)\).

\emph{(II) Descent and stability.}
For \(d=\vartheta-\beta\), the Taylor remainder is
\(\int_0^1(1-s)d'\nabla^2\Phi(\beta+sd)d\,ds\), which is at most
\((\delta/2)\|d\|_{Q_n}^2\). Thus
\(\Phi(\vartheta)\leq\Phi(\beta)+\nabla\Phi(\beta)'(\vartheta-\beta)
+(\delta/2)\|\vartheta-\beta\|_{Q_n}^2\).
The right side is a quadratic upper bound, minimized by the OLS update.
Indeed, if \(v=T^{(1)}(\beta)-\beta\), then
\(\nabla\Phi(\beta)=-\delta Q_nv\). Substitution yields
\begin{equation}
 \Phi(T^{(1)}(\beta))
 \leq\Phi(\beta)-\frac{\delta}{2}
       \|T^{(1)}(\beta)-\beta\|_{Q_n}^2.
 \label{eq:iols-power-mean-majorization-descent}
\end{equation}
To compare two starts, define
\(\overline H=\int_0^1\delta^{-1}Q_n^{-1/2}
\nabla^2\Phi(\vartheta+s(\beta-\vartheta))Q_n^{-1/2}\,ds\).
The curvature bound implies \(0\preceq\overline H\preceq I\).
Integration of the map's derivative gives
\(Q_n^{1/2}(T^{(1)}(\beta)-T^{(1)}(\vartheta))
=(I-\overline H)Q_n^{1/2}(\beta-\vartheta)\).
Writing \(v=Q_n^{1/2}(\beta-\vartheta)\), the squared norm of the
left side is \(v'(I-\overline H)^2v\). Since
\((I-\overline H)^2\preceq I-\overline H\), this proves firm
nonexpansiveness:
\(\|T^{(1)}(\beta)-T^{(1)}(\vartheta)\|_{Q_n}^2
\leq(T^{(1)}(\beta)-T^{(1)}(\vartheta))'Q_n(\beta-\vartheta)\).
The Cauchy--Schwarz inequality then gives ordinary nonexpansiveness.
These inequalities establish stability of the map. Existence of its
fixed point and a strict contraction factor require the sample conditions
used next.

\emph{(III) Equivalent power-order problem.}
The supporting results in Appendix~\ref{app:iols-phase1-supplement}
are stated at order one. Apply them to the outcome \(Y_i^\omega\)
and coefficient \(\gamma=\omega\beta\), and denote the resulting
pilot map and exact moment by \(\widetilde T^{(1)}\) and
\(\widetilde g_{\tau,n}\). Their working outcome at \(\gamma=\omega\beta\)
is \(\omega\) times \eqref{eq:iols-power-mean-pilot-response}; hence
linearity of OLS gives
\(\widetilde T^{(1)}(\omega\beta)=\omega T^{(1)}(\beta)\).
Similarly, \(\widetilde g_{\tau,n}(\omega\beta)
=\omega g_{\omega,\tau,n}(\beta)\).
Thus entire iteration sequences, fixed points and exact roots correspond
under this change of coordinates, with distances multiplied by \(\omega\)
and contraction factors unchanged. Since \(\omega>0\),
\(Y_i^\omega>0\) exactly when \(Y_i>0\); the design and positive-outcome
indicator \(D_Y\), and therefore all the sample conditions, are unchanged.

\emph{(IV) Existence, convergence and approximation.}
By the equivalence in (III), Proposition~\ref{prop:iols-phase1-global}
gives a unique pilot root and
Q-linear convergence from every finite start. More precisely, on the
invariant sublevel set reached from that start, the integrated derivative
bound gives
\(\|\beta^{(m+1)}-\widehat\beta^{(1)}_{\omega,\tau,\lambda,n}\|_{Q_n}
\leq\kappa\|\beta^{(m)}-\widehat\beta^{(1)}_{\omega,\tau,\lambda,n}\|_{Q_n}\)
for some \(\kappa<1\). The constant can depend on the starting value
at \(\tau=1\) and \(\tau=2\); for \(1<\tau<2\), the supporting
proposition gives a global bound independent of the start.
Proposition~\ref{prop:iols-exact-sample-root} gives the
unique original root. Proposition~\ref{prop:iols-raw-delta-path} gives the
\(O(\delta^{-1})\) distance between the two roots in \(\gamma\)
coordinates. Division by the fixed \(\omega>0\) and equivalence of the
Euclidean and \(Q_n\)-norms yield
\eqref{eq:iols-power-mean-root-rate}.
For any \(\varepsilon>0\), a fixed
\(\delta\geq\max\{\bar\delta_n,C_n/\varepsilon\}\)
therefore gives the approximation asserted in
Proposition~\ref{thm:iols-bounded-power-surrogate}.
\end{proof}

\subsection{Inference for a fixed pilot approximation}
\label{app:iols-fixed-lambda-inference}

Use the pilot contribution from
Section~\ref{sec:iols-power-mean-surrogate},
\(\psi_{\lambda i}(\beta)=X_i(H^{(1)}_{\omega,\tau,\lambda,i}(\beta)
-X_i'\beta)\), with the fixed \(\omega,\tau\) notation suppressed, and its
population root \(\beta_\lambda\) solving
\(\E(\psi_{\lambda i}(\beta_\lambda))=0\). The formal statement and its
proof are given here.

\begin{proposition}[Asymptotic properties of the approximate map]
\label{prop:iols-fixed-lambda-inference}
Fix \(\omega>0\), \(\tau\in[1,2]\) and \(0<\lambda<1\).
Suppose
Assumption~\ref{ass:iols-positive-overlap} holds with probability
approaching one, and Assumption~\ref{ass:iols-sampling-design} holds with
\(\psi_i=\psi_{\lambda i}\) and \(\beta_\star=\beta_\lambda\). Then the
Phase 1 fixed point
\(\widehat\beta^{(1)}_{\lambda,n}\) converges in probability to
\(\beta_\lambda\) and satisfies
\begin{equation}
 \sqrt n(\widehat\beta^{(1)}_{\lambda,n}-\beta_\lambda)
 =J_\lambda^{-1}\frac1{\sqrt n}\sum_i
 \psi_{\lambda i}(\beta_\lambda)+o_p(1),
 \label{eq:iols-fixed-lambda-expansion}
\end{equation}
where \(J_\lambda=-\E(\partial\psi_{\lambda i}(\beta_\lambda)/
\partial\beta')\). Consequently,
\begin{equation}
 \sqrt n(\widehat\beta^{(1)}_{\lambda,n}-\beta_\lambda)
 \Longrightarrow N(0,\Omega_\lambda),
 \label{eq:iols-fixed-lambda-normality}
\end{equation}
with \(\Omega_\lambda=J_\lambda^{-1}
\E(\psi_{\lambda i}(\beta_\lambda)\psi_{\lambda i}(\beta_\lambda)')
J_\lambda^{-1}\).
\end{proposition}

Thus Phase 1 consistently estimates its population transformed-moment
root, which approximates the coefficient \(\beta_0\) of the exact
model. Corollary~\ref{cor:iols-fixed-lambda-numerical} transfers the same
first-order distribution to an iteration, obtained after finitely many iterations, whose transformed
sample-moment residual is \(o_p(n^{-1/2})\).

\begin{samepage}
\noindent The following proof proceeds in two steps.
\begin{enumerate}[label=\textup{(\Roman*)},leftmargin=*,itemsep=1pt,topsep=3pt,parsep=0pt]
\item \emph{Root existence and consistency.} Integrating the moment
gives a convex criterion whose population minimum is identified by
positive curvature. The moment law implies criterion convergence, so
the convex consistency theorem gives an interior consistent minimizer.
Local strict curvature and global convexity make this the unique sample root.
\item \emph{Linear expansion and normality.} Consistency makes the
integrated Jacobian converge to \(J_\lambda\). Its inverse converts the
sample moment at \(\beta_\lambda\) into the coefficient error, and the
moment's central limit theorem gives the stated distribution.
\end{enumerate}
\end{samepage}

\begin{proof}[Proof of Proposition~\ref{prop:iols-fixed-lambda-inference}]
All limits are as \(n\to\infty\), with \(\omega,\tau,\lambda\) fixed.

Define \(h_\lambda(\beta)=\E(\psi_{\lambda i}(\beta))\) and
\(h_{\lambda,n}(\beta)=P_n\psi_{\lambda i}(\beta)\).
Denote the sample negative Jacobian by
\(J_{\lambda,n}(\beta)=-\partial h_{\lambda,n}(\beta)/\partial\beta'\)
and its population counterpart by
\(\mathcal J_\lambda(\beta)=-\E(\partial\psi_{\lambda i}(\beta)/\partial\beta')\).

\emph{(I) Root existence and consistency.}
The derivative envelope in Assumption~\ref{ass:iols-sampling-design}(iii)
permits differentiation under expectation and makes
\(\mathcal J_\lambda\) continuous.
Since \(\mathcal J_\lambda(\beta_\lambda)=J_\lambda\succ0\), a closed
ball \(\mathcal N_0\subset\mathcal N\cap\operatorname{int}(\mathcal B)\)
centered at \(\beta_\lambda\) and a constant \(c>0\) can be chosen so that
\(\mathcal J_\lambda(\beta)\succeq cI\) throughout \(\mathcal N_0\).
The moment and derivative are continuous and have integrable envelopes
on this compact ball. Applying \citet[Lemma 2.4]{newey_mcfadden94}
componentwise gives
\(\sup_{\beta\in\mathcal N_0}
\|h_{\lambda,n}(\beta)-h_\lambda(\beta)\|_2=o_p(1)\) and
\(\sup_{\beta\in\mathcal N_0}
\|J_{\lambda,n}(\beta)-\mathcal J_\lambda(\beta)\|_2=o_p(1)\).
In particular, the sample Hessian is positive definite on
\(\mathcal N_0\) with probability approaching one.

By \eqref{eq:iols-power-mean-pilot-weight},
\(J_{\lambda,n}(\beta)=P_nX_iX_i'd_{\omega,\tau,\delta,i}(\beta)\succeq0\)
on \(\mathbb R^K\), so the pilot criterion is globally convex.
Subtract its value at \(\beta_\lambda\) and define
\begin{equation}
 L_{\lambda,n}(\beta)
 =-\int_0^1 h_{\lambda,n}(\beta_\lambda+t(\beta-\beta_\lambda))'
       (\beta-\beta_\lambda)\,dt.
 \label{eq:iols-fixed-lambda-anchored-criterion}
\end{equation}
Let \(L_\lambda\) be the same integral with \(h_\lambda\) in place of
\(h_{\lambda,n}\). If \(r_0\) is the radius of \(\mathcal N_0\), then
\(\sup_{\mathcal N_0}|L_{\lambda,n}-L_\lambda|
\leq r_0\sup_{\mathcal N_0}\|h_{\lambda,n}-h_\lambda\|_2=o_p(1)\).
Moreover, \(h_\lambda(\beta_\lambda)=0\) and the curvature bound give
\(L_\lambda(\beta)\geq(c/2)\|\beta-\beta_\lambda\|_2^2\).
Thus \(-L_{\lambda,n}\) is concave on the convex ball, converges
pointwise to \(-L_\lambda\), and its limit is uniquely maximized at
the interior point \(\beta_\lambda\).
\citet[Theorem 2.7]{newey_mcfadden94} therefore
gives a consistent sample minimizer. With probability approaching one,
it is interior and hence solves \(h_{\lambda,n}(\beta)=0\).

Local strict curvature makes this root unique in \(\mathcal N_0\).
If a second root existed elsewhere, convexity would make the criterion
constant on the segment joining the two minima, contradicting strict
curvature near the first root. The unique root is therefore the
Phase~1 fixed point on the event in
Assumption~\ref{ass:iols-positive-overlap}, whose probability tends to one.
This proves \(\widehat\beta^{(1)}_{\lambda,n}\to_p\beta_\lambda\).

\emph{(II) Linear expansion and normality.}
Abbreviate \(\widehat\beta=\widehat\beta^{(1)}_{\lambda,n}\) and define
\(\overline J_{\lambda,n}=\int_0^1J_{\lambda,n}
(\beta_\lambda+s(\widehat\beta-\beta_\lambda))\,ds\).
Integration of the sample derivative gives the exact identity
\begin{equation}
 0=h_{\lambda,n}(\widehat\beta)
 =h_{\lambda,n}(\beta_\lambda)
 -\overline J_{\lambda,n}(\widehat\beta-\beta_\lambda).
 \label{eq:iols-fixed-lambda-integrated-identity}
\end{equation}
Consistency places the segment in \(\mathcal N_0\) with probability
approaching one and makes it shrink to \(\beta_\lambda\).
The uniform derivative law controls its sampling error, and continuity
of \(\mathcal J_\lambda\) controls its change along this shrinking segment. Thus
\(\overline J_{\lambda,n}\to_pJ_\lambda\) and
\(\overline J_{\lambda,n}^{-1}=J_\lambda^{-1}+o_p(1)\).
The iid vectors \(\psi_{\lambda i}(\beta_\lambda)\) are centered by the
population restriction and square-integrable by part~(iii). Their central
limit theorem makes \(\sqrt n h_{\lambda,n}(\beta_\lambda)=O_p(1)\).
Solving \eqref{eq:iols-fixed-lambda-integrated-identity}, replacing the
inverse Jacobian costs only \(o_p(1)O_p(1)=o_p(1)\), proving
\eqref{eq:iols-fixed-lambda-expansion}. Slutsky's theorem gives the
normal limit with covariance \(\Omega_\lambda\). This is the
estimating-equation argument of \citet[Section 3.1]{newey_mcfadden94},
applied to the transformed moment using the integrated Jacobian.
\end{proof}

\Needspace{8\baselineskip}
\subsection{Two-phase convergence and inference}
\label{app:iols-two-phase-proof}
\begin{samepage}
\noindent The following proof proceeds in five steps.
For computation, it suffices to reach a neighborhood in which the exact
iteration contracts; (I)--(III) show that the pilot and repeated trials
guarantee reaching this neighborhood. For inference, (IV)--(V) show that the resulting numerical output from finitely many iterations 
is consistent and that its remaining error is negligible at the sampling scale.
\begin{enumerate}[label=\textup{(\Roman*)},leftmargin=*,itemsep=1pt,topsep=3pt,parsep=0pt]
\item \emph{Local convergence.} Positive curvature at the exact root
gives an admissible step and an invariant contraction neighborhood;
every start inside it therefore converges geometrically.
\item \emph{Pilot initialization.} Approximation error plus numerical
pilot error bounds the distance to the exact root. Making their sum
smaller than the radius in (I) provides a valid local start from any
finite original start.
\item \emph{Finite completion.} Such choices exist by (II); revisiting
candidate choices with growing iteration budgets must therefore reach
the requested tolerance without knowing the sufficient constants.
\item \emph{Consistency.} Identification and sampling-error control,
extended by convexity, exclude small moments far from the population root.
The vanishing residual guaranteed in (III) therefore implies consistency.
\item \emph{First-order inference.} At the consistent output, the
moment expansion has an invertible limiting Jacobian; a residual of
\(o_p(n^{-1/2})\) contributes nothing to its limit. The score central
limit theorem and power-order rescaling then give the stated distribution.
\end{enumerate}
\end{samepage}

\begin{proof}[Proof of Theorem~\ref{thm:iols-fixed-tau}]
\emph{(I) Local convergence.}

For part \textnormal{(i)}, applying
Proposition~\ref{prop:iols-exact-sample-root} to \(Y_i^\omega\) and
\(\gamma=\omega\beta\) gives positive curvature at the unique exact root.
The derivative of the OLS map in \(Q_n^{1/2}\beta\) coordinates is
\(I-a_\rho Q_n^{-1/2}J_{\omega,\tau,n}(\beta)Q_n^{-1/2}\).
The negative score derivative is symmetric, and is positive definite at
the root. Thus \(0<a_\rho<2/\nu_{\max,n}\) places every eigenvalue of
the map derivative strictly between \(-1\) and \(1\). These are the
hypotheses verified in the
\hyperref[prop:iols-local-exact-completion]{local exact-completion argument}:
derivative continuity gives an invariant \(Q_n\)-ball with a strict
contraction factor. Its bound \eqref{eq:app-iols-phase2-geometric}
proves the stated local Q-linear convergence.

\emph{(II) Pilot initialization.}
For this map or another specified locally convergent solver, let \(r>0\)
be a convergence-ball radius and let \(\eta\) be the computed pilot's
distance from its fixed point in the \(Q_n\)-norm. The triangle
inequality bounds its distance to the exact root by
\(C_n(1-\lambda)/\lambda+\eta\), provided
\(\delta=\lambda/(1-\lambda)\geq\bar\delta_n\), where \(C_n\) is the
fixed-sample approximation constant. Requiring this bound to be below \(r\)
places it in the convergence ball. Proposition~\ref{thm:iols-bounded-power-surrogate}
and \eqref{eq:iols-power-mean-root-rate} provide a finite approximation
whose root error is below \(r/3\). Convergence of its fixed Phase 1 map
from any finite start provides a finite iterate with numerical error
below \(r/3\). Both steps are finite and keep \(\omega,\tau\) fixed.

\emph{(III) Finite completion.}
The choices in (II) need not be known in advance.
Proposition~\ref{prop:iols-generic-completion} and the OLS construction
in Appendix~\ref{app:iols-ols-step-trials} give the repeated-trial
implementation when the constants are unknown. Some fixed triple of
approximation level, pilot iteration count, and admissible OLS step
therefore succeeds at every chosen positive tolerance. The schedule
revisits that triple with increasing exact-iteration budget limits, so it reaches
the tolerance in finitely many iterations.

\emph{(IV) Consistency.}
For part \textnormal{(ii)}, first let \(\omega=1\).
Assumption~\ref{ass:iols-positive-overlap} holds with probability
approaching one, by its final sentence, so part \textnormal{(i)} applies
on that event. Lemma~\ref{lem:iols-score-separation}(ii) establishes
that every finite output with a vanishing moment residual is consistent:
uniform separation excludes such residuals outside any fixed neighborhood
of \(\beta_0\). The repeated-trial value \(\overline\beta_n\)
has residual at most
\(\varepsilon_n=o(n^{-1/2})\) on a sample event whose probability
approaches one. The lemma therefore gives
\(\overline\beta_n\to_p\beta_0\) without requiring compact containment
of the numerical procedure.

\emph{(V) First-order inference.}
Consistency implies \(\Pr(\overline\beta_n\in\mathcal B)\to1\)
because \(\beta_0\) is interior to \(\mathcal B\).
Together with the residual rate and the stated sampling conditions,
this verifies every hypothesis of the
\hyperref[prop:iols-asymptotic-linearity]{approximate-root expansion}.
Equation~\eqref{eq:app-iols-asymptotic-linearity} therefore gives the
linear representation, and \eqref{eq:app-iols-asymptotic-normality}
gives its normal limit. The integrated-Jacobian argument established
there also shows why the residual contributes only \(o_p(1)\) at the
\(\sqrt n\) scale.

For a fixed \(\omega>0\), apply the same argument to the nonnegative
outcome \(Y_i^\omega\) and coefficient \(\gamma=\omega\beta\).
Its exact moment in \(\gamma\) is \(\omega g_{\omega,\tau,n}(\beta)\),
so the fixed positive scaling preserves the numerical residual rate.
The parameter set and local neighborhood become \(\omega\mathcal B\)
and \(\omega\mathcal N\). Their compactness, interior containment and
envelopes are preserved by this invertible linear change of coordinates.
The assumed conditional mean is exactly the mean restriction for this
transformed outcome. The negative derivative in \(\beta\) is
\(J_{\omega,\tau,n}\) and conditioning at \(\beta_0\) gives
\(J_{\omega,\tau}=\E(X_iX_i'\mu_i(\beta_0)^{\omega(2-\tau)})\).
Division of the coefficient expansion by \(\omega\) gives
\eqref{eq:iols-asymptotic-linearity} with observation contribution
\(s_{\omega,\tau,i}(\beta_0)\). Its covariance is the one stated in
the theorem, proving \eqref{eq:iols-asymptotic-normality}.

Finally, the sample-condition event is measurable by
Lemma~\ref{lem:iols-sample-event}. Defining the output to be a fixed
vector on its vanishing-probability complement gives an everywhere
defined estimator without changing either probability limit.
\end{proof}

\section{Proofs for fixed effects}
\label{app:fixed-effects}
\label{app:fe-main-proofs}

This appendix proves Proposition~\ref{prop:fe-general-cancellation} and
Theorem~\ref{thm:fe-growing-general}.
Appendix~\ref{app:fe-cancellation-proof} gives the cancellation proof,
which uses the group laws verified in
Appendix~\ref{app:fe-weighted-regularity}.
Appendix~\ref{app:fe-growing-general} states the growing-effect setting
and Assumption~\ref{ass:fe-growing-regularity}, which enter the
theorem's statement, together with the feasible corrections of
Appendix~\ref{app:fe-growing-corrections}.
Appendix~\ref{app:fe-growing-proof} then proves the theorem. That proof
uses three supporting lemmas proved in Appendix~\ref{app:fe-supplement}:
Lemma~\ref{lem:fe-growing-normalized} gives the normalized derivative
bounds and fitted-index concentration,
Lemma~\ref{lem:fe-growing-quadratic} the target quadratic fluctuations,
and Lemma~\ref{lem:fe-growing-feasible} the feasible covariance and bias
substitutions. 
For the appendix calculations, write
\(R_i(\theta)=Y_i\exp(-B_i'\theta)\), with \(\theta=\beta\)
when only slopes remain, and collect these relative outcomes in
\(R(\theta)\). The design \(B\) is the joint, cancelled or mixed
design specified in the corresponding result.\footnote{Whenever a covariance formula below has an undefined inverse, assign
the covariance estimate a fixed finite matrix. Use the same convention
when a numerical estimator needs a value on an invalid sample.
These are asymptotically negligible events.}

\subsection{Cancellation with many short groups}
\label{app:fe-cancellation-proof}
\begin{samepage}
\noindent The following proof proceeds in four steps.
\begin{enumerate}[label=\textup{(\Roman*)},leftmargin=*,itemsep=1pt,topsep=3pt,parsep=0pt]
\item \emph{Separate the expected equation.}
Verify that cancellation gives mean-zero group scores at \(\beta_0\).
The pairwise identity \eqref{eq:fe-weighted-identification} and positive
information limit then show that, for all sufficiently large \(n\),
\(\beta_0\) is the only solution of the expected equation in \(\Theta\).
The resulting bound separates the expected moment from zero outside
each fixed neighborhood of \(\beta_0\).
\item \emph{Transfer separation to the sample.}
Show that any approximate sample root in \(\Theta\) must be close to
\(\beta_0\). The group uniform law in Lemma~\ref{lem:fe-group-laws}
makes the sampling error uniformly small. Combined with Step~(I), it
bounds the estimation error by the sample residual and sampling error.
A vanishing residual therefore gives consistency, conditional on existence.
\item \emph{Construct a local root.}
Prove that such a sample solution actually exists with probability
approaching one. A contraction that leaves a closed neighborhood of
\(\beta_0\) invariant gives a unique local root by Banach's theorem.
Step~(II) excludes roots outside this neighborhood in \(\Theta\),
so the sample solution is unique and consistent.
\item \emph{Expansion and inference.}
Use consistency to linearize the sample equation around \(\beta_0\).
The nonsingular limiting Jacobian gives
\eqref{eq:fe-general-cancel-expansion}, with a numerical residual of
\(o_p(n^{-1/2})\) negligible at this order. The group central limit
theorem in Lemma~\ref{lem:fe-group-laws} then gives asymptotic normality.
Consistency and the uniform Jacobian and score-product laws give a
consistent sandwich covariance estimate.
\end{enumerate}
\end{samepage}

\begin{proof}[Proof of Proposition~\ref{prop:fe-general-cancellation}]
\emph{(I) Separate the expected equation.}
Write \(w_i=w_i(\beta)>0\),
\(W_g=\sum_{i\in g}w_i\), and \(d=\beta-\beta_0\).
At the truth, \(\E(Y_i/\mu_i(\beta_0)\mid\mathcal H_g)=A_g\).
Because the weighted deviations satisfy \(\sum_{i\in g}w_i(B_i-\bar B_{g,w})=0\)
identically in \(\beta\), factoring out \(A_g\) gives
\(\E(\psi_{g,w}(\beta_0)\mid\mathcal H_g)=0\) for every
group, without imposing independence within the group.
To bound the expected score away from zero when \(d\neq0\), define the
within-group index error \(t_i=B_i'd\) and its weighted group average
\(\bar t_{g,w}=\sum_{i\in g}w_it_i/W_g\). Since
\(\E(Y_i/\mu_i(\beta)\mid\mathcal H_g)=A_g\exp(-t_i)\), the
conditional expected score at \(\beta\) is
\(A_g\sum_{i\in g}w_i(B_i-\bar B_{g,w})\exp(-t_i)\), and
its inner product with \(d\) is
\(A_g\sum_{i\in g}w_i(t_i-\bar t_{g,w})\exp(-t_i)\).
Substituting \(t_i-\bar t_{g,w}=W_g^{-1}\sum_{j\in g}w_j(t_i-t_j)\)
and averaging the resulting double sum with the one obtained by
interchanging \(i\) and \(j\) yields
\begin{equation}
 d'\E(\psi_{g,w}(\beta)\mid\mathcal H_g)
 =\frac{A_g}{2W_g}\sum_{i,j\in g}
 w_iw_j(t_i-t_j)(\exp(-t_i)-\exp(-t_j))\leq0.
 \label{eq:fe-weighted-identification}
\end{equation}
Bounded regressors and compact \(\Theta\) bound all fitted means,
weights, and \(\exp(-t_i)\) above and away from zero, uniformly
over groups and \(\beta\in\Theta\). By the integral identity
\(\exp(-t_j)-\exp(-t_i)
=(t_i-t_j)\int_0^1\exp(-t_j-s(t_i-t_j))\,ds\),
the negative of each summand in \eqref{eq:fe-weighted-identification}
is bounded below by a fixed positive constant times
\(w_iw_j(t_i-t_j)^2\).
Using the weighted-variance identity
\((2W_g)^{-1}\sum_{i,j\in g}w_iw_j(t_i-t_j)^2
=\min_c\sum_{i\in g}w_i(t_i-c)^2\),
we obtain a lower bound proportional to
\(A_g\min_c\sum_{i\in g}w_i(\beta)(t_i-c)^2\).
Since \(w_i(\beta)\) and \(w_i(\beta_0)\) are uniformly comparable on
\(\Theta\), replacing \(w_i(\beta)\) by \(w_i(\beta_0)\) before
minimizing over \(c\) (which attains its minimum at
\(c=\bar B_{g,w}(\beta_0)'d\)) bounds
\(-d'\E(\psi_{g,w}(\beta)\mid\mathcal H_g)\) below by a fixed
positive constant times
\(A_g\sum_{i\in g} w_i(\beta_0)
(B_i'd-\bar B_{g,w}(\beta_0)'d)^2\).
To connect this lower bound to the expected Jacobian at \(\beta_0\),
write \(c_i(\beta)=w_i(\beta)(B_i-\bar B_{g,w}(\beta))\) so that
\(\psi_{g,w}(\beta)=\sum_{i\in g}c_i(\beta)Y_i/\mu_i(\beta)\).
Because \(\sum_{i\in g} c_i(\beta)=0\) for every \(\beta\),
differentiating with respect to \(\beta'\) gives
\(\sum_{i\in g}\partial c_i(\beta)/\partial\beta'=0\). At \(\beta_0\),
\(\E(Y_i/\mu_i(\beta_0)\mid\mathcal H_g)=A_g\) is constant across
\(i\in g\), so the terms involving \(\partial c_i(\beta_0)/\partial\beta'\)
sum to zero in conditional expectation, leaving the negative conditional
expected derivative \(A_g\sum_{i\in g} c_i(\beta_0)B_i'\).
Since \(\sum_{i\in g} c_i(\beta_0)=0\), subtracting
\(\bar B_{g,w}(\beta_0)'\) from \(B_i'\) leaves this matrix unchanged
and shows that group \(g\)'s conditional expected Jacobian contribution is
\(A_g\sum_{i\in g}w_i(\beta_0)(B_i-\bar B_{g,w}(\beta_0))
(B_i-\bar B_{g,w}(\beta_0))'\).
Averaging over groups and taking expectations, the lower bound from the
previous paragraph is proportional to \(d'\E(J_{w,n}(\beta_0))d\).
Since \(\E(J_{w,n}(\beta_0))\to J_w\succ0\), there exists \(a>0\) such
that for all sufficiently large \(n\),
\(-d'\E(\Psi_{w,n}(\beta))\geq a\|d\|_2^2\) uniformly
on \(\Theta\).
This establishes global separation of the expected sample equation
without requiring identical group distributions.

\emph{(II) Transfer separation to the sample.}
Define the deterministic expected score
\(h_n(\beta)=\E(\Psi_{w,n}(\beta))\) and the uniform sample error
\(\Delta_n=\sup_{\beta\in\Theta}
\|\Psi_{w,n}(\beta)-h_n(\beta)\|_2\).
The group uniform law in Lemma~\ref{lem:fe-group-laws}
gives \(\Delta_n=o_p(1)\), and the Cauchy--Schwarz inequality applied to
Step~(I) gives \(\|h_n(\beta)\|_2\geq a\|\beta-\beta_0\|_2\) for all
\(\beta\in\Theta\).
By the triangle inequality, any measurable candidate
\(\widetilde\beta_n\in\Theta\) therefore satisfies
\(a\|\widetilde\beta_n-\beta_0\|_2
\leq\|h_n(\widetilde\beta_n)\|_2
\leq\|\Psi_{w,n}(\widetilde\beta_n)\|_2+\Delta_n\).
Consequently, any approximate root \(\widetilde\beta_n\in\Theta\) with
residual \(\|\Psi_{w,n}(\widetilde\beta_n)\|_2=o_p(1)\) satisfies
\(\|\widetilde\beta_n-\beta_0\|_2=o_p(1)\).
Moreover, for any fixed radius \(r>0\), the event \(\Delta_n<ar\) rules
out exact roots of \(\Psi_{w,n}(\beta)=0\) outside the closed ball
\(N_r=\{\beta:\|\beta-\beta_0\|_2\le r\}\).
This step establishes consistency conditional on root existence while
using the \(n\)-dependent expected moment \(h_n(\beta)\).

\emph{(III) Construct a local root.}
To show that an exact root exists in \(N_r\subset\operatorname{int}(\Theta)\) with
probability approaching one, define the auxiliary map
\(T_n(\beta)=\beta+J_w^{-1}\Psi_{w,n}(\beta)\), whose fixed points
coincide with roots of \(\Psi_{w,n}(\beta)=0\) because \(J_w\) is
invertible. Its derivative is
\(DT_n(\beta)=J_w^{-1}(J_w-J_{w,n}(\beta))\).
By the triangle inequality, \(\sup_{\beta\in N_r}\|J_w-J_{w,n}(\beta)\|\)
is bounded by the sum of \(\|J_w-\E(J_{w,n}(\beta_0))\|\),
\(\sup_{\beta\in N_r}\|\E(J_{w,n}(\beta_0))-\E(J_{w,n}(\beta))\|\), and
\(\sup_{\beta\in N_r}\|\E(J_{w,n}(\beta))-J_{w,n}(\beta)\|\).
The first and third terms vanish as \(n\to\infty\) by
Assumption~\ref{ass:fe-weighted-groups}(ii) and Lemma~\ref{lem:fe-group-laws},
while equicontinuity of
\(\E(J_{w,n}(\cdot))\) makes the second term small when \(r>0\) is
chosen sufficiently small.
Hence \(\sup_{\beta\in N_r}\|DT_n(\beta)\|_{\rm op}\le1/2\) with
probability approaching one.
By the mean-value inequality on \(N_r\), every \(\beta\in N_r\) then
satisfies
\(\|T_n(\beta)-\beta_0\|_2
\le\|T_n(\beta)-T_n(\beta_0)\|_2+\|T_n(\beta_0)-\beta_0\|_2
\le r/2+\|J_w^{-1}\Psi_{w,n}(\beta_0)\|_2\le r\)
with probability approaching one, since
\(\Psi_{w,n}(\beta_0)=o_p(1)\).
On the intersection of these two high-probability events, \(T_n\) is a
contraction with modulus at most \(1/2\) that leaves the complete ball
\(N_r\) invariant. Banach's theorem
\citep[Theorem 4.4.10]{corbae_stinchcombe_zeman2009} therefore gives a
unique fixed point \(\widehat\beta_n\in N_r\), which solves
\(\Psi_{w,n}(\widehat\beta_n)=0\).
Combined with the event \(\Delta_n<ar\) from Step~(II),
\(\widehat\beta_n\) is the unique root in \(\Theta\) with probability
approaching one, and is consistent by Step~(II). This fixed-slope map is
an analytical device for sample-root existence and uniqueness, not a
global convergence claim for the computational algorithm.

\emph{(IV) Expansion and inference.}
For any consistent estimator \(\widehat\beta_n\), integrating the
derivative along the line segment from \(\beta_0\) to \(\widehat\beta_n\)
gives
\(\Psi_{w,n}(\widehat\beta_n)=\Psi_{w,n}(\beta_0)
-\widetilde J_{w,n}(\widehat\beta_n-\beta_0)\),
where
\(\widetilde J_{w,n}=\int_0^1
J_{w,n}(\beta_0+s(\widehat\beta_n-\beta_0))\,ds\).
Consistency places this segment inside \(N_r\) with probability
approaching one and makes it shrink to \(\beta_0\).
The uniform derivative law controls the sampling error along the segment,
and equicontinuity of the expected Jacobians controls their change from
\(\beta_0\). Together with \(\E(J_{w,n}(\beta_0))\to J_w\), this gives
\(\widetilde J_{w,n}\to_p J_w\).
When the residual satisfies \(\Psi_{w,n}(\widehat\beta_n)=o_p(n^{-1/2})\)
(which holds with zero residual at the exact root), multiplying by
\(\sqrt n\) and inverting \(\widetilde J_{w,n}\) yields
\begin{equation}
 \sqrt n(\widehat\beta_n-\beta_0)
 =J_w^{-1}\frac1{\sqrt n}\sum_g\psi_{g,w}(\beta_0)+o_p(1).
 \label{eq:fe-general-cancel-expansion}
\end{equation}
Specifically, the exact rearrangement equals
\(\widetilde J_{w,n}^{-1}\sqrt n\Psi_{w,n}(\beta_0)
-\widetilde J_{w,n}^{-1}\sqrt n\Psi_{w,n}(\widehat\beta_n)\); the second
term is \(o_p(1)\), and since \(\sqrt n\Psi_{w,n}(\beta_0)=O_p(1)\) by
its bounded covariance, replacing \(\widetilde J_{w,n}^{-1}\) by
\(J_w^{-1}\) alters the first term by only \(o_p(1)\).

Lemma~\ref{lem:fe-group-laws} verifies the Lindeberg condition for the
independent, possibly differently distributed group scores and gives
\(n^{-1/2}\sum_g\psi_{g,w}(\beta_0)\Rightarrow N(0,\Sigma_w)\).
Slutsky's theorem applied to \eqref{eq:fe-general-cancel-expansion}
therefore gives \eqref{eq:fe-general-cancel-normality}.

For the covariance estimate, define
\(M_n(\beta)=n^{-1}\sum_g\psi_{g,w}(\beta)\psi_{g,w}(\beta)'\)
and its deterministic expectation \(\overline M_n(\beta)=\E(M_n(\beta))\).
The uniform law in Lemma~\ref{lem:fe-group-laws} gives
\(M_n(\widehat\beta_n)-\overline M_n(\widehat\beta_n)=o_p(1)\).
Equicontinuity and consistency give
\(\overline M_n(\widehat\beta_n)-\overline M_n(\beta_0)=o_p(1)\),
and \(\overline M_n(\beta_0)\to\Sigma_w\) by the assumed covariance
limit. Thus \(\widehat\Sigma_w=M_n(\widehat\beta_n)\to_p\Sigma_w\).
The same argument for the Jacobian gives
\(\widehat J_w=J_{w,n}(\widehat\beta_n)\to_p J_w\).
By the continuous mapping theorem,
\(\widehat\Omega_w=\widehat J_w^{-1}\widehat\Sigma_w
(\widehat J_w^{-1})'\to_p\Omega_w\); dividing by \(n\) estimates the
sampling covariance of \(\widehat\beta_n\). We use the inverse transpose
because \(J_{w,n}(\widehat\beta_n)\) need not be symmetric away
from \(\beta_0\). Here \(n\) counts total observations rather than
groups; the corresponding \(\sqrt{G_n}\)-scaled limit has covariance
\(\Omega_w/c\) when \(n/G_n\to c\).
This adapts the standard M-estimation argument
\citep{newey_mcfadden94} to centered short-group moments.
\end{proof}

\subsection{Growing retained effects}
\label{app:fe-growing-general}
This subsection provides the  sufficient conditions used in
Theorem~\ref{thm:fe-growing-general}.\footnote{Sample-size subscripts on coefficients and matrices are suppressed.}
The total number of observations is denoted $n$ in both sampling
formulations below. The target dimension is fixed even when the
retained coefficient dimension grows. The proof uses a second-order
estimating-equation expansion, as in the nonlinear fixed-effects
bias literature \citep{hahnnewey2004bias,fernandez2016individual},
and the empirical correction of \citet{kosmidislunardon2024empirical}.
The calculations below, together with the supporting lemmas in
Appendix~\ref{app:fe-supplement}, verify its derivative, information and
feasible-correction requirements for the stated joint and cancelled
equations.

Fix \(\omega\neq0\), set \(Z_i=Y_i^\omega\), and impose the conditional mean
for this transformed outcome. All coefficients in this growing-dimension calculation refer to \(Z\).
Their substantive component is \(\beta=L\theta\), where \(\theta\) includes
the retained effects and \(L\) selects a fixed number of coefficients.
The corresponding power-mean coefficient is \(\beta_\omega=\beta/\omega\).
All expectations below are
conditional on the regressors and the true parameter.
We write \(b_i\) for the design vector denoted \(B_i\) in
Theorem~\ref{thm:fe-growing-general}. It includes the normalized effects
under joint estimation and is centered within the cancelled groups
under cancellation, as specified next.

For joint estimation, let \(b_i\) contain the regressors and normalized
fixed-effect indicators, write \(\mu_i(\theta)=\exp(b_i'\theta)\), and
impose \(\E(Z_i)=\mu_i(\theta_0)\), conditional on the full design and effects.
With \(w_i=\mu_i^2/V(\mu_i)\) and \(\lambda_i=\mu_i/V(\mu_i)\), the sample moment equation where fixed-effects are jointly estimated corresponds to 
\begin{equation}
 S_J(\theta)=\sum_i b_i\lambda_i(\theta)
                 (Z_i-\mu_i(\theta))=0.
 \label{eq:fe-growing-joint-score}
\end{equation}

For cancellation, partition observations into groups \(g\). First subtract
each group's ordinary regressor means, keeping all other normalized
effects in \(b_i\), so \(\sum_{i\in g}b_i=0\). The mean restriction is
\(\E(Z_i)=A_g\mu_i(\theta_0)\), with an unrestricted positive group scale
\(A_g\). Define \(W_g=\sum_{i\in g}w_i\),
\(\bar b_{g,w}=W_g^{-1}\sum_{i\in g}w_i b_i\), and
\(u_i=b_i-\bar b_{g,w}\). The sample moment when group fixed-effects are cancelled is
\begin{equation}
 S_C(\theta)=\sum_g\psi_g(\theta)=0,
 \qquad \psi_g(\theta)=\sum_{i\in g}\lambda_i(\theta)u_i(\theta)Z_i.
 \label{eq:fe-growing-cancel-score}
\end{equation}
The equality \(S_C(\theta)=0\) defines the estimating equation.
This group
score can equivalently be written as
\(\psi_g(\theta)=\sum_{i\in g}\lambda_i(\theta)u_i(\theta)
(Z_i-A_g\mu_i(\theta))\) since \(\lambda_i(\theta)\mu_i(\theta)=w_i(\theta)\) and \(\sum_{i\in g}w_i(\theta)u_i(\theta)=0\) by weighted centering. The unknown group scale therefore drops out of the sample equation.
At \(\theta_0\), the conditional mean restriction gives
\(\E(\psi_g(\theta_0))=0\), although the realized sample score
need not vanish there.

The weighted mean is computed separately within every group of the selected
block. It does not cancel all the other blocks. Both equations permit zero
outcomes when \(\omega>0\).

\subsubsection{Sampling and information conditions}

We distinguish two sampling formulations. In the \emph{cell formulation},
errors are independent or have a known dependency graph of bounded
degree, and \(\mathcal P\) is the set of ordered potentially dependent cell
pairs, including the diagonal. In the \emph{short-group formulation},
the \(G_n\) sampling groups are conditionally independent, their lengths
are uniformly bounded, and \(n=\sum_{g=1}^{G_n}n_g\), with
\(n/G_n\to c\in(0,\infty)\). Within-group dependence remains
unrestricted. Group scores are used
as the sampling contributions. The algebraic cancellation group and the
independent sampling unit need not be the same in the cell formulation.
Here a dependency graph means that collections of cells with no edge
between them are independent, conditional on the complete design and
effects.

Let \(d=d_n\) be the number of estimated coefficients after normalization and,
if applicable, cancellation. Define \(H=-\E(D S(\theta_0))\), where \(D=\nabla_\theta\) denotes
the coefficient derivative used in the main-text correction. Joint estimation gives
\(H=\sum_i w_i b_i b_i'\). Cancellation gives
\(H=\sum_g A_g B_g'C_gB_g\), where \(B_g\) has rows \(b_i'\) and
\(C_g=\diag(w_g)-w_gw_g'/W_g\). These are expected information matrices.

\begin{assumption}[Sufficient conditions for the general expansion]
\label{ass:fe-growing-regularity}
\textup{(i) Mean, moments and smoothness.}
The chosen conditional mean holds given the complete regressor array and
effects. Modeled outcomes \(Z_i\), regressors and true fitted
indices are bounded. In the cancelled model, \(A_g\) is bounded above and
away from zero. On a neighborhood of the fitted-mean range, \(V\) is positive
and four times continuously differentiable with bounded derivatives.
The applicable cell or short-group dependence condition above holds.

\textup{(ii) Information and retained dimension.}
After removing redundant columns, \(H\succ0\), \(d\le C\sqrt n\), and
\(\ell_n=\max_i\norm{H^{-1/2}b_i}^2\le Cd/n\).
In the short-group formulation this is equivalently required of
\(\max_g\norm{B_gH^{-1/2}}_{\rm op}^2\), up to the bounded group size.
The fixed target satisfies \(\norm{LH^{-1/2}}\le C/\sqrt n\) and
\(\max_i\norm{b_i'H^{-1}L'}\le C/n\).

\textup{(iii) Target variance.}
For \(K=\Var(S(\theta_0))\),
\(nLH^{-1}KH^{-1}L'\to\Omega\succ0\).
All constants in these conditions are independent of sample size.

\textup{(iv) Feasible cell correction after cancellation.}
Only in this case, require
\(\min_g n_g/\log(G_{C,n}+1)\to\infty\), where \(G_{C,n}\) is the
number of cancellation groups and \(n_g\) is their length.
\end{assumption}

In the short-group formulation, if
\(G_nLH^{-1}KH^{-1}L'\to\Omega_G\), then part (iii) uses
\(\Omega=c\Omega_G\), since \(n/G_n\to c\).
Thus \(\Omega\) is the covariance of the root-\(n\) limit,
not the root-\(G_n\) limit.

Part (ii) permits proportional two-way growth. It does not require a fixed
retained dimension. The target-row bound rules out concentration of the
identifying regressor variation in a few cells; Appendix~\ref{app:fe-growing-design}
derives it for the complete designs in Table~\ref{tab:fe-practical-choice}. Part (iv) is used to
construct centered cell residuals.
Corollary~\ref{cor:fe-growing-tails} replaces bounded outcomes by sufficient tail
conditions.

\subsubsection{Feasible bias correction and covariance estimation}
\label{app:fe-growing-corrections}

We first summarize the second-order expansion behind the coefficient bias
correction. Evaluate population quantities at \(\theta_0\), set
\(H=-\E(DS)\), and write \(\delta_1=H^{-1}S\) for the mean-zero
first-order displacement. For any vector \(v\in\mathbb R^d\), let
\(D^2S[v,v]\) denote the vector whose \(r\)th component is the quadratic
form \(v'D^2S_r v=\sum_{j,k}(\partial^2S_r/\partial\theta_j\partial\theta_k)v_jv_k\),
where \(D^2S_r=\partial^2S_r/\partial\theta\partial\theta'\) is the
Hessian matrix of the \(r\)th score equation \(S_r\).
For \(\Delta=\widehat\theta-\theta_0\), a second-order Taylor expansion
of \(S(\widehat\theta)=0\) around \(\theta_0\) gives
\(0=S+DS\,\Delta+\tfrac12D^2S[\Delta,\Delta]+\text{remainder}\).
Decomposing \(DS=-H+(DS+H)\), solving for \(\Delta\), and replacing
\(\Delta\) by its first-order approximation \(\delta_1\) in the
second-order terms yields
\(\Delta=\delta_1+H^{-1}(DS+H)\delta_1
+\tfrac12H^{-1}D^2S[\delta_1,\delta_1]+\text{remainder}\).
Replacing \(D^2S\) by \(\E(D^2S)\) and taking the expectation of the
target projection \(L\Delta\) gives the population bias
\eqref{eq:fe-growing-population-bias}.
The proof in Appendix~\ref{app:fe-growing-proof} bounds the centered
quadratic fluctuations and substitution remainders. The formulas below
estimate these expectations.\footnote{All formulas below use unnormalized score sums, with derivatives
evaluated at the computed root \(\widehat\theta\). Set
\(\widehat H=-DS(\widehat\theta)\). Because this observed Jacobian need not be
symmetric, inverse transposes are kept in the
covariance formulas.}

The cell correction formula is constructed as follows.
Write the score as \(S=\sum_i(f_i Z_i-a_i)\). For joint estimation,
\(f_i=\lambda_i b_i\), \(a_i=w_i b_i\), and the fitted full mean is
\(\widehat M_i=\widehat\mu_i\). For cancellation,
\(f_i=\lambda_i u_i\), \(a_i=0\), and the fitted full mean is
\(\widehat M_i=\widehat A_g\widehat\mu_i\), where
\(\widehat A_g=n_g^{-1}\sum_{i\in g}Z_i/\widehat\mu_i\).
Define the centered cell residual \(\widehat e_i=Z_i-\widehat M_i\), the
score covariance estimator
\(\widehat K=\sum_{(i,j)\in\mathcal P}
\widehat f_i\widehat f_j'\widehat e_i\widehat e_j\), and the coefficient
covariance estimator
\(\widehat{\mathcal V}=\widehat H^{-1}\widehat K\widehat H^{-T}\) where $H^{-T} = (H^{-1})'$.
The fixed-target bias estimate is
\begin{equation}
 \widehat b_n=L\widehat H^{-1}
 \left(\sum_{(i,j)\in\mathcal P}
 D\widehat f_i\widehat H^{-1}\widehat f_j\widehat e_i\widehat e_j
 +\tfrac12 D^2S(\widehat\theta):\widehat{\mathcal V}\right).
 \label{eq:fe-growing-cell-correction}
\end{equation}
Here \(D^2S(\widehat\theta):\widehat{\mathcal V}\) extends the quadratic
form \(D^2S(\widehat\theta)[v,v]=D^2S(\widehat\theta):(vv')\) from a
rank-one outer product \(vv'\) to the general \(d\times d\) covariance
matrix \(\widehat{\mathcal V}\): its \(r\)th component is
\(\bigl(D^2S(\widehat\theta):\widehat{\mathcal V}\bigr)_r
=\operatorname{tr}(D^2S_r(\widehat\theta)\widehat{\mathcal V})
=\sum_{j,k}(\partial^2S_r(\widehat\theta)/\partial\theta_j\partial\theta_k)
\widehat{\mathcal V}_{jk}\).
In the joint equation,
\(D^2S=\sum_i(Z_iD^2f_i-D^2a_i)\), where the fitted-mean derivative
\(-D^2a_i\) must be kept.
In the cancelled cell equation, individual cell scores are not centered on
their own, so the centered products \(\widehat e_i\widehat e_j\) cannot
be replaced by raw outcome products \(Z_iZ_j\).

The short-group correction formula is constructed similarly, except that
each group score \(\psi_g\) in \eqref{eq:fe-growing-cancel-score} is
already centered at the group level.
Set \(\widehat K=\sum_g\widehat\psi_g\widehat\psi_g'\) and
\(\widehat{\mathcal V}=\widehat H^{-1}\widehat K\widehat H^{-T}\).
Since \(D^2S(\widehat\theta)=\sum_g D^2\widehat\psi_g\), the main-text
correction \eqref{eq:fe-growing-group-correction} can be written in the
equivalent covariance form
\begin{equation}
 \widehat b_n=L\widehat H^{-1}
 \left(\sum_g D\widehat\psi_g\widehat H^{-1}\widehat\psi_g
 +\tfrac12\sum_g D^2\widehat\psi_g:\widehat{\mathcal V}\right).
 \label{eq:fe-growing-group-covariance-form}
\end{equation}
To see why \eqref{eq:fe-growing-group-covariance-form} coincides with
\eqref{eq:fe-growing-group-correction}, write
\(\widehat v_g=\widehat H^{-1}\widehat\psi_g\) so that
\(\widehat{\mathcal V}=\sum_g\widehat v_g\widehat v_g'\).
Because \(\operatorname{tr}(D^2S_r(\widehat\theta)\widehat v_g\widehat v_g')
=\widehat v_g'D^2S_r(\widehat\theta)\widehat v_g
=\bigl(D^2S(\widehat\theta)[\widehat v_g,\widehat v_g]\bigr)_r\),
contracting \(D^2S(\widehat\theta)=\sum_{g'}D^2\widehat\psi_{g'}\)
against \(\widehat{\mathcal V}\) gives
\(D^2S(\widehat\theta):\widehat{\mathcal V}
=\sum_g D^2S(\widehat\theta)[\widehat v_g,\widehat v_g]\),
where the directions \(\widehat v_g\) are held fixed during
differentiation.
Note that the curvature term differentiates the full score \(S\) rather
than only group \(g\)'s contribution, because sampling variation from
each group passes through the curvature of the entire system.
This group formulation estimates the aggregate bias without requiring
consistent estimates of each short group's scale \(A_g\) or within-group
covariance.

Equation~\eqref{eq:fe-growing-group-covariance-form} coincides with the
explicit reduced-bias M-estimator of
\citet{kosmidislunardon2024empirical} applied
to independent group contributions. In their notation,
\(j=\widehat H\), \(e=\widehat K\), and the Hessian of component
\(r\) is \(D^2S_r(\widehat\theta)\). Their adjustment vector is
\(A_{\rm KL}=-\sum_gD\widehat\psi_g\widehat H^{-1}
\widehat\psi_g-\tfrac12D^2S(\widehat\theta):\widehat{\mathcal V}\),
so their corrected parameter \(\widehat\theta+\widehat H^{-1}A_{\rm KL}\)
has target subvector \(L\widehat\theta-\widehat b_n\), with the inverse
transpose in \(\widehat{\mathcal V}\) retained whenever \(\widehat H\) is
nonsymmetric. While their general independent-vector framework allows
dependence within each group contribution, it assumes that the
unadjusted estimator converges to a root of the expected estimating
equation.
Theorem~\ref{thm:fe-growing-general} establishes that premise for our context, together
with feasibility of the bias correction and centered fixed-target
inference, for the cancelled equations with growing retained effects and
for the corresponding cell formulation (whereas joint estimation of
fixed-length group effects does not generally satisfy that premise).

Substantively, the first term inside the parentheses of
\eqref{eq:fe-growing-cell-correction} and
\eqref{eq:fe-growing-group-covariance-form} estimates the bias from
covariance between the sample Jacobian and the score, while the second
term captures the nonlinearity bias from passing coefficient uncertainty
through the curvature of the estimating equation. Both terms are used to recenter the
point estimate \(\widehat\beta^{\rm bc}=L\widehat\theta-\widehat b_n\) for valid inference.
The first-order covariance estimator for \(\widehat\beta^{\rm bc}\)
remains
\begin{equation}
 \widehat{\operatorname{Cov}}(\widehat\beta^{\rm bc})
 =L\widehat H^{-1}\widehat K\widehat H^{-T}L'.
 \label{eq:fe-growing-coefficient-covariance}
\end{equation}
Theorem~\ref{thm:fe-growing-general} shows that estimating and
subtracting \(\widehat b_n\) does not alter the first-order asymptotic
variance under its conditions, and that
\(nL\widehat{\mathcal V}L'\to_p\Omega\) using the applicable definition
of \(\widehat{\mathcal V}\) above.
Whereas the short-group formula requires no consistent estimate of an
individual group's scale, the cell formula after cancellation requires
uniform recovery of \(A_g\), which is why
Assumption~\ref{ass:fe-growing-regularity}(iv) is imposed only in that
case.

\subsubsection{Proof of Theorem~\ref{thm:fe-growing-general}}
\label{app:fe-growing-proof}
For the growing-dimension proof, all population quantities are evaluated
at the true retained coefficient. The definitions and sampling cases
are in Appendix~\ref{app:fe-growing-general} and estimators are
given in Appendix~\ref{app:fe-growing-corrections}.
Let \(e_i=Z_i-\E(Z_i)\). In the cell formulation,
\(S=\sum_i f_i e_i\), and
\(DS+H=\sum_i Df_i e_i\). For cancellation, the first identity follows
from \(\sum_{i\in g}w_i u_i=0\) at the truth; the second follows by
centering the derivative and using \(H=-\E(DS)\).
Let \(\mathcal T=\E(D^2S)\),
\(\mathcal V=H^{-1}KH^{-1}\), and define the bias term 
\begin{equation}
 b_n=LH^{-1}
 \left(\sum_{(i,j)\in\mathcal P}
 Df_iH^{-1}f_j\E(e_ie_j)+\tfrac12\mathcal T:\mathcal V\right).
 \label{eq:fe-growing-population-bias}
\end{equation}
For independent short groups, replace the first sum inside parentheses by
\(\sum_g\E(D\psi_g H^{-1}\psi_g)\), and use
\(K=\sum_g\E(\psi_g\psi_g')\). These are the expectations of the same
two Taylor terms, expressed at the appropriate sampling level.

\begin{samepage}
\noindent The following proof proceeds in five steps.
The local root's fitted-index rate and target derivative bounds reduce
the target error to a mean-zero linear score term plus quadratic bias,
with a remainder negligible at the \(n^{-1/2}\) scale.
Estimating that bias with \(o_p(n^{-1/2})\) error then
permits subtraction without changing first-order variance, so the score
central limit theorem yields centered inference.
\begin{enumerate}[label=\textup{(\Roman*)},leftmargin=*,itemsep=1pt,topsep=3pt,parsep=0pt]
\item \emph{Control the normalized equation.}
Normalize by the expected information so the expected derivative is
\(-I\). Concentration bounds then make both the derivative perturbation
and the maximum fitted-index displacement small, which permits a local
contraction even though the information-norm displacement can grow.
\item \emph{Construct and localize a root.}
Construct an invariant ball around that displacement to obtain a
unique local root and a vanishing maximum fitted-index error, which
permits a controlled Taylor expansion along the segment to the truth.
\item \emph{Isolate the quadratic bias.}
Project the second-order expansion on the fixed target and replace
the displacement in its quadratic terms by its first-order approximation;
negligible substitution errors and centered fluctuations leave the
expectations of these terms as the bias \(b_n=O(d/n)\), yielding the
expansion and target consistency.
\item \emph{Estimate covariance and bias.}
Use the root's uniform fitted-index rate in the feasible-correction
lemma to obtain normalized covariance consistency and a bias estimate
with error \(o_p(n^{-1/2})\), allowing its feasible subtraction from the
expansion.
\item \emph{Obtain centered inference.}
Apply a central limit theorem to the remaining linear score term
to obtain centered inference with the estimated covariance, and omit the
correction when \(b_n=o(n^{-1/2})\).
\end{enumerate}
\end{samepage}

\begin{proof}[Proof of Theorem~\ref{thm:fe-growing-general}]
Lemmas~\ref{lem:fe-growing-normalized}--\ref{lem:fe-growing-feasible}
derive the concentration, quadratic fluctuation and substitution bounds
used below from Assumption~\ref{ass:fe-growing-regularity}.

\emph{(I) Control the normalized equation.}
Define \(p_i=H^{-1/2}b_i\) and
\(q_n=\sqrt{\ell_n\log n}+\ell_n\log n=o(1)\).
Lemma~\ref{lem:fe-growing-normalized} gives
\(\sum_i p_ip_i'\preceq CI\), the normalized derivative bound
and the first-order displacement rates in both sampling formulations.

In particular, for \(\delta_1=H^{-1}S\), that lemma gives
\(\|H^{1/2}\delta_1\|=O_p(\sqrt d)\),
\(\max_i|b_i'\delta_1|=O_p(q_n)\), and
\(\norm{H^{-1/2}(DS+H)H^{-1/2}}=O_p(q_n)\).
Thus the full displacement need not vanish in information norm,
but every fitted index does. On a fixed sufficiently small index neighborhood,
changing all fitted indices from their true values by at most \(r\)
changes the normalized derivative by at most \(Cr\).

\emph{(II) Construct and localize a root.}
Define
\(\Phi(z)=z+H^{-1/2}S(\theta_0+H^{-1/2}z)\) and
\(z_1=H^{1/2}\delta_1\). Its derivative is
\(I+H^{-1/2}DS(\theta_0+H^{-1/2}z)H^{-1/2}\).
Since \(S(\theta_0)=H\delta_1\), integration along the segment to
\(\theta_0+\delta_1\) gives
\(S(\theta_0+\delta_1)=(DS(\theta_0)+H)\delta_1
+\int_0^1(DS(\theta_0+t\delta_1)-DS(\theta_0))\,dt\,\delta_1\).
The normalized derivative and index bounds above therefore imply
\(\|\Phi(z_1)-z_1\|=O_p(q_n\sqrt d)\).

For a fixed constant \(K_0\), consider the ball with center \(z_1\)
and radius \(R_n=K_0q_n\sqrt d\). Every point in it has additional
maximum index displacement at most
\(\sqrt{\ell_n}R_n=K_0q_n\sqrt{d\ell_n}=O(q_n)\), because
\(d\ell_n\le C\). Thus \(\sup\|D\Phi\|\le1/2\) on this ball with
probability approaching one. Given any \(\epsilon>0\), choose \(K_0\)
large enough that \(\|\Phi(z_1)-z_1\|\le R_n/2\) with probability at
least \(1-\epsilon\) for all large \(n\). On the intersection,
\(\|\Phi(z)-z_1\|\le\|\Phi(z)-\Phi(z_1)\|+\|\Phi(z_1)-z_1\|
\le R_n/2+R_n/2\).
The closed ball is complete, invariant and the map is a contraction.
Banach's theorem \citep[Theorem 4.4.10]{corbae_stinchcombe_zeman2009}
therefore gives a unique fixed point in that ball, equivalently a
root of the sample equation. Since \(\epsilon\)
is arbitrary, a regular local root exists with probability approaching
one and
\(\norm{H^{1/2}(\widehat\theta-\theta_0)-z_1}=O_p(q_n\sqrt d)\),
with \(\max_i|b_i'(\widehat\theta-\theta_0)|=O_p(q_n)\).
For a sufficiently small fixed \(r>0\), the fitted-index neighborhood
\(\{\theta:\max_i|b_i'(\theta-\theta_0)|\leq r\}\) is convex,
and its normalized Jacobian is uniformly within \(1/2\) of \(-I\)
with probability approaching one. For two roots in this convex
neighborhood, integration of the
normalized Jacobian along their connecting segment gives
\(0=\bar M H^{1/2}(\theta'-\theta)\), where
\(\bar M=\int_0^1H^{-1/2}DS(\theta+t(\theta'-\theta))H^{-1/2}\,dt\).
The bound \(\|\bar M+I\|\le1/2\) makes \(\bar M\) invertible,
so the roots coincide.

\emph{(III) Isolate the quadratic bias.}
Define \(\Delta=\widehat\theta-\theta_0\).
Taylor expansion on the localized segment gives
\(0=S+DS\,\Delta+\tfrac12D^2S[\Delta,\Delta]+R_3\), where
\(R_3=\tfrac12\int_0^1(1-t)^2
D^3S(\theta_0+t\Delta)[\Delta,\Delta,\Delta]\,dt\).
After writing \(DS=-H+(DS+H)\) and
\(D^2S=\mathcal T+(D^2S-\mathcal T)\), the target error is the sum of
\(LH^{-1}S\), \(LH^{-1}(DS+H)\Delta\),
\(\tfrac12LH^{-1}\mathcal T[\Delta,\Delta]\),
\(\tfrac12LH^{-1}(D^2S-\mathcal T)[\Delta,\Delta]\), and
\(LH^{-1}R_3\). We now justify replacing \(\Delta\) by \(\delta_1\)
in the two retained quadratic terms.

For target coordinate \(k\), let \(e_k\) denote its unit vector and define
\(u=H^{-1}L'e_k\). The target-row condition gives
\(\max_i|b_i'u|=O(n^{-1})\). The normalized projected mean Hessian
\(Q=H^{-1/2}(u'\mathcal T)H^{-1/2}\) has operator norm
\(O(n^{-1})\) and Frobenius norm \(O(\sqrt d/n)\).
Its centered sample counterpart
\(H^{-1/2}\{u'(D^2S-\mathcal T)\}H^{-1/2}\)
has norm \(O_p(q_n/n)\). The normalized random derivative row
\(u'(DS+H)H^{-1/2}\) has norm \(O_p(\sqrt d/n)\).
These are the target-specific conclusions
of Lemma~\ref{lem:fe-growing-normalized}, including growing
cancellation groups.

The derivative replacement is bounded by
\(\|LH^{-1}(DS+H)H^{-1/2}\|\,
\|H^{1/2}(\Delta-\delta_1)\|=O_p(dq_n/n)\).
This includes the iterated derivative and derivative--curvature terms
obtained by further substitution of the root equation.

For the other replacements, define \(z=H^{1/2}\Delta\).
Since \(Q\) is symmetric,
\(z'Qz-z_1'Qz_1=(z-z_1)'Q(z+z_1)\).
The expected-Hessian change is therefore at most
\((C/n)(\|z\|+\|z_1\|)\|z-z_1\|=O_p(dq_n/n)\).
The random Hessian contributes at most
\(O_p(q_n/n)\|z\|^2=O_p(dq_n/n)\).
The third-derivative bound in
Lemma~\ref{lem:fe-growing-normalized} makes the cubic remainder
at most \((C/n)\max_i|b_i'\Delta|\|z\|^2=O_p(dq_n/n)\).
Since \(d/\sqrt n\) is bounded and \(q_n=o(1)\), these
errors are all \(o_p(n^{-1/2})\).
It remains to identify the retained quadratic means.
The derivative term is
\(LH^{-1}\sum_{i,j}Df_iH^{-1}f_j e_ie_j\).
Its expectation keeps only pairs in \(\mathcal P\), because
\(\E(e_ie_j)=0\) outside that set. The curvature term has expectation
\(\tfrac12LH^{-1}\mathcal T:\mathcal V\), since
\(\E(\delta_1\delta_1')=\mathcal V\).
For independent short groups, off-diagonal terms in
\(\sum_{g,h}\E((D\psi_g-\E(D\psi_g))H^{-1}\psi_h)\)
vanish by independence and \(\E(\psi_h)=0\); the diagonal terms are
\(\E(D\psi_gH^{-1}\psi_g)\).
Thus the two means are exactly \eqref{eq:fe-growing-population-bias}.
Lemma~\ref{lem:fe-growing-quadratic} bounds these means by \(O(d/n)\)
and their centered fluctuations by
\(O_p(\sqrt d/n)=o_p(n^{-1/2})\).
Together with the preceding remainder bounds, this proves
\eqref{eq:fe-growing-general-expansion}, with \(b_n=O(d/n)\),
without assuming either the nuisance rate or a negligible remainder.
In particular the target is consistent, since its linear variance is
\(O(n^{-1})\), \(d/n\to0\), and the remainder vanishes.

\emph{(IV) Estimate covariance and bias.}
The local solution's uniform index rate satisfies the hypothesis of
Lemma~\ref{lem:fe-growing-feasible}. With the additional scale-recovery
condition only for cancelled cell inference, that lemma proves
normalized covariance consistency and
\(\widehat b_n-b_n=o_p(n^{-1/2})\).
It first centers an oracle calculation with deterministic population
covariance. The pathwise tensor bound
\eqref{eq:fe-growing-tensor-bound} then permits substitution of
the empirical covariance even though it uses the same observations
as the empirical curvature. No independence of fitted contributions
or consistency of an individual short-group covariance is invoked.
For the covariance conclusion, write
\(E_n=H^{1/2}(\widehat{\mathcal V}-\mathcal V)H^{1/2}\).
The lemma gives \(\|E_n\|=o_p(1)\), and therefore
\(n\|L(\widehat{\mathcal V}-\mathcal V)L'\|
\le n\|LH^{-1/2}\|^2\|E_n\|=o_p(1)\).
This explains why a fixed-target covariance is consistently estimated
although the full coefficient dimension increases.

\emph{(V) Obtain centered inference.}
Fix a nonzero target vector \(a\).
The scalar linear term is
\(\sqrt n\,a'LH^{-1}S=\sum_i c_i e_i\), where
\(c_i=\sqrt n\,a'LH^{-1}f_i\).
The target-row bound and bounded working weights imply
\(|c_i|\le C/\sqrt n\); after cancellation, the weighted group
mean in \(f_i=\lambda_i u_i\) preserves this bound.
For independent short groups, the scaled whole-group contributions
are also bounded by \(C/\sqrt n\), so the Lindeberg condition holds.
For bounded-degree dependence, a joint cumulant vanishes when its
indices split into independent collections. Only \(O(n)\) connected
\(r\)-tuples remain, and bounded errors give cumulants of order
\(O(n^{1-r/2})\to0\) for every fixed \(r\ge3\).
The variance tends to \(a'\Omega a>0\), so the scalar central limit
theorem, followed by Cram\'er--Wold, gives the vector normal limit.
For the dependency-graph argument, the bounded summands also bound
each fixed moment of the normalized sum. The vanishing higher
cumulants and convergent second cumulant consequently give the
Gaussian moments, which determine the Gaussian distribution.

Subtracting the feasible correction from the expansion yields
\(\sqrt n(\widehat\beta_n-\widehat b_n-\beta_0)
=\sqrt n\,LH^{-1}S+\sqrt n(b_n-\widehat b_n)+o_p(1)\).
The second term is \(o_p(1)\), which proves
\eqref{eq:fe-growing-corrected-limit}. The feasible lemma also gives
\(nL\widehat{\mathcal V}L'\to_p\Omega\).
If \(d_n=o(\sqrt n)\), then \(b_n=O(d_n/n)=o(n^{-1/2})\).
Under this condition, or directly when \(b_n=o(n^{-1/2})\), the same
expansion and central limit theorem give the stated centered limit
without correction.
\end{proof}

\section{Instrumental variables: constructions, results and proofs}
\label{app:iv-pilot-proof}

This appendix provides the definitions, algorithm and formal results
described in Section~\ref{sec:extension-iv}, then proves
Propositions~\ref{prop:iv-i2sls-two-phase}
and~\ref{prop:iv-pilot-initialization}.
The first proof establishes global convergence of the ordinary I2SLS
pilot under a sufficient weighted first-stage condition, then uses its
approximation to initialize exact completion. The second treats the
local GMM pilot under overidentification.
Appendix~\ref{app:iv} explains the condition and the role of
\(\lambda\), provides regression identities and examples, and states
the statistical conditions and further extensions.

\subsection{Definitions and results}
\label{app:iv-statements}

\subsubsection{Additive and multiplicative instrument restrictions}

Endogeneity requires
an instrument restriction as well as a mean specification.
\citet{mullahy1997instrumental} develops
instrumental-variable estimation for a multiplicative disturbance,
whereas \citet{windmeijersantossilva1997} studies additive-error
moments.

Write \(\mu_i(\beta)=\exp(X_i'\beta)\) and
\(U_i(\beta)=Y_i/\mu_i(\beta)\), with \(X_i\in\R^K\).
Let \(Z_i\in\R^L\) be instruments, with \(L\geq K\). With exogenous regressors,
the exponential conditional mean restriction centers both relative and levels
errors. With endogenous regressors, the multiplicative IV model
instead specifies
\begin{equation}
 \E(Z_i(U_i(\beta_0)-1))=0,
 \label{eq:iv-moment}
\end{equation}
as implied by \(\E(U_i(\beta_0)\mid Z_i)=1\). However, the additive alternative is
\(\E(Z_i(Y_i-\mu_i(\beta_0)))=0\), implied by
\(\E(\mu_i(\beta_0)(U_i(\beta_0)-1)\mid Z_i)=0\).
A valid instrument for one need not be valid for the other. Changing
between the corresponding relative-error and levels-error moments
can therefore change validity, not merely precision.

The multiplicative restriction is natural when unobserved productivity,
demand or other determinants scale the outcome.
For a structural equation \(Y(x,u)=\exp(x'\beta_0)u\), changing
a regressor while holding \(u\) fixed has a common proportional
effect wherever the baseline outcome is positive. Its coefficient
therefore preserves the familiar semielasticity or proportional interpretation. This
motivates the relative-error moments used here.\footnote{
As an alternative, the control-function approach uses additional restrictions
under which a first-stage residual or other control restores the
conditional mean \citep{imbenswooldridge2007control,wooldridge_cf}.}

This distinction also governs variance weighting. Multiplying
\eqref{eq:iv-moment} by an endogenous fitted-mean weight generally
changes its validity. The exogenous power-score family is therefore
not automatically a family of valid IV estimators of the same
parameter. Similarly, a model for \(U\) does not by itself imply
an instrument restriction for \(U^\omega\). Additional power orders
or weights require their own justified moments.

\subsubsection{Iterated 2SLS under exact identification}

The two-phase construction of Section~\ref{sec:iols} extends to
endogenous regressors by replacing OLS with ordinary 2SLS on unchanged
regressors and instruments, giving iterated two-stage least squares
(I2SLS). Define \(r_i(\beta)=U_i(\beta)-1\),
\(m_n(\beta)=Z'r(\beta)/n\), \(D_n=Z'X/n\), and
\(G_n(\beta)=Z'\diag(U(\beta))X/n=-\partial m_n(\beta)/\partial\beta'\).
Throughout this construction \(L=K\), \(D_n\) is nonsingular, and
the observations and coefficients are finite.

For \(0<\lambda<1\), define the raw logarithmic correction
\(\ell_\lambda(u)=\log(\lambda+(1-\lambda)u)\) and its normalized
version \(r_\lambda(u)=\ell_\lambda(u)/(1-\lambda)\).
The pilot equation is
\(m_{\lambda,n}(\beta)=Z'r_\lambda(U(\beta))/n=0\).
It approaches \(m_n(\beta)=0\) as \(\lambda\uparrow1\).
The normalization is useful for comparing equations and checking
accuracy, but the working regression uses the raw correction:
\begin{equation}
 T_{\lambda,n}(\beta)
 =\beta+(1-\lambda)D_n^{-1}m_{\lambda,n}(\beta)
 =\beta+D_n^{-1}Z'\ell_\lambda(U(\beta))/n.
 \label{eq:iv-raw-pilot-map}
\end{equation}
This is exactly 2SLS of \(X\beta+\ell_\lambda(U(\beta))\) on \(X\)
using \(Z\). There is no separately chosen pilot step size.
Zeros are allowed because \(\ell_\lambda(0)=\log(\lambda)\) is finite.

\begin{algorithm}[Exactly identified I2SLS]
\label{alg:iv-i2sls}
Choose a finite starting vector, a finite increasing grid
\(0<\lambda_1<\cdots<\lambda_J<1\), pilot tolerances
\(\varepsilon_{1,j}>0\), an exact-completion step
\(a_{\mathrm{step}}>0\), and a final tolerance \(\varepsilon_n>0\).
A single pilot value (\(J=1\)) is allowed.
\textit{Phase 1 (pilot).} At each \(\lambda_j\), starting from the
preceding output, repeatedly regress
\(X\beta+\ell_{\lambda_j}(U(\beta))\) on \(X\) by ordinary 2SLS
with instruments \(Z\), until
\(\norm{m_{\lambda_j,n}(\beta)}_2\leq\varepsilon_{1,j}\).
\textit{Phase 2 (exact completion).} From the final pilot output,
repeatedly regress \(X\beta+a_{\mathrm{step}}r(\beta)\) on \(X\)
by 2SLS using \(Z\), until
\(\norm{m_n(\beta)}_2\leq\varepsilon_n\).
\end{algorithm}

Both phases reuse the same linear system. Phase 2 has map
\(T_{{\rm IV},n}(\beta)=\beta+a_{\mathrm{step}}D_n^{-1}m_n(\beta)\)
and its fixed points solve the original IV equation.
The pilot tolerance uses the normalized moment: a small raw correction
near \(\lambda=1\) can reflect the factor \(1-\lambda\), rather
than proximity to a root.

To state sufficient conditions for convergence from any finite start,
separate the first-stage fitted regressors from their residuals.
Let \(P_Z=Z(Z'Z)^{-1}Z'\), \(\widehat X=P_ZX\),
\(E_n=X-\widehat X\), and \(Q_{Z,n}=\widehat X'\widehat X/n\succ0\).
Define \(\mathsf C_n=\widehat XQ_{Z,n}^{-1/2}\) and
\(\mathsf R_n=E_nQ_{Z,n}^{-1/2}\).
Thus \(\mathsf C_n'\mathsf C_n/n=I\) and
\(\mathsf C_n'\mathsf R_n=0\).
Let \(W_{\lambda,n}(\beta)\) be diagonal with entries
\(w_{\lambda i}(\beta)=(1-\lambda)U_i(\beta)/
(\lambda+(1-\lambda)U_i(\beta))\), and define
\(\mathsf A_{\lambda,n}=\mathsf C_n'W_{\lambda,n}\mathsf C_n/n\),
\(\mathsf B_{\lambda,n}=\mathsf C_n'W_{\lambda,n}\mathsf R_n/n\),
and \(\underline a_{\lambda,n}
=\lambda_{\min}(\mathsf A_{\lambda,n})\).
These matrices depend on \(\beta\); \(\lambda_{\min}\) denotes the
smallest eigenvalue of a symmetric matrix.

\begin{assumption}[Weighted first-stage condition]
\label{ass:iv-global-pilot}
For the fixed sample and specified \(\lambda\in(0,1)\),
\textup{(i)} the positive-outcome fitted regressors span all
coefficient directions:
\(\mathsf C_n'D_Y\mathsf C_n/n\succ0\), where
\(D_Y=\diag(\ind(Y_i>0))\);
\textup{(ii)} there is a constant \(\kappa\in[0,1)\) such that
\begin{equation}
 \norm{\mathsf B_{\lambda,n}(\beta)}_2
 \leq\kappa\,\underline a_{\lambda,n}(\beta)
 \qquad\text{for every }\beta\in\R^K.
 \label{eq:iv-global-alignment}
\end{equation}
Here \(\norm{\cdot}_2\) is the matrix operator norm.
\end{assumption}

This assumption is needed to guarantee global numerical convergence because of the nonlinearities. The first-stage residuals are orthogonal to the fitted regressors
before weighting. Yet, the pilot weights depend on the residual
ratios which may remove that orthogonality.
Condition~\eqref{eq:iv-global-alignment} bounds this weighted
cross-product relative to the smallest information eigenvalue from the
fitted regressors. It prevents the residual component of the first
stage from overturning the contraction provided by its fitted
component. This is a sufficient numerical condition, distinct from
instrument validity and relevance.

\begin{proposition}[Global pilot convergence and exact completion]
\label{prop:iv-i2sls-two-phase}
Fix a sample with \(L=K\), nonsingular \(D_n\), and nonnegative outcomes.

\textup{(i) Global pilot.}
If Assumption~\ref{ass:iv-global-pilot} holds at a specified
\(\lambda\in(0,1)\) and \(m_{\lambda,n}\) has a finite unique root
\(\widehat\beta_{\lambda,n}\).
Iteration of \eqref{eq:iv-raw-pilot-map} converges to it from every
finite starting vector, geometrically in the \(Q_{Z,n}\)-norm.
The contraction factor may depend on the sample, \(\lambda\), and
starting vector.

\textup{(ii) Two-phase convergence.}
Suppose \(m_n\) has a finite root \(\widehat\beta_n\), and
Assumption~\ref{ass:iv-global-pilot} holds with a common
\(\kappa<1\) for all \(\lambda\) sufficiently close to one.
Then \(D_n^{-1}G_n(\widehat\beta_n)\) is positive stable.
For any fixed
\(0<a_{\mathrm{step}}<\min_z2\operatorname{Re}(z)/|z|^2\),
where \(z\) ranges over its eigenvalues, there are a norm
\(\norm{\cdot}_*\) and sample-dependent constants \(C_n<\infty\),
\(r_n>0\), and \(\lambda_{0,n}<1\) such that
\(\norm{\widehat\beta_{\lambda,n}-\widehat\beta_n}_*
\leq C_n(1-\lambda)\) for \(\lambda_{0,n}<\lambda<1\).
Each of these pilot roots exists and is globally reachable by part
\textup{(i)}. If its computed approximation \(b\) satisfies
\(\norm{b-\widehat\beta_{\lambda,n}}_*\leq\eta\) and
\(C_n(1-\lambda)+\eta<r_n\), Phase 2 converges geometrically to
\(\widehat\beta_n\). Consequently, a sufficiently 
accurate pilot initializes exact completion from any finite original
starting vector.
\end{proposition}

With strictly positive outcomes,
\(\lambda=0\) gives ordinary 2SLS on \(\log(Y)\) in one step.
For small positive \(\lambda\), the pilot approaches this map on
every fixed compact coefficient set. With zeros, that endpoint is
undefined and positive-outcome variation remains essential.
Near \(\lambda=1\), the normalized equation approaches the exact
IV equation, while the raw pilot step becomes small.
There is no general monotone ordering of the global condition in
\(\lambda\).

Current-Jacobian 2SLS is an alternative for exact completion: using
\(\diag(U(\beta))X\) as regressors gives increment
\(a_{\mathrm{step}}G_n(\beta)^{-1}m_n(\beta)\).
With unit step this is Newton's method, locally quadratically
convergent at a regular root.
For either completion, Assumption~\ref{ass:iv-statistics} and
\(\varepsilon_n=o(n^{-1/2})\) give consistency and the expansion
\eqref{eq:iv-expansion} when the output lies in the identified
parameter set with probability approaching one.

\subsubsection{Overidentification and nonlinear GMM}

With \(L>K\), we consider the nonlinear GMM criterion
\begin{equation}
 J_n(\beta)=\tfrac12m_n(\beta)'W_nm_n(\beta),
 \label{eq:iv-gmm-criterion}
\end{equation}
where \(W_n\succ0\) is held fixed during minimization.
When \(G_n(\beta)\) has full column rank, linearizing the
moments gives the correction
\begin{equation}
 \beta^{\rm new}
 =\beta+a_{\mathrm{step}}
 (G_n(\beta)'W_nG_n(\beta))^{-1}G_n(\beta)'W_nm_n(\beta).
 \label{eq:iv-main-linear-gmm}
\end{equation}
It is linear GMM on the current regressors \(\diag(U(\beta))X\),
and a 2SLS regression when \(W_n=(Z'Z/n)^{-1}\). The pilot criterion is
\(J_{\lambda,n}=m_{\lambda,n}'W_nm_{\lambda,n}/2\). Its
current-Jacobian linearization uses
\(G_{\lambda,n}=-\partial m_{\lambda,n}/\partial\beta'\).

\begin{proposition}[A local shifted-log GMM pilot]
\label{prop:iv-pilot-initialization}
Fix a sample with \(L\geq K\) and \(W_n\succ0\). Let
\(\widehat\beta_n\) be a finite stationary point of \(J_n\) with
positive-definite Hessian. For \(\lambda\) sufficiently close to
one from below, \(J_{\lambda,n}\) has a locally unique stationary
point \(\widehat\beta_{\lambda,n}^{(1)}\), also a strict local
minimum, and a finite constant \(C_n\) gives
\(\norm{\widehat\beta_{\lambda,n}^{(1)}-\widehat\beta_n}
\leq C_n(1-\lambda)\).
Suppose a specified exact solver converges to \(\widehat\beta_n\)
from every start within radius \(r_n>0\). A computed pilot with
error at most \(\eta\) initializes that solver whenever
\(C_n(1-\lambda)+\eta<r_n\).
\end{proposition}

Thus a sufficiently accurate pilot on the regular branch can initialize
any exact GMM solver with a known local basin, such as the
current-Jacobian step under the local curvature and step-size conditions
in Appendix~\ref{app:iv-numerics}. Under exact identification, a
regular moment root satisfies the Hessian condition; when
Proposition~\ref{prop:iv-i2sls-two-phase} also applies, the pilot
branches coincide and that proposition provides global pilot
initialization under its additional condition. The overidentified
GMM result remains local and does not establish global minimization.%

\subsection{Main proofs}
\label{app:iv-main-proofs}

\subsubsection{Global pilot convergence and exact completion}
\label{app:iv-two-phase-proof}
\begin{samepage}
\noindent The following proof proceeds in five steps.
\begin{enumerate}[label=\textup{(\Roman*)},leftmargin=*,itemsep=1pt,topsep=3pt,parsep=0pt]
\item \emph{Derivative bound.} Decompose the pilot derivative into the
fitted-regressor information and the weighted first-stage residual
term. Assumption~\ref{ass:iv-global-pilot} makes its norm smaller than
one at every finite coefficient vector.
\item \emph{Global pilot convergence.} On a closed ball containing the
starting vector and centered at a finite pilot root, continuity makes
that bound uniform. The root makes the ball invariant, which proves
convergence from the arbitrary start and uniqueness of the root.
\item \emph{Exact Jacobian.} Let \(\lambda\uparrow1\) in the weighted
condition after removing its common vanishing scale. The limiting
bound gives positive stability of the exact preconditioned Jacobian,
which makes the assumed finite exact root regular.
\item \emph{Pilot existence and approximation.} Smoothly extend the
normalized moments to \(\lambda=1\). The nonsingular Jacobian from
Step~(III) gives nearby pilot roots and an order \(1-\lambda\) distance bound,
so Step~(II) now applies without separately assuming pilot existence.
\item \emph{Exact completion.} The admissible step makes the exact map
contract near its root. The distance bound from Step~(IV), plus the
pilot's numerical error, places its output in that neighborhood.
\end{enumerate}
\end{samepage}

\begin{proof}[Proof of Proposition~\ref{prop:iv-i2sls-two-phase}]
\emph{(I) Derivative bound.}
Exact identification gives
\(D_n^{-1}Z'/n=Q_{Z,n}^{-1}\widehat X'/n\).
Since
\(\partial\ell_\lambda(U_i(\beta))/\partial\beta'
=-w_{\lambda i}(\beta)X_i'\), differentiation of
\eqref{eq:iv-raw-pilot-map} yields
\begin{equation}
 Q_{Z,n}^{1/2}DT_{\lambda,n}(\beta)Q_{Z,n}^{-1/2}
 =I-\mathsf A_{\lambda,n}(\beta)-\mathsf B_{\lambda,n}(\beta).
 \label{eq:iv-pilot-derivative-decomposition}
\end{equation}
Here \(X=\widehat X+E_n\) gives the two matrices on the right.
Every weight lies in \([0,1)\) and is strictly positive when
\(Y_i>0\). Therefore
\(0\prec\mathsf A_{\lambda,n}(\beta)\preceq I\):
positive definiteness follows from the positive-outcome rank condition
for every finite \(\beta\). The induced matrix norm in
\(\norm{v}_{Q_{Z,n}}=(v'Q_{Z,n}v)^{1/2}\) is the Euclidean
operator norm after this conjugation. Because
\(\mathsf A_{\lambda,n}\) is symmetric,
\(\norm{I-\mathsf A_{\lambda,n}}_2=1-\underline a_{\lambda,n}\).
The triangle inequality and \eqref{eq:iv-global-alignment} give
\begin{equation}
 \norm{DT_{\lambda,n}(\beta)}_{Q_{Z,n}}
 \leq 1-(1-\kappa)\underline a_{\lambda,n}(\beta)<1.
 \label{eq:iv-pilot-contraction-bound}
\end{equation}
No symmetry of the IV Jacobian or of \(\mathsf B_{\lambda,n}\) is used.

\emph{(II) Global pilot convergence.}
For part \textup{(i)}, let \(\widehat\beta_{\lambda,n}\) be the
assumed finite root and choose any finite starting vector \(b\).
Choose \(R>0\) such that
\(\norm{b-\widehat\beta_{\lambda,n}}_{Q_{Z,n}}\leq R\), and define
the closed ball
\(\mathcal B_R=\{\beta:
\norm{\beta-\widehat\beta_{\lambda,n}}_{Q_{Z,n}}\leq R\}\).
Continuity of the smallest eigenvalue and compactness give
\(a_R=\min_{\beta\in\mathcal B_R}
\underline a_{\lambda,n}(\beta)>0\).
Thus \eqref{eq:iv-pilot-contraction-bound} is at most
\(q_R=1-(1-\kappa)a_R<1\) throughout the ball.

For \(\beta,\vartheta\in\mathcal B_R\), their segment lies in the
ball. Integrating the derivative on this segment gives
\(\norm{T_{\lambda,n}(\beta)-T_{\lambda,n}(\vartheta)}_{Q_{Z,n}}
\leq q_R\norm{\beta-\vartheta}_{Q_{Z,n}}\).
Taking \(\vartheta=\widehat\beta_{\lambda,n}\), a fixed point,
shows that the image of every point remains within distance
\(q_RR\leq R\) of the root. The ball is therefore invariant.
It is complete, so the contraction theorem of
\citet[Theorem~4.4.10]{corbae_stinchcombe_zeman2009} gives convergence
from every point of the ball, with
\(\norm{\beta^{(t)}-\widehat\beta_{\lambda,n}}_{Q_{Z,n}}
\leq q_R^t\norm{b-\widehat\beta_{\lambda,n}}_{Q_{Z,n}}\).
Any second finite root would belong to a larger ball of the same
kind, where uniqueness of the fixed point excludes it.
This proves part \textup{(i)}.

\emph{(III) Exact Jacobian.}
For part \textup{(ii)}, write \(a=1-\lambda\).
At the exact root, \(W_{\lambda,n}/a\to
\diag(U(\widehat\beta_n))\).
Define
\(\mathsf A_{*,n}=\mathsf C_n'\diag(U(\widehat\beta_n))\mathsf C_n/n\)
and
\(\mathsf B_{*,n}=\mathsf C_n'\diag(U(\widehat\beta_n))\mathsf R_n/n\).
The positive-outcome rank condition gives
\(a_*=\lambda_{\min}(\mathsf A_{*,n})>0\).
Dividing \eqref{eq:iv-global-alignment} by \(a\) and taking limits
gives \(\norm{\mathsf B_{*,n}}_2\leq\kappa a_*\).
The matrix
\(\mathsf M_n=\mathsf A_{*,n}+\mathsf B_{*,n}\) is similar to
\(D_n^{-1}G_n(\widehat\beta_n)\).
For every real vector \(v\),
\(v'(\mathsf M_n+\mathsf M_n')v/2
\geq (1-\kappa)a_*\norm{v}_2^2\).
Consequently the symmetric part of \(\mathsf M_n\) is positive
definite, which gives a positive real part for every eigenvalue of
\(\mathsf M_n\). Similarity transfers positive stability to
\(D_n^{-1}G_n(\widehat\beta_n)\). In particular,
\(G_n(\widehat\beta_n)\) is nonsingular.

\emph{(IV) Pilot existence and approximation.}
Define
\(f_i(\beta,a)=a^{-1}\log(1+a(U_i(\beta)-1))\) for \(a\ne0\)
and \(f_i(\beta,0)=U_i(\beta)-1\).
On a compact neighborhood of \(\widehat\beta_n\), all
\(U_i(\beta)\) and their derivatives are bounded. For sufficiently
small \(|a|\), every logarithm argument is greater than \(1/2\).
The representation
\(f_i(\beta,a)=\int_0^{U_i(\beta)-1}(1+av)^{-1}\,dv\)
therefore gives a smooth extension through \(a=0\), including
observations with \(Y_i=0\).
Define \(M(\beta,a)=\Pn Z_if_i(\beta,a)\).
Then \(M(\beta,0)=m_n(\beta)\),
\(M(\beta,1-\lambda)=m_{\lambda,n}(\beta)\), and
\(\partial_\beta M(\widehat\beta_n,0)=-G_n(\widehat\beta_n)\)
is nonsingular by Step~(III).

The extension is defined on an open neighborhood of
\((\widehat\beta_n,0)\), so
\citet[Corollary~6.7.13]{corbae_stinchcombe_zeman2009} gives a continuously differentiable
branch \(\beta(a)\) satisfying \(M(\beta(a),a)=0\) and
\(\beta(0)=\widehat\beta_n\) for small \(|a|\).
Its derivative
\(\beta'(a)=-(\partial_\beta M)^{-1}\partial_aM\)
is bounded on a sufficiently small closed interval.
Integration therefore gives
\(\norm{\beta(a)-\widehat\beta_n}_*\leq C_n|a|\)
in the norm used in Step~(V), after enlarging \(C_n\) if necessary.
For small positive \(a\), set
\(\widehat\beta_{\lambda,n}=\beta(1-\lambda)\).
Step~(II) makes each root globally unique and reachable, since the
weighted condition holds for these \(\lambda\)'s.

\emph{(V) Exact completion.}
Let \(A_n=D_n^{-1}G_n(\widehat\beta_n)\) and
\(B_n=I-a_{\mathrm{step}}A_n\).
For each eigenvalue \(z\) of \(A_n\), the admissible step gives
\(\lvert1-a_{\mathrm{step}}z\rvert^2
=1-2a_{\mathrm{step}}\operatorname{Re}(z)
+a_{\mathrm{step}}^2|z|^2<1\).
Thus \(B_n\) has spectral radius below one.
By \citet[Lemma~5.6.10 and Theorem~5.6.7]{horn_johnson2013},
there is an induced norm \(\norm{\cdot}_*\) in which
\(\norm{B_n}_*<1\).
Continuity gives a closed ball of some radius \(r_n>0\) about
\(\widehat\beta_n\) on which the exact derivative norm is bounded
by a fixed \(q<1\). Segment integration and the fixed-point property
make this ball invariant. The same contraction theorem as in
Step~(II) then gives geometric convergence of Phase 2.

Finally,
\(\norm{b-\widehat\beta_n}_*
\leq\norm{b-\widehat\beta_{\lambda,n}}_*
+\norm{\widehat\beta_{\lambda,n}-\widehat\beta_n}_*
\leq\eta+C_n(1-\lambda)\).
The stated inequality places \(b\) inside that exact-completion
ball. For a sufficiently advanced fixed \(\lambda\), Step~(II)
can make \(\eta\) arbitrarily small from any finite start.
Locally, nonsingularity of the pilot Jacobian also bounds root
distance by a constant times the normalized moment residual.
Thus sufficiently accurate Phase 1 stopping provides the required
initialization.
\end{proof}

\subsubsection{The local shifted-log GMM pilot}
\label{app:iv-gmm-pilot-proof}
\begin{samepage}
\noindent The following proof proceeds in five steps.
We first construct the stationary branch and verify that it consists
of strict local minima. A bound on its displacement then proves the
initialization assertion for the specified solver. The final step
explains its agreement with the moment-root construction when both
propositions apply.
\begin{enumerate}[label=\textup{(\Roman*)},leftmargin=*,itemsep=1pt,topsep=3pt,parsep=0pt]
\item \emph{Stationary branch.} Use the smooth moment extension to
write pilot and exact stationarity as one equation. Its coefficient
derivative is the stipulated positive-definite full Hessian, so the
implicit-function theorem gives a locally unique stationary branch.
\item \emph{Strict local minima.} Continuity preserves a positive
lower eigenvalue bound near the branch. The corresponding quadratic
increase of the criterion verifies that these stationary points are
strict local minima, not merely solutions of the first-order equation.
\item \emph{Branch displacement.} The same Hessian bound controls the
inverse in the derivative of the stationary branch. Integrating this
bounded derivative gives the order \(1-\lambda\) approximation needed
for initialization.
\item \emph{Initialization.} Adding the computed pilot's error to this distance places it
inside the specified exact solver's assumed convergence neighborhood
under the stated inequality.
\item \emph{Coinciding branches.} Under exact identification, nearby stationarity implies zero
moments, so local uniqueness makes this branch coincide with the
I2SLS pilot-root branch when both results apply.
\end{enumerate}
\end{samepage}

\begin{proof}[Proof of Proposition~\ref{prop:iv-pilot-initialization}]
\emph{(I) Stationary branch.} The smooth extension from the proof of
Proposition~\ref{prop:iv-i2sls-two-phase} uses only the finite sample
and requires neither exact identification nor a condition on \(D_n\).
With
\(M(\beta,a)=\Pn Z_if_i(\beta,a)\) as there, define
\(J(\beta,a)=M(\beta,a)'W_nM(\beta,a)/2\).
At \(a=0\), these are the exact moments and criterion.
Define \(G(\beta,a)=-\partial M(\beta,a)/\partial\beta'\).
Because \(W_n\) is held fixed,
\(\nabla_\beta J(\beta,a)=-G(\beta,a)'W_nM(\beta,a)\).
At \((\widehat\beta_n,0)\) this gradient vanishes by stationarity.
Its coefficient derivative is the full Hessian
\(\mathcal H_n=\nabla^2J_n(\widehat\beta_n)\), positive definite
by hypothesis. For a direction \(v\in\R^K\),
\(dG_n(\beta)[v]=-\Pn Z_iU_i(\beta)(X_i'v)X_i'\), so differentiating
\(-G_n'W_nm_n\) gives
\(\nabla^2J_n(\beta)=G_n(\beta)'W_nG_n(\beta)
+\Pn U_i(\beta)(Z_i'W_nm_n(\beta))X_iX_i'\).
The second term need not vanish at an overidentified stationary
point, so full column rank of \(G_n\) alone does not replace the
Hessian hypothesis. This smooth stationarity equation is defined on
an open neighborhood, vanishes at \((\widehat\beta_n,0)\), and has
invertible coefficient derivative \(\mathcal H_n\). Thus
\citet[Corollary~6.7.13]{corbae_stinchcombe_zeman2009} gives a
continuously differentiable stationary branch \(\beta(a)\) through
\(\widehat\beta_n\).

\emph{(II) Strict local minima.} By continuity, the Hessian differs
from \(\mathcal H_n\) in operator norm by less than half its smallest eigenvalue
on a smaller convex neighborhood and an interval of \(a\)'s.
\citet[Corollary~4.3.15]{horn_johnson2013} therefore gives
\(\nabla_\beta^2J(\beta,a)\succeq cI\) for some \(c>0\).
Shrink the interval so that the branch remains inside this neighborhood.
Strict convexity makes its stationary point unique there.
For sufficiently small \(d\), the integral Taylor formula at a branch
point gives
\(J(\beta(a)+d,a)-J(\beta(a),a)
=\int_0^1(1-t)d'\nabla_\beta^2J(\beta(a)+td,a)d\,dt
\geq(c/2)\norm{d}_2^2\).
Thus every point on the branch is a strict local minimum.

\emph{(III) Branch displacement.}
The implicit derivative is
\[
 \beta'(a)=-(\nabla_\beta^2J(\beta(a),a))^{-1}
 \partial_a\nabla_\beta J(\beta(a),a).
\]
Step~(II) bounds the Hessian inverse by \(c^{-1}\) in Euclidean
operator norm. The remaining derivative is continuous and bounded on
a smaller closed interval along the branch, so
\(\sup_a\norm{d\beta(a)/da}\leq C_n<\infty\).
Integrating it from zero to \(a\) gives
\(\norm{\beta(a)-\widehat\beta_n}\leq C_n|a|\).
For \(a=1-\lambda>0\), this is the claimed pilot approximation.

\emph{(IV) Initialization.} If the computed pilot \(b_\lambda\) is within \(\eta\) of
\(\beta(1-\lambda)\), the triangle inequality gives
\(\norm{b_\lambda-\widehat\beta_n}\leq
\eta+C_n(1-\lambda)\). If this upper bound is less than \(r_n\),
the computed pilot belongs to the specified solver's assumed
convergence neighborhood, so its exact-completion iterates converge to
\(\widehat\beta_n\). Neither
the construction of this branch nor the inequality proves that an
arbitrary original start finds it.

\emph{(V) Coinciding branches.} Under exact identification, a regular root has Hessian
\(G_n'W_nG_n\succ0\). Continuity also makes \(G(\beta,a)\) nonsingular near
\((\widehat\beta_n,0)\). Hence \(G_{\lambda,n}'W_n\) is nonsingular
there, and pilot stationarity
\(G_{\lambda,n}'W_nm_{\lambda,n}=0\) implies \(m_{\lambda,n}=0\).
Both this branch and that in the proof of
Proposition~\ref{prop:iv-i2sls-two-phase} are then locally unique roots
of \(M(\cdot,a)\) through \(\widehat\beta_n\), so they coincide for
small \(a\).
\end{proof}

\clearpage
\bibliographystyle{bbpplainnat}
\bibliography{references,references_additions}
\end{document}